\documentclass[a4paper,11pt]{article}
\pdfoutput=1 

\usepackage{jheppub} 

\usepackage[T1]{fontenc} 

\usepackage{amsmath}
\usepackage{amsfonts}
\usepackage{amssymb}
\usepackage{bbm}

\usepackage{cancel}
\usepackage[normalem]{ulem}

\usepackage{hyperref}

\usepackage{graphicx} 
\graphicspath{{figures/}}

\usepackage{xcolor}
\usepackage{tikz}
\usepackage{caption}
\usepackage{subcaption}

\newcommand{\bJ}{{\mathbb{J}}}
\newcommand{\bR}{{\mathbb{R}}}
\newcommand{\bZ}{{\mathbb{Z}}}
\newcommand{\cD}{{\mathcal{D}}}
\newcommand{\cN}{{\mathcal{N}}}
\DeclareMathOperator{\Tr}{Tr}
\DeclareMathOperator{\sign}{sgn}
\newcommand{\vev}[1]{\left\langle #1\right\rangle}

\title{\bf Twisted times, the Schwarzian and its deformations in DSSYK}

\author[a]{Micha Berkooz}
\author[a]{, Ronny Frumkin}
\author[b,c]{, Ohad Mamroud}
\author[a]{, Josef Seitz}

\affiliation[a]{Department of Particle Physics and Astrophysics, Weizmann Institute of Science, Rehovot
7610001, Israel}
\affiliation[b]{SISSA, via Bonomea 265, 34136 Trieste, Italy}
\affiliation[c]{INFN, Sezione di Trieste, Via Valerio 2, 34127 Trieste, Italy}

\emailAdd{micha.berkooz@weizmann.ac.il}
\emailAdd{ronny.frumkin@weizmann.ac.il}
\emailAdd{omamroud@sissa.it}
\emailAdd{josef-emanuel.seitz@weizmann.ac.il}

\abstract{The IR dynamics of SYK is that of the Schwarzian theory, the effective theory of broken reparametrization invariance. In the double scaling limit, SYK is completely solvable by chord diagrams, whose generating functional is a bilocal Liouville theory. At low temperatures a set of modes in this description becomes soft. We interpret them as reparametrization of some twisted time coordinates, and show explicitly that they lead to the nonlinear Schwarzian theory. We further consider deformations of the theory in the double scaling limit, giving rise to diagrams with multiple species of chords, and show that the generating functional is now a Liouville theory with multiple fields. These deformations can be tracked to the IR and we discuss how they affect the Schwarzian.}

\begin{document} 
\maketitle
\flushbottom

\section{Introduction}

The Sachdev-Ye-Kitaev (SYK) model \cite{sachdev1993, sachdev2010,kitaev2015simple, Maldacena:2016hyu} is a disordered, chaotic \cite{kitaev2015simple,Maldacena:2015waa, Maldacena:2016hyu, garcia2016,Cotler:2016fpe,you2017sachdev}, many--body quantum mechanical model of $N$ interacting Majorana fermions. In the large $N$ limit, the IR of the model becomes strongly coupled and can be described by the Schwarzian theory -- the effective theory of the broken reparametrization symmetry of the thermal circle. This effective description has been instrumental in learning about holography from SYK \cite{kitaev2015simple, Maldacena:2016upp, Cotler:2016fpe, Saad:2018bqo, Maldacena:2018lmt, Goel:2018ubv, Jensen:2016pah, Polchinski:2016xgd, sachdev2010,Sarosi:2017ykf,Blommaert:2024whf,Blommaert:2024ydx}, as the Schwarzian also describes the (perturbative) dynamics of JT gravity in two dimensional anti-de Sitter.

The emergence of the Schwarzian theory from the IR of SYK had been established by comparing the predictions of the two theories for various observables: the partition function, thermal correlators, and out-of-time-ordered correlators. 
Beyond comparisons, one can try and derive it from more complete descriptions such the $G\Sigma$ approach. This is only partially doable, as it requires some arbitrary non-local regularization and numerics \cite{Sarosi:2017ykf,Trunin:2020vwy}. In the large $p$ limit (where $p$ is the number of interacting fermions in each term of the Hamiltonian) the situation is better in general, and one of the goals of this paper is to fully derive the Schwarzian from the Liouville $G\Sigma$ action.  

There is a related limit, the double scaling limit, where one keeps $\lambda = 2p^2/N$ fixed \cite{Cotler:2016fpe,erdHos2014phase,Berkooz:2018jqr,Berkooz:2018qkz} while taking $N\to\infty$. In the double scaling limit, the SYK model becomes solvable, as it is described by a diagrammatic expansion in chord diagrams, and its dynamics are governed by a new effective Hamiltonian acting on the chord Hilbert space \cite{Berkooz:2018jqr,Berkooz:2018qkz} (see also the review \cite{Berkooz:2024lgq}). In the triple scaling limit ($\lambda \to 0$ while focusing on the IR), one can show that this Hamiltonian becomes exactly the one that governs the Schwarzian theory \cite{Berkooz:2018jqr,Berkooz:2018qkz,Lin:2022rbf}. Yet, as it employs the canonical formalism, its relation to reparametrizations is not immediately evident. It also provides a less convenient starting point to study deformations of the model and understand how they affect the IR.

The current work starts with a UV complete path integral description of double scaled SYK (DSSYK), identifies the reparametrization modes, and derives the Schwarzian path integral from them. This description is the bi--local Liouville theory\footnote{This action is formally unbounded from below, and one might worry it is ill defined. One can find a different action that is well defined and has the same saddle point expansion that we will use here \cite{Berkooz:2024evs,Berkooz:2024ofm}, however here we will just stick to the Liouville and assume that these illnesses are cured by a normalization condition regarding the size of the Hilbert space, as signaled by $\eta$ in \eqref{eq:2D Liouville action0}.} \cite{Cotler:2016fpe,Stanford-talk-kitp,Lin:2023trc},
\begin{equation}\label{eq:2D Liouville action0}
    Z = \eta \int Dg \, e^{-I[g]} \,, \quad I =\frac{1}{4\lambda}\int_{0}^{\beta}d\tau_{1}\int_{0}^{\beta}d\tau_{2} \left[\frac{1}{2}\partial_{\tau_{1}}g\partial_{\tau_{2}}g - 2\bJ^{2}e^{g}\right] \,, \quad g(\tau,\tau) = 0  \,,
\end{equation}
with the boundary conditions specified explicitly, and where $\eta$ is a normalization constant that enforces the normalization of the trace, $\Tr(\mathbbm{1}) = 1$. The perturbative expansion around\footnote{Rescaling the integrals in the action to be $\beta$-independent, we see that the proper expansion parameter is $\beta \bJ$, and the described expansion is simply a high temperature expansion.} $\bJ = 0$ generates the chord diagrams order by order \cite{Stanford-talk-kitp,Lin:2023trc}. This action can also be derived from the $G\Sigma$ formalism in the large $p$ limit \cite{Cotler:2016fpe,Stanford-talk-kitp,Goel:2023svz}, and when $\lambda \to 0$ the two limits agree.

Observables in SYK are readily computable in this framework. For example, two-point functions of operators with conformal dimension\footnote{For the original fermions of the model, $\Delta = 1/p$.} $\Delta$ are given by $\langle e^{\Delta g(\tau_1,\tau_2)}\rangle$. We therefore expect that reparametrizations $\tau \mapsto \phi(\tau)$ act as 
\begin{equation}
    g(\tau_1,\tau_2) \mapsto g(\phi(\tau_1),\phi(\tau_2)) + \log(\phi'(\tau_1)\phi'(\tau_2)) \qquad \text{(naive).}    
\end{equation} 
These are the reparametrizations that are used as a starting point for the effective Schwarzian description of \cite{Maldacena:2016hyu,Maldacena:2016upp,kitaev2015simple,Kitaev:2017awl}. Unfortunately, from the get-go they violate the boundary conditions, and so they are not allowed modes in the full theory at all. If one insists on using them, one needs to resort to using the arbitrary regulator mentioned before.

To find the actual reparametrizations we need to concentrate on a saddle point of the model. In Section~\ref{sec:quick review} we will briefly review some known facts about DSSYK, the bilocal Liouville formulation, and the Schwarzian theory. In particular, we review the known saddle point in the $\lambda \to 0$ limit, which reproduces the two-- and four--point functions of SYK \cite{Maldacena:2016upp,Streicher:2019wek,Choi:2019bmd,Goel:2023svz,Mukhametzhanov:2023tcg,Okuyama:2023bch}. We also compute the partition function in the Liouville theory to 1--loop order.

In Section~\ref{sec: Schwarzian computation} we will see that when $\lambda \to 0$ and at low temperatures the tension between the boundary conditions (coming from short timescales) and the naive reparametrization invariance of \eqref{eq:2D Liouville action} leads to the emergence of the Schwarzian action. This is expected, but it happens in a peculiar way. One can define twisted reprameterizations -- soft modes that are contained in the theory and do act as arbitrary reparametrizations, $s_i\rightarrow \phi(s_i)$-- of some twisted coordinates
\begin{equation} \label{eq: intro twisted coordinates}
    \begin{split}
        s_1 &= \frac{1}{2}\left((1+v) \tau_1 + (1-v)\tau_2 +\frac{\beta(1-v)}{2}\right)  \\ 
    s_2 &= \frac{1}{2}\left((1-v) \tau_1 + (1+v)\tau_2 - \frac{\beta(1-v)}{2}\right) ,
    \end{split}
\end{equation}
where $v$ is related to the temperature. The action of these modes will be the Schwarzian, and thus we identify it within the full theory. The use of these coordinates also has a physical implication for the Schwarzian -- the bulk distance between any two boundary points is always positive, which is similar to what happens in the fake disk \cite{Lin:2023trc} and to a proposed holographic dual of DSSYK \cite{Blommaert:2024whf}. We discuss it in Section \ref{sec: gen functionals}. 

There are also other saddles\footnote{At the final stages of this work, discussion of these saddles appeared in \cite{Blommaert:2024whf} in the context of the sine dilaton theory dual to DSSYK.} for the bilocal Liouville besides the one discussed above, and we turn to them in Section~\ref{sec: other saddles}. The IR theory describing them is a Schwarzian with large conical defects, effectively looking at reparametrizations that wind around the thermal circle several times (the corresponding contributions in the exact DSSYK are reviewed in Appendix~\ref{app:Other saddles from DSSYK}).

Now that we have a UV formulation from which we can directly derive the IR theory, we can start deforming our model. A very general class of deformations is by random polarized operators \cite{Berkooz:2024ofm,Berkooz:2024evs}. Their form can be similar to that of the SYK Lagrangian (but with different, independent random couplings), or they could preserve additional symmetries, see Section~\ref{sec: multifield Liouville}. Their dynamics are also described by chord diagrams, but now with multiple types of chords -- one for each set of independent couplings \cite{Berkooz:2024ofm,Berkooz:2024evs}.

In Section~\ref{sec: multifield Liouville} we argue that the dynamics of all these multi-chord models whose Hamiltonian is $H = \sum_i \kappa_i H_i$, for some independent random Hamiltonians $H_i$, are described by a multi--Liouville theory,
\begin{equation}
    \int \left(\prod_{i=1}^K \mathcal{D}[ g_{i}]\right) \exp\left[\frac{1}{4\lambda}\int_0^\beta d\tau_{1}\int_0^{\beta}d\tau_{2}\left[\sum_{i,j=1}^K\frac{1}{2}\alpha_{ij}g_{i}\partial_{1}\partial_{2}g_{j} + \sum_{i=1}^K 2 \kappa_{i}^2 \bJ^2 e^{\sum_{j=1}^K\alpha_{ij}g_{j}}\right] \right] \,,
\end{equation}
where $K$ is the number of types of chords, and $e^{-\lambda \alpha_{ij}}$ are the weights of intersections of different chords.

In Section \ref{generic deformations}, we then focus on a few specific examples for deformations and use the multi--Liouville formalism to track the effect of such deformations into the IR. In Section \ref{subsec: spectrum at kappa^2}, we focus on effects at lowest order in the deformation. We confirm that at the quadratic level \cite{Maldacena:2016upp} the ostensibly irrelevant deformations of mass dimension $1<\Delta < 3/2$ lead to a nonlocal theory of reparametrizations. However, we can control the endpoint of the flow and for example show that the spectrum is bounded from below. We also show that deformations of $\Delta > 3/2$ modify the Schwarzian coefficient of the IR theory. In Section \ref{subsec: the role of gz}, we highlight some features of the next to leading order in the deformation. For example, we show that at this order, generic deformations lead to contact 4-point interactions whose strength depend on the multi--chord intersections.

\textbf{Note:} As we finalized this work we became aware of \cite{MezeiBucca}, which provided two derivations of the Schwarzian from the bilocal Liouville, one of which is morally similar to the one given in our Section~\ref{sec: Schwarzian computation}. They do not use the twisted coordinates and deal with the boundary conditions in a different way.

\section{DSSYK, bilocal Liouville, and the Schwarzian}
\label{sec:quick review}
In this Section we summarize known results and conventions from the literature regarding DSSYK, the bilocal Liouville formulation, and the Schwarzian theory, focusing on the chord perspective of DSSYK \cite{Berkooz:2018jqr,Berkooz:2018qkz,Lin:2022rbf,Berkooz:2024ofm,Berkooz:2024lgq,Lin:2023trc,Berkooz:2024evs}. The reader familiar with the subject can skip to the next section. In Section~\ref{sec:Liouville saddle point} we place some emphasis on the saddle point and 1-loop for the bilocal Liouville, and compute the partition function in that language. 

The SYK model is a disordered quantum mechanical model of $N$ Majorana fermions $\psi_i$, $\{\psi_i,\psi_j\}=2\delta_{ij}$, with all-to-all, random, $p$-local interactions between them,
\begin{equation}
    \label{eq:DSSYK Hamiltonian}
    H = i^{p/2} \sum_{1\le i_1 < \cdots < i_p \le N} J_{i_1, \cdots , i_p} \psi_{i_1}\cdots \psi_{i_p} .
\end{equation}
The couplings $J_{i_1,\cdots,i_p}$ are random Gaussian variables,
\begin{equation}
    \vev{J_{i_1,\cdots,i_p}}_J=0,\qquad \vev{J_{i_1,\cdots,i_p} J_{j_1,\cdots,j_p}}_J = \frac{\bJ^2}{\lambda} {N \choose p}^{-1} \delta^{i_1}_{j_1} \cdots \delta^{i_p}_{j_p} \, , \qquad \lambda \equiv \frac{2p^2}{N} ,
\end{equation}
where $\langle\cdots\rangle_J$ stands for ensemble average. The trace is normalized such that 
\begin{equation}
    \label{eq:trace normalization}
    Z(0) = \Tr (\mathbbm{1}) = 1 \,,
\end{equation} 
and our conventions imply ${\left\langle\Tr\left(H^2\right)\right\rangle = \frac{\bJ^2}{\lambda}}$. 

The observables in the model are annealed average quantities, such as the partition function $Z(\beta) = \langle\Tr\left(e^{-\beta H}\right)\rangle_J$. They are amenable to a diagrammatic expansion using \emph{chord diagrams}, which capture the pattern of Wick contractions between the $J$'s. We are interested in the double scaling limit, in which $N,p\to\infty$ while $\lambda$ is kept fixed. In that limit, the contribution of each diagram can be found combinatorially  \cite{erdHos2014phase,Cotler:2016fpe,Berkooz:2018jqr,Berkooz:2018qkz,Garcia-Garcia:2018fns,Garcia-Garcia:2017pzl} (see the review \cite{Berkooz:2024lgq} for further details). The partition function, for example, can then be written as a sum over all chord diagrams, where each intersection between the chords is weighted by $e^{-\lambda}$,
\begin{equation}
    \label{eq:chord expansion}
    Z(\beta) = \sum_{k=0}^\infty \frac{(-\beta)^{2k}}{(2k)!} \left( \frac{\bJ^2}{\lambda} \right)^k \sum_{\substack{\text{chord diagrams} \\ {\text{with $k$ chords}}}}  e^{-\lambda \cdot \text{\# intersections}} .
\end{equation}

\begin{figure}[t]
    \centering
    \includegraphics[scale=0.25]{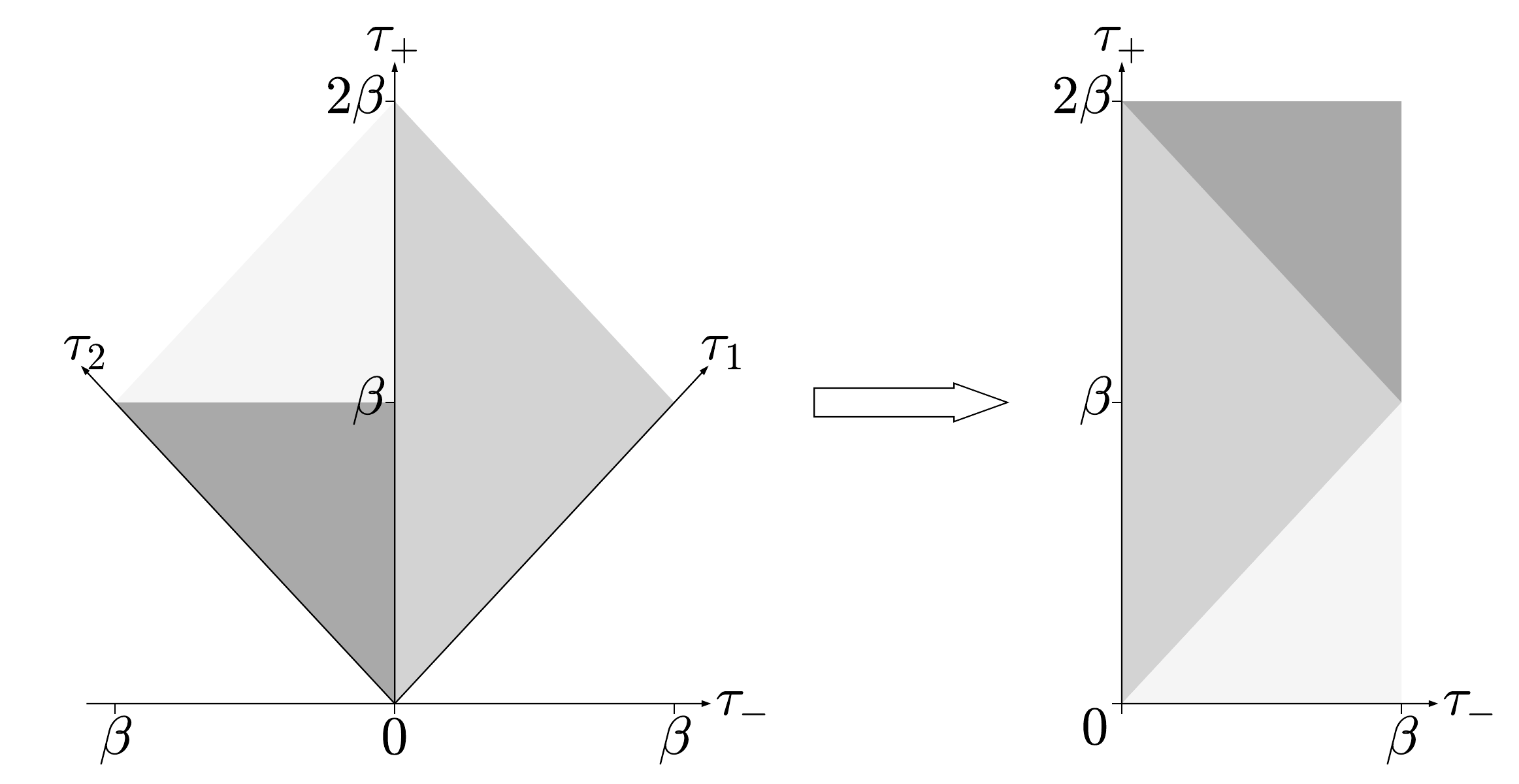}
    \caption{Integration region reordering.}
    \label{fig:integ-reg-reorder2}
\end{figure}

DSSYK can also be analyzed in the $G\Sigma$ approach \cite{Maldacena:2016hyu}, where the dynamical variable is the bi-local field $G(\tau_1,\tau_2) = \frac{1}{N} \sum_i \psi_i(\tau_1)\psi_i(\tau_2)$. In the double scaling limit the theory can be written in terms of a new symmetric bi-local field $g$, defined as $G(\tau_1,\tau_2)=\sign(\tau_1-\tau_2)e^{g(\tau_1,\tau_2)/p}$ \cite{Cotler:2016fpe,Stanford-talk-kitp}
\begin{equation}\label{eq:2D Liouville action}
    Z = \eta \int Dg \, e^{-I[g]} \,, \qquad I =\frac{1}{4\lambda}\int_{0}^{\beta}d\tau_{1}\int_{0}^{\beta}d\tau_{2} \left[\frac{1}{2}\partial_{\tau_{1}}g\partial_{\tau_{2}}g - 2\bJ^{2}e^{g}\right] ,
\end{equation}
with a flat integration measure over $g$. $\eta$ is a normalization constant\footnote{We could have also absorbed it by redefining the measure.}, determined by \eqref{eq:trace normalization}. We will write its explicit form below. The boundary conditions on $g$ are
\begin{equation}
\label{eq:boundary conditions g}
    \forall\tau : \qquad g(\tau,0)=g(\tau,\beta), \qquad g(0,\tau)=g(\beta,\tau), \qquad g(\tau,\tau) = 0 \,,
\end{equation}
where the first two originate from the periodicity of $g$ and the third from the fact that for every Majorana fermion $\psi_i(\tau)\psi_i(\tau) = 1$. Later, we will find it convenient to work with the variables
\begin{equation} \label{eq: tau_- and tau_p definition}
    \tau_\pm = \tau_1 \pm \tau_2, \, \qquad \tau_- \in [0,\beta], \quad \tau_+ \in [0,2\beta] \,,
\end{equation}
where we used the periodicity of the field $g(\tau_1,\tau_2)$ to rearrange the integration region in a rectangle, as shown in Figure~\ref{fig:integ-reg-reorder2}. 
In these variables the boundary conditions for $g(\tau_-,\tau_+)$ are
\begin{equation} \label{eq:boundary conditions g after rearange}
     g(\tau_-,\tau_+) = g(\beta-\tau_-,\tau_+ - \beta), \quad g(\tau_-,0) = g(\tau_-,2\beta),
        \quad  g(0,\tau_+) = g(\beta,\tau_+) = 0 \, .
\end{equation}

This $G\Sigma$ (or bi-local Liouville) description is intimately related to the chord picture, as can be seen by expending around $\bJ = 0$ (or equivalently, at high temperatures), where the $k$-th term in the expansion generates all chord diagrams with $k$ chords, reproducing \eqref{eq:chord expansion} \cite{Stanford-talk-kitp} (see Appendix H of \cite{Lin:2023trc} for the full derivation). Ultimately, this is because the free propagator is 
\begin{equation}
    \label{eq:high temp propagator} \vev{g(\tau_1,\tau_2)g(\tau_3,\tau_4)}_{\bJ=0} = \begin{cases}
        -\lambda & (\tau_1,\tau_2) \text{ crosses } (\tau_3,\tau_4) \,,\\
        0 & \text{otherwise,}
    \end{cases}
\end{equation}
and it indicates whether the chord connecting the points $\tau_1,\tau_2$ on the thermal circle crosses the chord connecting $\tau_3,\tau_4$. In this way, the field $g(\tau_1,\tau_2)$ can be though of as counting the number of chords crossing the line connecting $\tau_1$ with $\tau_2$ \cite{Berkooz:2024ofm,Berkooz:2024evs}. We will use this idea in Section~\ref{sec: multifield Liouville}, although in the rest of the paper we will primarily focus on $\lambda\rightarrow 0$.

\subsection{The Schwarzian theory} \label{subsec:the_schwarzian theory}
The Schwarzian theory is that of broken reparametrizations,
\begin{equation}
    \label{eq:Schwarzian path integral}
    Z_{\rm Sch} = \int \cD \phi \, \exp\left[ C\int_0^\beta d\tau\left\{\tan\left(\frac{\pi \phi(\tau)}{\beta}\right),\tau\right\}\right] \,,
\end{equation}
where $\{f(\tau),\tau\} = \frac{f'''}{f'} - \frac{3}{2}\left(\frac{f''}{f'}\right)^2$ is the Schwarzian derivative and $C$ is an inverse coupling constant. The path integration is over $\text{Diff}(S^1)/SL(2,\bR)$, i.e. all reparametrizations $\phi(\tau)$ that increase monotonically and wind once around the thermal circle, $\phi(\tau + \beta) = \phi(\tau) + \beta$, moded out by $SL(2,\bR)$ transformations that act as $\tan(\frac{\pi \tau}{\beta}) \mapsto \frac{a \tan(\frac{\pi \tau}{\beta}) + b}{c \tan(\frac{\pi \tau}{\beta}) + d}, \, ad-bc=1$. The integration measure is $\cD \phi = \prod_\tau \frac{d\phi}{\phi'(\tau)}$ \cite{Bagrets:2016cdf}. For infinitesimal fluctuations around the saddle, $\phi(\tau) = \tau + \varepsilon(\tau)$, the measure can be expressed in terms of the Fourier modes of $\varepsilon$ \cite{Stanford:2017thb,Moitra:2021uiv,Mertens:2022irh}
\begin{equation}
    \cD\phi = \prod_{n \ge 2} 4\pi (n^3-n)d\varepsilon_n d\varepsilon_{-n} \,, \qquad \varepsilon(\tau) = \frac{\beta}{2\pi} \sum_{|n|\ge 2} \varepsilon_n e^{-2\pi in \tau/\beta} \,,
\end{equation}
and the quadratic action is 
\begin{equation}
    \label{eq:Schw quadratic action}
    -\frac{2\pi^2 C}{\beta} + \frac{4\pi^2 C}{\beta}\sum_{n\ge2} n^2(n^2-1)\bar\varepsilon_n \varepsilon_n.
\end{equation}
The Schwarzian is 1-loop exact \cite{Stanford:2017thb}, giving rise to the partition function
\begin{equation}
    \label{eq:Schwarzian partition func}
    Z_{\rm Sch} 
    = e^{-\beta E_0 +\frac{2\pi^2 C}{\beta}} \prod_{n\ge 2} \frac{\beta}{C n} 
    = a_0 \left(\frac{C}{\beta}\right)^{3/2} e^{-\beta E_0 + \frac{2\pi^2 C}{\beta}} \,.
\end{equation}
where $E_0$ is the ground state energy, and $a_0$ is some scheme dependent constant that is independent of $\beta$.

The Schwarzian theory shows up in DSSYK and describes its semi-classical ($\lambda \to 0$) low temperature ($\beta\bJ \to \infty$) limit, where the exact partition function \eqref{eq:chord expansion} becomes \cite{Cotler:2016fpe,Berkooz:2018jqr,Berkooz:2018qkz,Berkooz:2024lgq}
\begin{equation}
    \label{eq:DSSYK low temp Z}
    Z(\beta) = \frac{\pi \sqrt{2}}{(\beta\bJ)^{3/2}}\exp\left[\left(\frac{2}{\lambda} + \frac{1}{2}\right)\beta\bJ - \frac{\pi^2}{2\lambda} + \frac{\pi^2}{\lambda \beta\bJ}\right] \left(1 + O(\lambda, 1/\beta\bJ)\right)\,.
\end{equation}
The power law and the last term in the exponent are the typical signatures of the Schwarzian, with $C = \frac{1}{2\lambda \bJ}$, the ground state energy ${E_0 = -\frac{2\bJ}{\sqrt{\lambda(1-e^{-\lambda})}} \approx -\left(\frac{2}{\lambda} + \frac{1}{2}\right)\bJ}$, and $a_0 = 2\pi \lambda^{3/2} e^{-\frac{\pi^2}{2\lambda}}$, all determined by the UV theory.

\subsection{Saddle point in bilocal Liouville}
\label{sec:Liouville saddle point}
Let us now see how to get these results from the bilocal Liouville theory, which we will analyze in the semi-classical limit $\lambda \to 0$ throughout this and the next section. The action is dominated by the saddle \cite{Maldacena:2016hyu}
\begin{equation} \label{eq:saddle point g}
    g_0(\tau_1,\tau_2) = 2 \log \left[ \frac{\cos \left(\frac{\pi v}{2}\right)}{\cos (x)} \right], \qquad \frac{\pi v}{\cos \left(\frac{\pi v}{2}\right)} = \beta \bJ \, , \qquad  x \equiv \pi v\left(\frac{1}{2} - \frac{|\tau_1-\tau_2|}{\beta}\right) ,
\end{equation}
whose action is
\begin{equation} \label{eq: action in the saddle}
    I\left[g_{0}\right] = -\frac{1}{\lambda}\left[2\pi v\tan\left(\frac{\pi v}{2}\right) - \frac{\pi^2 v^2}{2}\right] .
\end{equation}
We will use $v\in[0,1]$ interchangeably with the temperature from now on, where $v\to 1$ ($v\to0$) corresponds to low (high) temperatures. In fact, there is a family of such saddles, coming from different $v$'s solving the second equation, as discussed in Section~\ref{sec: other saddles}. At any finite $v$ the two- and four-point functions can be derived via saddle point techniques, and their low temperature behavior matches that of the Schwarzian theory \cite{Maldacena:2016hyu,Okuyama:2023bch,Goel:2023svz,Mukhametzhanov:2023tcg,Choi:2019bmd,Streicher:2019wek}. Here we will concentrate on the low temperature limit of the partition function, which requires some extra care.

We now turn to the fluctuations around the saddle \eqref{eq:saddle point g}, $g = g_0 + \delta g$. To leading order in $\lambda$, it is enough to consider the quadratic Lagrangian,
\begin{equation}
    I_{\rm quad} = \int_{-\frac{\pi v}{2}}^{+\frac{\pi v}{2}} \frac{dx}{v} \int_0^{2\pi} dy \, \frac{1}{16\lambda} \delta g\left[-\partial_y^2 + v^2 \partial_x^2 - \frac{2v^2}{\cos^2(x)}  \right] \delta g \,, \qquad y = \pi\tau_+ / \beta \,.
\end{equation}
It can be diagonalized by solving for the eigenfunctions $\Psi_{n,k}(x,y)$ and eigenvalues $\lambda_{n,k}$ of the differential operator \cite{Choi:2019bmd} 
\begin{subequations} \label{eq: quadratic fluctuations}
\begin{gather}
    \Psi_{n,k}(x,y) = \begin{cases}
        \cN_{n,k} e^{iny} \left(\cos(k x)\tan(x) - k \sin(kx)\right)\, & n \in 2\bZ+1\,, \\
        \cN_{n,k} e^{iny} \left(\sin(k x)\tan(x) + k \cos(kx)\right)\, & n \in 2\bZ\,,
    \end{cases} \\ 
    \lambda_{n,k} = \frac{n^2 - v^2 k^2}{16\lambda} \,,
\end{gather}    
\end{subequations}
and $\cN_{n,k}$ are normalizations that guarantee that the functions are orthonormal. By convention, we choose the $k$'s to be positive, as flipping their sign corresponds to the same eigenfunction. Each eigenvalue appears once for $|n|=0$ and twice for $|n|\neq 0$. The exact values of $k$ are determined by the boundary conditions $\Psi_{n,k}\left(x=\pm\frac{\pi v}{2},y\right)=0$,
\begin{equation}\label{eq:constraints on k}
    \begin{split}
        \cos(\tfrac{k \pi v}{2})\tan(\tfrac{\pi v}{2}) &= \phantom{-}k \sin(\tfrac{k \pi v}{2}) \,, \qquad n\in 2\bZ + 1\,, \\
    \sin(\tfrac{k \pi v}{2})\tan(\tfrac{\pi v}{2}) &= - k \cos(\tfrac{k \pi v}{2}) \,, \qquad n\in 2\bZ \,.
    \end{split}
\end{equation}
To 1-loop order, the partition function \eqref{eq:2D Liouville action} is therefore
\begin{equation}
    Z(\beta) = e^{-I[g_0(v)]} \eta \prod_{n=-\infty}^\infty \prod_{k(v)} \sqrt{\frac{2\pi}{\lambda_{n,k(v)}(v)}} \,,
\end{equation}
where we explicitly denoted the dependence of the eigenvalues on the temperature $v$. We can now impose \eqref{eq:trace normalization} in order to determine the constant $\eta$ in the Liouville path integral \eqref{eq:2D Liouville action}. Note that when $\beta\bJ = 0$ ($v=0$) the on--shell action \eqref{eq: action in the saddle} vanishes, and so
\begin{equation}
    \label{eq:part function 1 loop schematic}
    \eta = \prod_{n=-\infty}^\infty \prod_{k(0)} \sqrt{\frac{\lambda_{n,k(0)}(0)}{2\pi}} \quad \implies \quad Z(\beta) = e^{-I[g_0(v)]} \prod_{n=-\infty}^\infty  \sqrt{\frac{\prod_{k(0)} \lambda_{n,k(0)}(0)}{\prod_{k(v)}\lambda_{n,k(v)}(v)}}\,.
\end{equation}

We can be slightly more explicit than that. At high temperatures, $\beta\bJ \approx \pi v \ll 1$, the $k$'s that solve the boundary conditions are parameterized by integers\footnote{One might be worried by the existence of an infinite number of tachyonic modes, whenever $m > n$. Should they all be taken into account? The answer is positive. We have argued for the Liouville action based on the high-temp propagator \eqref{eq:high temp propagator}. One can verify (at least numerically) that in order to reproduce it in the theory for the fluctuations, all the modes need to be taken into account.}
\begin{equation}
    \label{eq: high temp k}
    k_m = \frac{m}{v} + \frac{v}{m} + O(v^3) \,, \quad \begin{cases}
        1 < m \in 2\bZ  & n\in 2\bZ + 1 \,, \\
         1 \le m \in 2\bZ+1 & n\in 2\bZ \,.
    \end{cases} 
    \qquad \text{(high temp).}
\end{equation}

As we change $v$, the $k$'s change smoothly. Let us denote by $k_{m}^{e}\left(v\right)$ and $k_{m}^{o}\left(v\right)$ those that are associated with even and odd values of $n$ and approach $\frac{2m-1}{v}$ and $\frac{2m}{v}$ as $v\to 0$, respectively. The exact 1-loop contribution at any finite $v$ is then the product over the eigenvalues times the normalization factor $\eta$, 
\begin{multline}
    \label{eq:finite v 1-loop}
    Z_{\text{1-loop}}(\beta) = \prod_{m=1}^{\infty}\left[\frac{2m-1}{vk_{m}^{e}(v)}\prod_{n=1}^{\infty}\left(\frac{(2n)^2-(2m-1)^{2}}{(2n)^2-(vk_{m}^{e}\left(v\right))^2}\right)\prod_{n=0}^{\infty}\left(\frac{(2n+1)^2-(2m)^2}{(2n+1)^2-(vk_{m}^{o}\left(v\right))^2}\right)\right] \\
     = \prod_{m=1}^{\infty}\left[-\frac{1}{\sin\left(\frac{\pi vk_{m}^{e}\left(v\right)}{2}\right)\cos\left(\frac{\pi vk_{m}^{o}\left(v\right)}{2}\right)}\right] \,.
\end{multline}
where the first product comes from the eigenvalues that appear only once, and we used $\prod_{n=1}^\infty \left(1-\frac{z^2}{n^2}\right) = \frac{\sin(\pi z)}{\pi z}$ and $\prod_{n=0}^\infty \left(1-\frac{z^2}{(n+\frac12)^2}\right) = \cos(\pi z)$.

At low temperatures, 
\begin{equation}
 \delta v \equiv 1 - v \approx \frac{2}{\beta\bJ} \ll 1,   
\end{equation}
the on--shell action of the Schwarzian already appears through the saddle $g_0$ \eqref{eq:saddle point g} of the bilocal Liouville, as its action is  
\begin{equation}
    \label{eq:saddle action low temp}
    I[g_0] = -\frac{2\beta\bJ}{\lambda} + \frac{\pi^2}{2\lambda} - \frac{\pi^2}{\lambda\beta\bJ}+\mathcal{O}( (\beta\bJ)^{-2}) \,.
\end{equation}
The other factors in \eqref{eq:DSSYK low temp Z} come from the 1-loop around that saddle. The boundary conditions \eqref{eq:constraints on k} can be solved perturbatively in $\delta v$ for $k \ll 
\frac{1}{\delta v}$,
\begin{equation}
    \label{eq:low temp k}
    k_m = m + \frac{1}{12}m(m^2-1)\pi^2 (\delta v )^3 + O(\delta v ^4) \,, \quad m-n \in 2\bZ, \; m > 1 \quad \text{(low temp),}
\end{equation}
which give the eigenvalues
\begin{equation}
    \label{eq:low temp eigenvalues}
    \lambda_{n,k_m} = \frac{n^2 - m^2}{16\lambda} + \frac{m^2}{8\lambda}\delta v + O(\delta v^2) \,, \quad n-m \in 2\bZ , \quad m > 1, \qquad \text{(low temp).}
\end{equation}
For the modes with $m = 0, 1$ the relevant eigenfunction vanishes, and so they are not to be taken into account. The modes with $n = m$ are the soft modes, as their eigenvalues are of order $O(1/\beta\bJ)$. We will see below that they are the ones responsible for the Schwarzian behavior. 

Plugging these values of $k$ into \eqref{eq:finite v 1-loop} we find that the 1-loop contribution is
\begin{equation}
    Z_{\text{1-loop}}(\beta) = \prod_{m = 1} \frac{4}{(2m)(2m+1)\pi^2 \delta v^2} = \prod_{n \ge 2} \frac{\beta\bJ}{\pi n}
\end{equation}
which is the 1-loop factor\footnote{Up to an overall $\beta$ independent constant, which we can take to be part of the regularization prescription.} of the Schwarzian theory!

However, note that we did not write an upper bound on this product. In the Schwarzian theory, it should be infinity and we should prescribe a regularization procedure. In the bilocal Liouville theory, though, when $k$ is very large ($k \gg 1/\delta v \sim \tan(\frac{\pi v}{2})$), our low-temperature expression \eqref{eq:low temp k} breaks down, and the soft modes are no longer soft. In this regime the high-temperature expansion \eqref{eq: high temp k} applies again\footnote{In fact, the solution is $k_m = \frac{m}{v} + \frac{2\tan\left(\frac{\pi v}{2}\right)}{\pi m} + \cdots$, which holds at any $v$ as long as $m \gg \tan(\frac{\pi v}{2})$.}, such that for large enough $k$ the eigenvalues $\lambda_{n,k}$ approach those of the high temperature expansion and cancel between the numerator and denominator of \eqref{eq:part function 1 loop schematic}, truncating the product and giving a finite answer. This amounts to fixing the regularization scheme in the Schwarzian theory. 

In order to understand the effect of this regularization on the partition function \eqref{eq:finite v 1-loop}, let us first look at a toy model: a Schwarzian theory with a hard cutoff over the modes at scale of the inverse temperature $\beta = 2/\delta v$, 
\begin{equation}
    Z_{\text{toy,1-loop}}(v) = \prod_{n=2}^{\beta} \frac{2}{\pi\delta v n} = \left(\frac{2}{\pi\delta v}\right)^{\beta - 1} \frac{1}{\beta\Gamma(\beta)} \approx \left(\frac{\beta}{\pi}\right)^{\beta-1} \sqrt{\frac{1}{\pi\beta}} \left(\frac{2e}{\beta}\right)^{\beta} = \frac{\sqrt{\pi}}{\beta^{3/2}} e^{\beta \log\left(\frac{2e}{\pi}\right)} \,.
\end{equation}
We see that the power law $\beta^{-3/2}$ is unchanged by this regularization, but we also find an exponential piece which can be interpreted as a shift in the ground state energy. 

Now, going back to the exact partition function \eqref{eq:finite v 1-loop}, we have to plug in the exact solutions to \eqref{eq:constraints on k} which we find numerically. The 1-loop contribution then amounts to
\begin{equation}
    Z_{\text{1-loop}}(\beta) = \frac{\pi}{8}\delta v^{3/2} e^{\frac{1}{\delta v}} = \frac{\pi}{8}\delta v^{3/2} e^{\frac{1}{\delta v}} \,,
\end{equation}
and in terms of the temperature $\beta\bJ = 2/\delta v$ the partition function is
\begin{equation}
    Z(\beta) = e^{-I[g_0]} Z_{\text{1-loop}}(\beta) = 
    \frac{\sqrt{2}\pi}{(2\beta\bJ)^{3/2}} \exp\left[\left(\frac{2}{\lambda} + \frac{1}{2}\right)\beta\bJ - \frac{\pi^2}{2\lambda} + \frac{\pi^2}{\lambda\beta\bJ}\right],
\end{equation}
which gives\footnote{There is a discrepancy by a factor of $2^{3/2}$ between the exact SYK result \eqref{eq:DSSYK low temp Z} and this 1-loop result. We do not understand its origin, but as it is independent of $\beta$ and $\lambda$ it does not affect the discussion.} the typical Schwarzian temperature dependence $(\beta\bJ)^{-3/2}$, but also a shift in the zero point energy, which is exactly the one anticipated by the exact DSSYK expression \eqref{eq:DSSYK low temp Z}. The physical interpretation of this regularization scheme within the Schwarzian theory will be discussed in Section~\ref{sec: gen functionals}.

Overall, we see that the bilocal Liouville theory is indeed a UV completion of the Schwarzian -- it reproduces its partition function with the correct regularization scheme dictated by DSSYK. But we have yet to relate any of this to any description resembling an effective action for reparametrization modes.  In the following section we will see how the path integral form of the Schwarzian \eqref{eq:Schwarzian path integral} emerges from the bi-local Liouville formulation, and how the reparametrization modes are embedded in the full, UV complete theory.

\section{The Schwarzian in DSSYK}
\label{sec: Schwarzian computation}

In this section, we identify the field configurations of the 2D Liouville theory \eqref{eq:2D Liouville action} which dominate in the IR limit, and we explicitly recover the Schwarzian theory of reparametrizations.
In Section \ref{sec:naive_verses_twisted}, we see that the former are `twisted' reparametrizations of the saddle, with twisted coordinates $s_1$, $s_2$. These twisted reparametrizations slightly violate the boundary condition, in a way which can be corrected by an effective boundary term, as explained in Section \ref{sec: effective boundary term}. In Section \ref{sec: summerize_schwarzian_computaiton}, we show how the full nonlinear Schwarzian emerges as reparametrizations of the twisted coordinates.
We also show how the correct measure emerges in Section \ref{sec: schwarzian measure}. Finally, in Section \ref{sec: other saddles}, we comment on other saddles of the Liouville theory and their interpretation as spaces with conical defects.

\subsection{Twisted reparametrizations} \label{sec:naive_verses_twisted}
We would like to understand how the Schwarzian \eqref{eq:Schwarzian path integral} emerges from the Liouville theory \eqref{eq:2D Liouville action}.  
That is, we are looking to obtain an effective action (in the IR) of 
reparametrizations\footnote{Here the second term vanishes on the saddle point, when $\phi \left( \tau \right) = \tau$, as the on--shell action of the Schwarzian is already contained in $I[g_0]$.}, with the action
\begin{equation} \label{eq:schwartian action}
    I[\tilde{g}_{\phi}] = I[g_0] -\frac{1}{2\lambda \bJ}\int_{0}^{\beta}d\tau \left[ \left\{  \tan\left(\frac{\pi\phi\left(\tau\right)}{\beta}\right),\tau\right\} - \frac{2\pi^2}{\beta^2} \right]  \,.
\end{equation}
But the first question is what are these reparametrizations, and on what do they act. The $G\Sigma$ action has a naive reparametrization symmetry in the IR \cite{Maldacena:2016hyu,Sarosi:2017ykf}, which in the 2D Liouville picture amounts to the transformation
\begin{equation} \label{eq:naive reparams}
    g_0(\tau_1,\tau_2) \mapsto  g_0(\phi(\tau_1),\phi(\tau_2)) + \log \phi'(\tau_1) + \log \phi'(\tau_2)\,, \qquad \text{(naive),}
\end{equation}
where $\phi(\tau)$ denotes the reparametrization.  It is tempting to plug \eqref{eq:naive reparams} into the action \eqref{eq:2D Liouville action} to see if it results in the Schwarzian action over $\phi$. However, there are two issues: first, the reparametrized field configuration is not allowed, as it violates the UV boundary condition; second, ignoring the boundary conditions, the action itself is unchanged under \eqref{eq:naive reparams} (and does not give rise to a Schwarzian).\footnote{One can show that a simple regulation, in the form of multiplying \eqref{eq:naive reparams} by a generic smooth function that vanishes at the boundaries \textbf{$\tau_1=\tau_2$}, does not yield \eqref{eq:schwartian action}, and so a more subtle approach is needed.}
In fact, the tension between the boundary condition and the naive reparametrization is the reason that the latter (slightly tweaked) are lifted by the Schwarzian action in the IR. The correct IR description should involve a reparametrization structure which is deformed as $\tau_1\rightarrow \tau_2$.

A first hint towards the reparametrization structure is then provided by looking at the quadratic fluctuations around the saddle $g = g_{0}+\delta g$, which should resemble an infinitesimal version of reparametrizations. Indeed, when computing the quadratic spectrum around the saddle as in \eqref{eq: quadratic fluctuations}, one finds a set of soft modes ($m=|n| > 1$ for $n\ll \frac{1}{\delta v}$ in \eqref{eq:low temp eigenvalues})
\begin{equation} \label{eq: specific light modes}
    \delta g(x,y) =\sum_{|n|\geq2}\epsilon_{n} \Psi_{n,k_n}(x,y) , \quad \text{with} \quad  k_n = n + \mathcal{O}((n \delta v )^3) ,
\end{equation}
where $\epsilon_{n}^* = \epsilon_{-n}$ for odd $n$ and $\epsilon_{n}^* = -\epsilon_{-n}$ for even $n$, which result in the quadratic action
\begin{equation}\label{eq: quadratic schwarzian action}
    I[g] = I[g_0] + \frac{\pi^2 \delta v}{2\lambda} \sum_{n \;, |n|\geq 2} n^2 (n^2-1) |\epsilon_n|^2+\mathcal{O}\left(\delta v ^2\right) .
\end{equation}
The leading term of order $\mathcal{O} \left( \delta v \right)$ is the infinitesimal version of the Schwarzian in \eqref{eq:schwartian action}, for reparametrizations of the form $\phi \left( \tau \right) = \tau + \sum_{ |n|\geq 2} \frac{\beta}{2\pi} \epsilon_n e^{2 \pi i n \tau /\beta}$.

At zero temperature $\beta \bJ = \infty$, $v=1$, and thus $k_n=n$, the light fluctuations \eqref{eq: specific light modes} are massless and indeed correspond to infinitesimal reparametrizations of the $\left(\tau_1,\tau_2 \right)$ coordinates in $g_0 = -2 \log [ \frac{\beta \bJ}{\pi}\sin \frac{\pi \tau_-}{\beta}]$. The absence of the modes with $n = 0,1$ corresponds to the absence (or gauging) of the reparameterizations associated with $SL(2,\bR)$, which are exact symmetries of the saddle.
However, in this strict case $\delta v=0$, and so we cannot see the contribution of the second term in \eqref{eq: quadratic schwarzian action}. We therefore need the leading correction in $1/\beta\bJ$. In this case, the light modes do \textit{not} describe reparametrizations of the saddle with respect to $\tau_1,\tau_2$ anymore.

\begin{figure}
\centering
\includegraphics[width=0.3\linewidth]{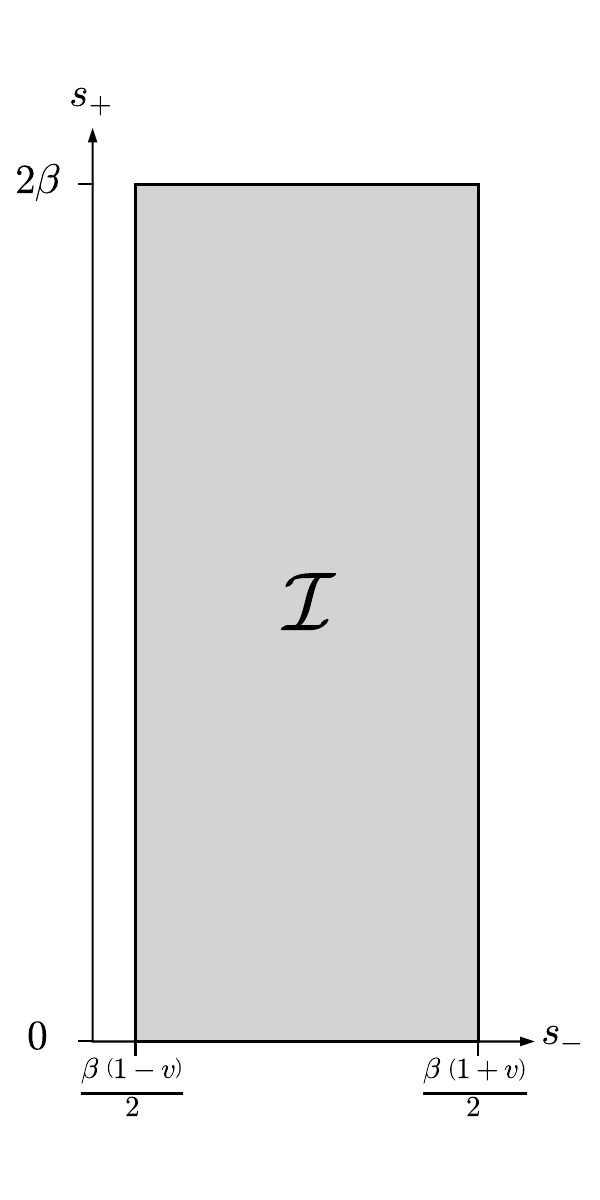}
\quad \quad \quad \quad \quad \quad 
\includegraphics[width=0.3\linewidth]{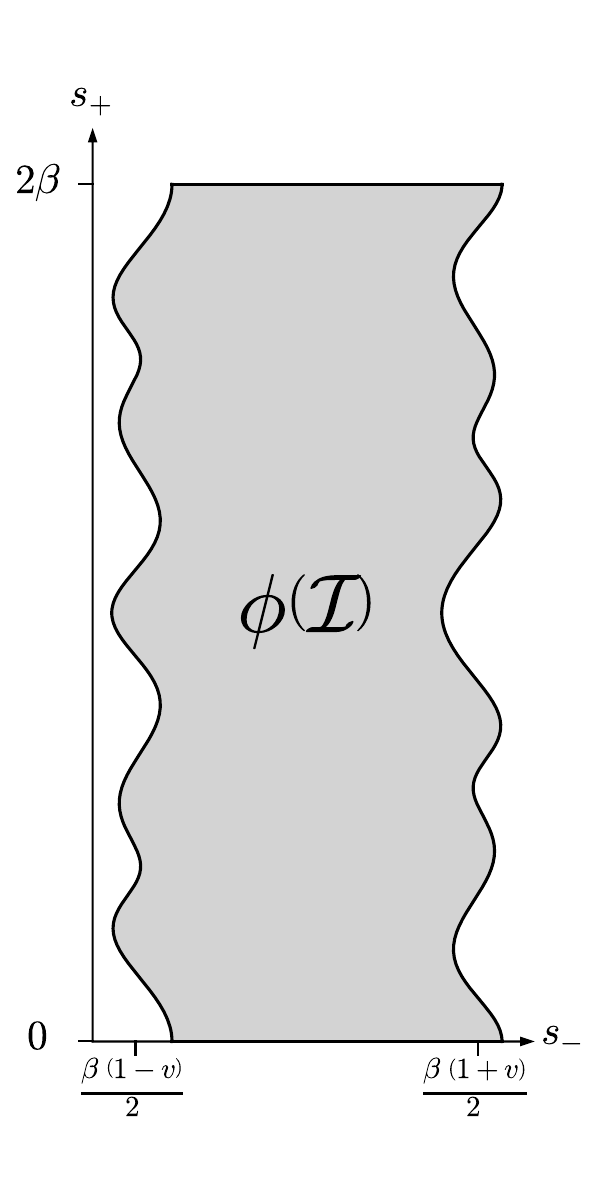}
\caption{\textbf{Left:} The range of integration in the $\left(s_-,s_+\right)$ coordinate system. \textbf{Right:} Illustration of the region $\phi\left(\mathcal{I}\right) $ -- the  range of integration after reparametrization as used in  \eqref{eq: potential term intermidiate}.}
\label{fig: splus sminus range} 
\end{figure}

We can still use the approximation $k_n \approx n$ without changing the action at order $O(\delta v)$. 
A straightforward calculation using the explicit wavefunctions of \eqref{eq: quadratic fluctuations}  shows that the light modes \eqref{eq: specific light modes} can then be rewritten as
\begin{equation} \label{eq: specific light modes 2}
    \delta g(s_1,s_2) \approx \sum_{|n|\geq 2} \epsilon_n \frac{\beta}{2\pi} \Bigg[ 
    e^{\frac{2\pi i n s_1}{\beta}} \partial_{s_{1}} g_0
    + e^{\frac{2\pi i n s_2}{\beta}} \partial_{s_{2}} g_0 + \partial_{s_{1}} e^{\frac{2\pi i n s_1}{\beta}} + \partial_{s_{2}} e^{\frac{2\pi i n s_2}{\beta}} \Bigg] \, ,
\end{equation}
where we introduced the \textit{twisted coordinates}
\begin{equation} \label{eq: twisted coords}
    s_{1} \equiv \frac{1}{2}\left(\tau_+ + v\tau_- + \frac{\beta}{2}(1-v)\right) , \qquad 
     s_{2} \equiv \frac{1}{2}\left(\tau_+ - v\tau_- - \frac{\beta}{2}(1-v)\right) . 
\end{equation}
In \eqref{eq: specific light modes 2}, $g_0$ is the exact solution of the saddle point for finite $v$ \eqref{eq:saddle point g}, which in terms of the twisted coordinates is
\begin{equation} \label{eq: g0_ins1_s2}
    g_0(s_{1},s_{2}) = 2\log\left(\frac{\cos\left(\frac{\pi v}{2}\right)}{\sin\left(\frac{\pi}{\beta}\left(s_1-s_2\right)\right)}\right) .
\end{equation}
Moreover, \eqref{eq: specific light modes 2} can be rewritten at the linearized level as
\begin{equation} \label{eq: specific light modes 3}
    g_{\Tilde{\epsilon}}(s_{1},s_{2}) = g_{0}(s_{1},s_{2})+\delta g(s_{1},s_{2}) \approx g_{0}(s_{1}+\Tilde{\epsilon}(s_{1}),s_{2}+\Tilde{\epsilon}(s_{2}))+\log(1+\Tilde{\epsilon}'(s_{1}))+\log(1+\Tilde{\epsilon}'(s_{2})),
\end{equation}
with $\Tilde{\epsilon}(s) = \sum_{|n|\geq 2} \epsilon_n \frac{\beta}{2\pi} e^{2\pi i n s / \beta}$.
The quadratic fluctuations are thus approximately infinitesimal reparametrizations of the twisted coordinates \eqref{eq: twisted coords}. The coordinate range of these twisted coordinates, after rearranging the time coordinates as in Figure \ref{fig:integ-reg-reorder2}, is 
\begin{equation} \label{eq: integraiton region s coordiantes}
        s_+ \equiv s_1 + s_2 \in [0,2\beta] \,, \qquad  
         s_- \equiv s_1 - s_2 \in \left[\frac{\beta (1-v)}{2},\frac{\beta (1+v)}{2}\right]\,,
\end{equation}
as sketched in the left of Figure \ref{fig: splus sminus range}.

Assuming that this idea of reparametrizations of twisted coordinates can be lifted to finite reparametrizations, we would have
\begin{equation} \label{eq: twisted coordinate reparametrization of g}
    g_{\phi}(s_{1},s_{2}) \equiv 2\log\left(\frac{\cos\left(\frac{\pi v}{2}\right)}{\sin\left(\frac{\pi}{\beta}\left|\phi(s_{1})-\phi(s_{2})\right|\right)}\right)+\log(\partial_{s_{1}}\phi(s_{1})\partial_{s_{2}}\phi(s_{2})) \, .
\end{equation}
In the remainder of this section, we will use \eqref{eq: twisted coordinate reparametrization of g}, rather than \eqref{eq:naive reparams}, as a better starting point to reproduce the Schwarzian action. As it stands, this reparametrization still violates the boundary conditions, but in a mild way which can be corrected, which is the subject of the next subsection.

It is important to highlight that the derivation above relies on the assumption that $n\ll \frac{1}{\delta v}$. This is not too surprising, as modes with large momentum shouldn't be described by the IR of the theory. Indeed, for momenta $n \sim 1/\delta v$, the action \eqref{eq: quadratic schwarzian action} is no longer $\mathcal{O}(\delta v)$, and as we saw in Section~\ref{sec:quick review}, the UV theory smoothly regulates these modes. Since the length of the thermal circle is $\beta$, such $n$ corresponds to momenta that probe distances of $\mathcal{O}(1/\bJ)$. The significance of this scale will be addressed in more detail in Section \ref{sec: gen functionals}.

\subsection{Effective boundary term}
\label{sec: effective boundary term}

\eqref{eq: twisted coordinate reparametrization of g} still violates the boundary conditions, but more softly: at $\tau_1=\tau_2$, $s_1$ and $s_2$ differ at order $\delta v$. The first term in \eqref{eq: twisted coordinate reparametrization of g} then becomes $-2\log \phi'$ at leading order, cancelling away the second term and leaving only a violation of $\mathcal{O}(\delta v^2)$. 
We now want to correct \eqref{eq: twisted coordinate reparametrization of g} in such a way that the boundary condition $g(\tau_-=0)=g(\tau_-=\beta)=0$ is exactly fulfilled. One way to do this which turns out to be convenient is
\begin{equation} \label{eq: twisted coordinate reparametrization of g regularized}
    \Tilde{g}_{\phi} \equiv  g_0 + f(\tau_-) (g_{\phi}-g_0) \, ,
\end{equation}
where $f$ is a regulating function which is $1$ everywhere except in an interval $\mathcal{D}$ of size $\delta v^2$ around $\tau_-=0,\beta$, and equal to $0$ at $\tau_-=0,\beta$.
By construction, $\Tilde{g}_{\phi}$ fulfills the UV boundary condition.\footnote{By choosing  $f$ to be symmetrical in $\tau_- \rightarrow \beta - \tau_-$, it is easy to see that $\Tilde{g}_{\phi}$ also fulfills the KMS condition and the $\beta$--periodicity \eqref{eq:boundary conditions g after rearange}, which in terms of $s_1$ and $s_2$ can be written as
$ \tilde g_\phi(s_2,s_1-\beta) = \tilde g_\phi(s_1,s_2), \,
    \tilde g_\phi(s_1+\beta,s_2+\beta) = \tilde g_\phi(s_1,s_2)\, $. 
} We will show that plugging $\Tilde{g}_{\phi}$ into the action reproduces the Schwarzian action with the correct coefficient. 

In practice, it is easier to work with the non-regularized version \eqref{eq: twisted coordinate reparametrization of g}.
The difference between them in leading order can be compensated by adding an effective boundary term: at leading order in $\delta v$, a direct computation shows that
\begin{equation}
    \begin{split}
    I[\Tilde{g}_{\phi}]-I[g_{\phi}] =& -\frac{1}{8 \lambda} \int_0^{2\beta} d\tau_+ \int_{\mathcal{D}} d\tau_- (\partial_- g_0)  \partial_- \left[ (f-1) (g_\phi -g_0) \right] + \mathcal{O} (\delta v ^2) \nonumber \\
    =& -\frac{1}{8 \lambda} \int_0^{2\beta} d\tau_+ \int_{\mathcal{D}} d\tau_- \partial_-   \left[ (\partial_-g_0) (f-1) (g_\phi -g_0) \right] + \mathcal{O} (\delta v ^2).    
    \end{split}
\end{equation}

Here we have used that in the boundary region $\cal D$, where $\tau_- \sim \delta v^2$ (or $\beta-\tau_- \sim \delta v^2$), \eqref{eq: twisted coordinate reparametrization of g} implies that\footnote{This is an important difference to the naive (non-twisted) reparametrization: in that case, the difference between the saddle and its reparametrization at the boundary is $\mathcal{O}(1)$.} $g_\phi - g_0 \sim \delta v^2$.
Additionally, since $f$ is a smooth function that interpolates from 1 to 0 in the region $\cal D$, its derivative scales as $\partial_- f \sim \delta v^{-2}$. We have also used the fact that in the region $\cal D$, $\partial_- g_0 \sim \delta v^{-1}, \partial_-^2 g_0 \sim \delta v^{-2}, \partial_- (g_{\phi}-g_0) \sim 1$, as can be computed directly from \eqref{eq:saddle point g}, \eqref{eq: twisted coordinate reparametrization of g}, as well as the equation of motion for $g_0$. Note that to derive this scaling, one needs to use the restriction that $\phi$ describes IR fluctuations, specifically that $\phi'\ll \frac{1}{\delta v}$ and $\phi^{(n)}/\phi' \ll \beta \bJ^{n}$ for any $n\ge 2 $. These relations are compatible with the infinitesimal restriction that $n\ll \frac{1}{\delta v}$.

Integrating over $\tau_-$, and taking into account that $f=0$ or $f=1$ at the boundaries of $\mathcal{D}$, we find that\footnote{Notice that a factor of 2 arises because $\mathcal{D}$ consists of two parts.}
\begin{equation}
    I[\Tilde{g}_{\phi}] = I[g_{\phi}] + \frac{1}{4 \lambda} \int_0^{2\beta} d\tau_+    \left( (\partial_-g_0) (g_\phi -g_0) \right)|_{\tau_-=0} + \mathcal{O} (\delta v ^2) .
\end{equation} 
Finally, at the boundary $\tau_-=0$ we have $g_0=0$ and, at leading order in $\mathcal{O}(\delta v)$, $\partial_-g_0$ can be replaced by $\partial_-g_\phi$. Thus, when calculating the action at $\mathcal{O} \left( \delta v \right)$, we can use the unregulated version $g_{\phi}$ instead of $\Tilde{g}_{\phi}$, as long as we add  the boundary term\footnote{A less rigorous but quicker way to obtain \eqref{eq: I_boundary} is to notice that for $\delta g|_{\tau_- = 0/\beta}\neq0$, the variational principle is violated by a term of the form $\delta I \supset -\frac{1}{8\lambda} \int_0^{2\beta} d\tau_+ \big[ \delta g \partial_- g \big]\Big|_{\tau_-=0}^{\tau_-=\beta}$. Adding the boundary term $I_{\text{boundary}}$ restores the variational principle, at least at leading order in $\delta v$.} $I_{\text{boundary}}$:
\begin{equation} \label{eq: I_boundary}
      I[\Tilde{g}_{\phi}]= I[g_{\phi}] + I_{\text{boundary}}[g_{\phi}], \quad I_{\text{boundary}}[g_{\phi}] \equiv \frac{1}{4 \lambda} \int_0^{2\beta} d\tau_+  g_\phi \partial_-g_\phi|_{\tau_-=0}.
\end{equation}

There are other ways to modify \eqref{eq: twisted coordinate reparametrization of g} such that it obeys the boundary conditions, and reproduces the Schwarzian. For example, the authors of \cite{MezeiBucca} find another explicit deformation that does so, however it is not connected smoothly to the soft modes at finite $\delta v$. In addition, in Appendix \ref{sec: finite temperatures}, we present an alternative suggestion in which the derivatives in  \eqref{eq: twisted coordinate reparametrization of g} are replaced by finite difference operators.

\subsection{Computing the Schwarzian action} \label{sec: summerize_schwarzian_computaiton}

We now compute the action of the regularized twisted reparametrization modes in \eqref{eq: twisted coordinate reparametrization of g regularized}, and show that it approximates to the saddle point action plus a Schwarzian at leading order in $\delta v$, with the expected coefficient \cite{Maldacena:2016hyu}.

In terms of $s_1,s_2$, the action \eqref{eq:2D Liouville action} can be written as
\begin{equation} \label{eq: definition of potantial and kinetic terms}
    \begin{split}
    I[\tilde{g}_\phi] &= I_{\text{boundary}}[g_\phi]+ \\ &+ \underbrace{\frac{1}{8\lambda v}\int_{\mathcal{I}}ds_{1}ds_{2}\left[-4 \bJ^{2}e^{g_\phi}\right]}_{I_{\text{pot}}[g_{\phi}]} + \\ 
    &+\underbrace{\frac{1}{8\lambda v}\int_{\mathcal{I}}ds_{1}ds_{2}\left[\frac{1}{2}(1+v^{2})\partial_{s_{1}}g_\phi\partial_{s_{2}}g_\phi+\frac{1}{4}(1-v^{2})\partial_{s_{1}}g_\phi\partial_{s_{1}}g_\phi+\frac{1}{4}(1-v^{2})\partial_{s_{2}}g_\phi\partial_{s_{2}}g_\phi\right]}_{I_{\text{kin}}[g_{\phi}]} ,
    \end{split}
\end{equation}
where $\mathcal{I}$ is the integration region defined in \eqref{eq: integraiton region s coordiantes}. The full calculation of each term is left to Appendix \ref{sec: Schwarzian details}. Here, we highlight the method and the main result.

\paragraph{Potential term} Substituting \eqref{eq: twisted coordinate reparametrization of g} into the potential term, we find that
\begin{equation} 
    I_{\text{pot}}[g_{\phi}] = -\frac{v}{2\lambda} \frac{\pi^2}{\beta^2} \int_{\mathcal{I}} ds_1 ds_2 \phi'(s_1)\phi'(s_2) \frac{1}{\sin^2 \left(\frac{\pi}{\beta}\left(\phi(s_1)-\phi(s_2)\right)\right)} \,. 
\end{equation}
Changing to the integration variables $\phi_1 = \phi (\tau_1),\phi_2 = \phi (\tau_2)$, we integrate over the parametrized region, which we denote as $\phi (\mathcal{I})$ and is illustrated in Figure \ref{fig: splus sminus range}. We find that
\begin{equation} \label{eq: potential term intermidiate}
    I_{\text{pot}}[g_{\phi}] = -\frac{\pi^2 v}{2\lambda \beta^2} \int_{\phi(\mathcal{I})} \frac{d\phi_1 d\phi_2}{\sin^2 \left( \frac{\pi}{\beta}(\phi_1-\phi_2) \right) } \, .
\end{equation}
It is now possible to use Stokes' theorem and integrate only over the boundaries of the region $\phi(\mathcal{I})$. Approximating the result to leading order in $\delta v$ gives
\begin{equation} \label{eq: final potential term}
    I_{\text{pot}}\left[g_{\phi}\right]-I_{\text{pot}}\left[g_0\right] =-\frac{1}{\lambda}\frac{v}{12\bJ}\int_{0}^{\beta}d\tau\left[\frac{\phi^{\left(3\right)}}{\phi'}-\left(\frac{2\pi}{\beta}\right)^{2}\phi'{}^{2}+\left(\frac{2\pi}{\beta}\right)^{2}\right] + \mathcal{O}(\delta v^3).
\end{equation}

\paragraph{Kinetic term}
The full simplification of the kinetic term is technical, and can be done using tricks similar in nature to those described in the potential term. We refer the reader to Appendix \ref{sec: Schwarzian details} for more details. The end result, at leading order in $\mathcal{O} (\delta v)$, is 
\begin{equation} \label{eq: final kinetic term}
    I_{\text{kin}}[g_{\phi}]-I_{\text{kin}}[g_0]=\frac{5}{12\lambda \bJ}\intop_{0}^{\beta}d\tau\left[\left(\frac{\phi''}{\phi'}\right)^{2}-\left(\frac{2\pi}{\beta}\right)^{2}\phi'{}^{2}+\left(\frac{2\pi}{\beta}\right)^{2}\right] + \mathcal{O}(\delta v^2).
\end{equation}

\paragraph{Boundary term}
The boundary term is already written as only one integration. Plugging in \eqref{eq: twisted coordinate reparametrization of g}, the end result is
\begin{equation} \label{eq: final boundary term}
    I_{\text{boundary}}[g_{\phi}] = - \frac{1}{12\lambda \bJ}\int_0^{\beta} d\tau \Bigg[ \left( \frac{\phi''}{\phi'}\right) ^2-\left(\frac{2\pi}{\beta}\right)^2 \phi'^2 + \left(\frac{2\pi}{\beta}\right)^2 \Bigg] + \mathcal{O}(\delta v^2).
\end{equation}
The final result is just the sum of Eqs. \eqref{eq: final potential term}, \eqref{eq: final kinetic term} and \eqref{eq: final boundary term}\footnote{Above, we used $\frac{\phi^{(3)}}{\phi'}=\left(\frac{\phi''}{\phi'}\right)^{\prime}+\left(\frac{\phi''}{\phi}\right)^{2}$. The term $\left(\frac{\phi''}{\phi'}\right)^{\prime}$ is the total derivative of a periodic function in $\beta$, so it does not contribute to the action.}:
\begin{equation} \label{eq: result of schwarzian}  
    I[\tilde{g}_{\phi}]-I[g_0] = \frac{1}{4\lambda \bJ} \int_0^{\beta} d\tau \left[ \left(\frac{\phi''}{\phi'}\right)^2-\left(\frac{2\pi}{\beta}\right)^2 \phi'^2 + \left(\frac{2\pi}{\beta}\right)^2 
    \right]  \, .
\end{equation}
Which is precisely the Schwarzian action \eqref{eq:schwartian action}. We see that the action of the reparametrization modes $\tilde{g}_{\phi}$ is given by the saddle point value plus a contribution of order $\sim \frac{1}{\lambda \beta \bJ}$, which justifies the association of those modes with the IR dynamics of the theory. The resulting Schwarzian action and its coefficient agree with the results by Maldacena and Stanford \cite{Maldacena:2016hyu} in the large $p$ limit.

\paragraph{Cutout region of $\text{AdS}_{2}$}

Notice that the contribution to the action of each term is, by itself, a Schwarzian. For the potential term, there is a natural interpretation of this fact: the variation is a fluctuation of a cutout region of $\text{AdS}_{2}$.

To see this, we denote $\phi_+=\phi_1+\phi_2$, $\phi_-=\phi_1-\phi_2$. The integrand in \eqref{eq: potential term intermidiate} can then be written as
\begin{equation}\label{Ads coordinate frame}
    \frac{d\phi_1 d\phi_2}{\sin^2 \left(\frac{\pi}{\beta}\left(\phi_1-\phi_2\right)\right)} =  \frac{d\phi_+ d\phi_-}{2\sin^2 \left(\frac{\pi}{\beta}\phi_- \right)}
\end{equation}
which is (up to a proportionality constant) the same expression as $\sqrt{-h}R$, with $h_{\mu \nu}$ is the metric of $AdS_2$ in the coordinate frame where we interpret \eqref{Ads coordinate frame} as a metric (the curvature is constant).

Before and after the reparametrization, the induced metric at the boundaries is the same (at leading order) and proportional to $h_{\xi, \xi} \propto \frac{1}{\delta v^2} $\footnote{This can be shown by explicit computation of the reparametrized boundary.}, where $\xi$ is the coordinate at the boundaries. Since the sum of \eqref{eq: potential term intermidiate} and the boundary integral over the extrinsic curvature $\int_{\partial \phi(I)} K$ is equal to the Euler characteristic, we can express \eqref{eq: potential term intermidiate} as a difference of extrinsic curvatures. For $\delta v \rightarrow 0 $, we can then repeat the usual calculation \cite{Maldacena:2016upp} to see that the extrinsic curvature is at leading order given by a Schwarzian.

\subsection{The Schwarzian measure} \label{sec: schwarzian measure}

At this point we have shown that the action for our twisted reparametrizations is the full nonlinear Schwarzian action. To reproduce the full theory, we also need to show that the measure is the correct one. Note that our twisted reparametrizations \eqref{eq: twisted coordinate reparametrization of g} form a group, and so by the arguments in \cite{Bagrets:2016cdf,Stanford:2017thb}, the invariant measure is the natural one. However, the point of this work is to derive these features explicitly, from the UV description.

To do so, we restrict ourselves to the measure around the identity element, i.e. the one that comes from the infinitesimal reparametrizations \eqref{eq: specific light modes 3}, which are of the form $\phi(s) = s + \frac{\beta}{2\pi} \sum_{|n|\ge 2} \varepsilon_n e^{2\pi i n s/\beta}$. There are going to be two parts that contribute to the measure, the first coming from the change of variables to the $\varepsilon$'s, and the second from integrating out the non--zero modes.

As a starting point our original bilocal Liouville path integral had a flat measure over the function $g$, and therefore also over the fluctuations $\delta g$. Expanding the latter in the basis of orthonormal eigenfunctions \eqref{eq: quadratic fluctuations}, $\delta g = \sum_{n,k} a_{n,k} \Psi_{n,k}$, this statement amounts to saying that the measure is $\prod_{n=-\infty}^\infty \prod_{k(v)} da_{n,k}$.

We now concentrate on the zero modes (the modes with $k = |n| \ge 2$), which we can write as
\begin{multline}
    \Psi_{n,k_n}(s_1,s_2) =
        \cN_{n,k_n} \frac{\beta}{4\pi i}\Bigg[\partial_{s_{1}}e^{\frac{2\pi ins_{1}}{\beta}}+e^{\frac{2\pi ins_{1}}{\beta}}\partial_{s_{1}}g_{0}\left(s_{1},s_{2}\right) \\
        +\partial_{s_{2}}e^{\frac{2\pi is_{2}}{\beta}}+e^{\frac{2\pi ins_{2}}{\beta}}\partial_{s_{2}}g_{0}\left(s_{1},s_{2}\right)\Bigg] + O(\delta v^3) \,, 
\end{multline}
where $g_0(s_1,s_2) $ is our saddle \eqref{eq: g0_ins1_s2}. Demanding that the eigenfunctions are orthonormal implies that at low temperatures the normalization is $\cN_{n,k_n} = \frac{1}{\sqrt{2\pi^3(n^2-1)}}$. We can now expand a fluctuation $\delta g$ originating from a reparameterization $\varepsilon(s)$ in the eigenfunction basis,
\begin{multline}
    \delta g(s_1,s_2) = \varepsilon(s_1) \partial_{s_1} g_0(s_1,s_2) + \varepsilon'(s_1) + \varepsilon(s_2) \partial_{s_2} g_0(s_1,s_2) + \varepsilon'(s_2) \\
    = -i\sum_{|n|\ge 2} \sqrt{32\pi^3(n^2-1)}\varepsilon_n \Psi_{n,k_n}(s_1,s_2) \,,
\end{multline}
and infer that the contribution to the measure over the space of twisted reparameterizations is
\begin{equation}
        \label{eq:zero mode measure change of base}
    D\varepsilon = \prod_{n\ge 2} (32\pi^3(n^2-1)) d\varepsilon_n d \varepsilon_{-n} \,, \qquad \text{(change of basis).}
\end{equation}

But we are not yet done -- we also have the other, non-zero modes, and the normalization $\eta$ that contribute to the path integral. The easiest way to track their contribution is to use \eqref{eq:finite v 1-loop}, and divide by the contribution of the zero modes, which leaves us with
\begin{equation}
    \prod_{m=1}^{\infty}\left[-\frac{\lambda_{2m,k_{m}^e(v)}\lambda_{2m+1,k_{m}^o(v)}}{\sin\left(\frac{\pi vk_{m}^{e}\left(v\right)}{2}\right)\cos\left(\frac{\pi vk_{m}^{o}\left(v\right)}{2}\right)}\right] \,.
\end{equation}
We can use the low temperature approximations \eqref{eq:low temp k} and \eqref{eq:low temp eigenvalues} to find that the additional contribution to the measure is 
\begin{equation}
    \prod_{m=1}^{\infty}\left[-\frac{\delta v^2(2m)^2(2m+1)^2}{64\lambda^2 \sin\left(\frac{\pi vk_{m}^{e}\left(v\right)}{2}\right)\cos\left(\frac{\pi vk_{m}^{o}\left(v\right)}{2}\right)}\right]  = \prod_{m=1}^{\infty}\left[\frac{(2m)(2m+1)}{16\lambda^2 \pi^2} + O(\delta v^2)\right] = \prod_{m=2}^{\infty}\frac{m}{4\pi\lambda}\,.
\end{equation}
Combining this with \eqref{eq:zero mode measure change of base} we find that the measure becomes
\begin{equation}
    D\varepsilon = \prod_{n\ge 2} \frac{8\pi^2n(n^2-1)}{\lambda} d\varepsilon_n d \varepsilon_{-n} \,,
\end{equation}
thus reproducing the measure of the Schwarzian theory\footnote{In fact, we only reproduce the standard formulae to a multiplicative constant which is independent of $n$ or $\beta$. Such a constant depends on the regularization scheme of the Schwarzian.}. We stress that we had to integrate out the other modes in order to reproduce the correct measure, and couldn't simply ignore them.

In fact, we have cheated a bit, since the low temperature approximations only apply when $m \ll 1/\delta v$. This creates a natural regularization procedure, precisely as discussed at the end of Section~\ref{sec:quick review}. We will see the physical implications of this procedure below in Section~\ref{sec: gen functionals}.

\subsection{Additional saddles and Schwarzian with defects}
\label{sec: other saddles}
So far, we have considered one saddle for the Liouville action \eqref{eq:saddle point g}, and studied the soft modes around it at low temperature. However, we will now argue that at low temperatures the bilocal Liouville theory has additional (subleading) saddle points, parameterized by positive integers $k$. The fluctuations around each saddle are given by a Schwarzian theory with a conical defect,  creating a deficit angle of $\alpha=2\pi\left(2k+1\right)$, defined by the action \cite{Mertens:2019tcm}
\begin{equation}
    I[\phi] = -\frac{1}{2\lambda\bJ}\int_{0}^{\beta}d\tau \, \left\{ \tan\left(\frac{\pi(2k+1)\phi(\tau)}{\beta}\right), \tau\right\} \,,
\end{equation}
where $\{\cdot,\cdot\}$ denotes the Schwarzian derivative, and the path integral is over ${\text{Diff}(S^{1})/SL^{2k+1}\left(2,\bR\right)}$. These are reparameterizations that wind $2k+1$ times around the thermal circle. There are also saddles corresponding to reparameterizations that change the orientation, given formally by the Schwarzian of temperature $-\beta$ and conical defect of angle $\alpha=2\pi\left(1-2k\right)$. All these saddles have recently been discussed in the context of the holographic dual to DSSYK, the sine-dilaton theory, in \cite{Blommaert:2024whf}. Here we only show the on-shell action of these saddles, and in Appendix~\ref{app:Other saddles from DSSYK} we show that these saddles also arise in the exact DSSYK partition function, where their 1-loop contribution also matches that of the Schwarzian.

Let us begin by describing these saddles. Up to now, we have concentrated on the saddle
\begin{equation}
    \label{eq:basic saddle}
    g_0\left(\tau\right) = 2\log\left[\frac{\cos\left(\frac{\pi v}{2}\right)}{\cos\left[\pi v\left(\frac{1}{2}-\frac{\tau}{\beta}\right)\right]}\right] \,, \qquad \frac{\pi v}{\cos\left(\frac{\pi v}{2}\right)} = \beta\bJ\,, \quad v \in [-1,1] \,,
\end{equation}
whose on--shell action was
\begin{equation}
    \label{eq:on shell Liouville finite v}
    I = \frac{\pi^{2}v^{2}}{2}-2\pi v\tan\left(\frac{\pi v}{2}\right) \,.
\end{equation}
But the equations of motion also admit other saddles. Their functional form is identical to that of the saddle we studied, the only difference being the value of $v$, which can now be outside $[-1,1]$. In fact, for low temperatures, $\beta\bJ \gg 1$, there is a single solution \eqref{eq:basic saddle} in the range $v \in [-1,1]$, and two solutions in the range ${[(-1)^k 2k - 1, (-1)^k 2k + 1]}$ for any positive integer $k$, given by 
\begin{subequations}
\begin{align}
    v_{k}^{\text{lower}} &= \left(-1\right)^{k}\left(2k+1\right)\left(1-\frac{2}{\beta\bJ}+\frac{4}{\left(\beta\bJ\right)^{2}}-\frac{8+\frac{\pi^{2}}{3}\left(1+2k\right)^{2}}{\left(\beta\bJ\right)^{3}}\right) + O((\beta\bJ)^{-4}) \,, \\
    v_{k}^{\text{upper}} &= \left(-1\right)^{k}\left(2k-1\right)\left(1+\frac{2}{\beta\bJ}+\frac{4}{\left(\beta\bJ\right)^{2}}+\frac{8+\frac{\pi^{2}}{3}\left(1-2k\right)^{2}}{(\beta\bJ)^{3}}\right) + O((\beta\bJ)^{-4})  \,,
\end{align}
\end{subequations}
where $k \ge 0$ for the lower and $k > 0$ for the upper. Later we will see that these saddles are associated with contributions that are close to the lower and upper edge of the spectrum, respectively. 

There is a slight issue, though: for solutions where $v_k\notin\left[-1,1\right]$, $g\left(\tau\right)$ diverges for some values of $\tau\in\left(0,\beta\right)$,
\begin{equation}
    \tau_m = \frac{1}{2} \pm \frac{2m+1}{2v_k} \,,\quad m = 0,\cdots, k-1 \,.
\end{equation}
Yet, this only happens when $v\in\bR$ (equivalently, $\beta\in\bR$), and one can regularize the divergence by starting with a slightly imaginary temperature $\beta+i\epsilon$, evaluating the action, and then taking $\epsilon\to0$. The resulting on--shell action\footnote{The first terms, those that are linear in $\beta\bJ$, can be identified as the contribution of the ground state (or maximal) energy to the partition function.} for each of these saddles is the same as \eqref{eq:on shell Liouville finite v}, with the appropriate value of $v$, 
\begin{subequations}
\begin{align}
    I_k^{\rm lower} &= -\frac{2\beta\bJ}{\lambda} + \frac{\left(1+2k\right)^{2}\pi^{2}}{2\lambda} -\frac{\left(1+2k\right)^{2}\pi^{2}}{\lambda\beta\bJ} \,,  \\ 
    I_k^{\rm upper} &= \phantom{-}\frac{2\beta\bJ}{\lambda} + \frac{\left(1-2k\right)^{2}\pi^{2}}{2\lambda} + \frac{\left(1-2k\right)^{2}\pi^{2}}{\lambda\beta\bJ} \,.
\end{align}
\end{subequations}
Since our regularization prescription involves the complexification of $\beta$ and results in a divergent $g(\tau)$, one may wonder if we should consider other complexified or divergent saddles. We will only consider solutions that give real (even if divergent) saddles for $g(\tau)$ and real, finite action in the limit $\epsilon \to 0$, which are precisely the ones described above for non-negative integer $k$'s. 

The last terms in these actions correspond to the on--shell action of Schwarzian theories with conical defects, as described at the beginning of this subsection. In Appendix~\ref{app:Other saddles from DSSYK} we show that these saddles also appear in the exact DSSYK partition function as derived by diagonalizing the effective Hamiltonian acting on the chord Hilbert space. From the exact DSSYK approach one can also derive the one-loop contribution around them, which also matches this description. 

While we do not try to directly reproduce the Schwarzian path integral for these saddles, we note that a regularized twisted reparameterization of the form \eqref{eq: twisted coordinate reparametrization of g regularized} might do the trick. We need the reparameterization to wind several times around the thermal circle, and to choose the regularizing function $f$ to vanish not just at the edges but at all points that map to the same point on the thermal circle (up to winding).

\section{The multifield Liouville}
\label{sec: multifield Liouville}

In the previous sections, we showed explicitly how a Schwarzian theory arises in the IR of DSSYK. There we relied on the Liouville description of DSSYK, given by the action
\begin{equation}
    I[g] =-\frac{1}{4\lambda}\int_{0}^{\beta}d\tau_{1}\int_{0}^{\beta}d\tau_{2} \left[-\frac{1}{2}\partial_{\tau_{1}}g\partial_{\tau_{2}}g+2\bJ^{2}e^{g}\right] \,,
\end{equation}
which can be understood as the generating functional of chords, in an expansion around $\bJ^2=0$ \cite{Stanford-talk-kitp,Lin:2023trc}.  Starting from the exact chord side, we get a slightly different, and somewhat better defined, action \cite{Berkooz:2024evs,Berkooz:2024ofm}.  

However, many interesting problems have to do with several operators, and several types of random couplings\footnote{For example, systems such as the SYK chain \cite{Gu:2016oyy,Altland:2019lne} whose double scaling limit is discussed in \cite{MezeiBucca}, RG flows such as those discussed in \cite{Jiang:2019pam,Anninos:2020cwo,Anninos:2022qgy}, transitions to quasi-integrable behavior as in \cite{Berkooz:2024evs,Berkooz:2024ofm,Gao:2024lem}, and more.}, whether these are part of the Hamiltonian or probe operators. One such problem is deforming the theory and tracking the deformation to the IR. The relevant microscopic Hamiltonians are of the form $H = \sum_{i=1}^K \kappa_i H_i$, where each $H_i$ is a random operator, resulting in $K$ types of chords in the double--scaled limit, with chord crossing weight between type $i$ and type $j$ given by $q^{\alpha_{ij}}$.

In Subsection \ref{sec: multi-liouville derivation} we provide a proof, along the lines of \cite{Stanford-talk-kitp,Lin:2023trc}, for the Liouville action generalized to multiple types of chords, given in \cite{Berkooz:2024lgq} as: 
\begin{equation}\label{eq:multiLiouville}
    Z = \int \left(\prod_{i=1}^K \mathcal{D}[ g_{i}]\right) \exp\left[\frac{1}{4\lambda}\int_0^\beta d\tau_{1}\int_0^{\beta}d\tau_{2}\left[\sum_{i,j=1}^K\frac{1}{2}\alpha_{ij}g_{i}\partial_{1}\partial_{2}g_{j} + \sum_{i=1}^K 2 \kappa_{i}^2 \bJ^2 e^{\sum_{j}\alpha_{ij}g_{j}}\right] \right] \,.
\end{equation} 
A related formula for a specific set of weights for chord intersections appears in \cite{Gao:2024lem}.

Readers interested in applications can jump to Subsection \ref{sec: gen functionals} where we also write down the generating functional for correlation functions of random operators in the Liouville language, and use it to suggest a heuristic connection to the fake disk of \cite{Lin:2023trc}, as well as the length positivity constraint of \cite{Blommaert:2024ydx}.

\subsection{The multi--Liouville action}
\label{sec: multi-liouville derivation}

We consider the DSSYK partition function for a Hamiltonian of the form $H = \sum_{i=1}^K \kappa_i H_i\,$. Here $H_i$ are appropriate random operators.
In the double scaling limit, intersections of chords associated with these operators are weighted by parameters $\lambda_{ij}$:
\begin{equation}\label{eq: multi-SYK conventions}
    \frac{\langle \Tr(H_i H_j H_i H_j) \rangle_J}{\langle \Tr(H_i H_i)\rangle_J \langle \Tr( H_j H_j) \rangle_J} = e^{-\lambda_{ij}}\,, \qquad  \ \langle \Tr(H_i H_j) \rangle_J = \frac{\bJ^2}{\lambda}\delta_{ij}\,.
\end{equation}
Next we write $\lambda_{ij}=\lambda\alpha_{ij}$ ($\lambda$ sets the scaling of $\lambda_{ij}$, and $q=e^{-\lambda})$. It is also often convenient to take $\alpha_{11}=1$ and think of $H_1$ as the undeformed Hamiltonian, and the other $H_i$'s as deformations. This way of writing enables us to take the limit $\lambda\rightarrow 0$ uniformly. 

Specific examples  \cite{Berkooz:2018qkz,Berkooz:2018jqr,Gao:2024lem} of different $\lambda_{ij}$ are SYK--like operators that take the form of the Hamiltonian \eqref{eq:DSSYK Hamiltonian} with different length $\tilde p$, or they could be more general polarized operators \cite{Berkooz:2024ofm,Berkooz:2024evs} that might have\footnote{They could also not preserve any symmetries, but have non-uniform distribution.} some additional symmetries, such as
\begin{equation}
    \sum_{\substack{\ \ \ 1 \le j_{1} < \cdots < j_{k^\prime} \le M \\ 2M < i_1 < \cdots < i_k \le N}} \tilde J_{i_1,\cdots,i_k}^{j_{1},\cdots,j_{k^\prime}} X_{j_{1}} \cdots X_{j_{k^\prime}} \psi_{i_1} \cdots \psi_{i_k} \,, \qquad X_j = \psi_{2j-1} \psi_{2j} ,
\end{equation}
for some $M \le N/2$. Insertions of such operators are also described by (matter) chords, and their intersections with themselves and with the Hamiltonian chords might be weighted by different factors. For further details see Appendix B of \cite{Berkooz:2024ofm}.

Computing the partition function $\langle \Tr(e^{-\beta H})\rangle_J$ amounts to summing over all chord diagrams with $K$ types of chords and intersection numbers $\alpha_{ij}$ \cite{Berkooz:2024ofm,Berkooz:2024evs}:
\begin{equation}
    \label{eq:moments as crossings}
    Z = \sum_{n=0}^{\infty}\frac{\bJ^{2n}\beta^{2n}}{\lambda^{n}(2n)!}\sum_{m_{1}+\dots m_{K}=n}\left(\prod_{a=1}^{K}\kappa_{a}^{2m_{a}}\right) \sum_{\text{CD}} \prod_{j<i} e^{-\lambda_{ij} \cdot \#\text{crossings between $i$- and $j$-chords}} \, .
\end{equation}
Note that at infinite temperature $Z(\beta=0)=1$, as we work in the normalization $\Tr(1)=1$. We now show how to reproduce these chord rules from the multi--Liouville theory, given by the action \eqref{eq:multiLiouville}.  

As before, the path integral integrates over $g_i$ which are symmetric and fulfill the boundary condition $g_{i}(\tau,\tau)= g_i(\tau,\tau+\beta) = 0$.
Since the $g_i$ are symmetric in their arguments, we restrict the integration region in the action to the triangle $\tau_1 \geq \tau_2$ (denoting this region as $\Delta$), such that the partition function becomes\footnote{We suppress an additional constant imposing $Z(\beta=0) = \Tr(\mathbbm{1}) = 1$ as before.} 
\begin{equation}\label{eq:Mult}
    Z = \int \left(\prod_{i=1}^K \mathcal{D}[g_{i}]\right) \exp\left[\frac{1}{2\lambda}\iint_{\Delta} d\tau_{1}d\tau_{2}\left[\sum_{i,j=1}^K\frac{1}{2}\alpha_{ij}g_{i}\partial_{1}\partial_{2}g_{j} + \sum_{i=1}^K 2 \kappa_{i}^2 \bJ^2 e^{\sum_{j}\alpha_{ij}g_{j}}\right] \right] \,.
\end{equation}
Perturbatively expanding in $\bJ^2$ gives
\begin{equation} \label{eq: perturbative expansion of n-Liouville}
    Z=\sum_{n=0}^{\infty}\frac{\bJ^{2n}}{\lambda^{n}n!}\iint_{\Delta_{1}}\dots\iint_{\Delta_{n}}\left\langle \prod_{i=1}^{n}\left(\sum_{a=1}^{K}\kappa_{a}^{2}e^{\sum_{b}\alpha_{ab}g_{b}(\tau_{1}^{i},\tau_{2}^{i})}\right)\right\rangle _{0}\,  .
\end{equation}
Here the correlators are computed in the free $\bJ^2=0$ theory. Taking into account the boundary conditions, the free propagator in this theory is given by
\begin{equation}
    \begin{split}
        \langle g_i(\tau_1^{i},\tau_2^{i}) g_j(\tau_1^{j},\tau_2^{j}) \rangle = & - \lambda (\alpha^{-1})_{ij} \big[ \theta(\tau_1^{i}-\tau_1^{j})\theta(\tau_2^{i}-\tau_2^{j}) +  \theta(\tau_1^{j}-\tau_1^{i})\theta(\tau_2^{j}-\tau_2^{i}) \\  - & \theta(\tau_2^{i}-\tau_1^{j})\theta(\tau_1^{i}-\tau_2^{j}) - \theta(\tau_1^{j}-\tau_2^{i})\theta(\tau_2^{j}-\tau_1^{i})\big] \\  \equiv & - \lambda (\alpha^{-1})_{ij} |i\cap j|,
    \end{split}
\end{equation}
with the prescription $\theta (0)=0$ \cite{Lin:2023trc}.
In other words, the propagator is $-\lambda (\alpha^{-1})_{ij}$ when an $i$--chord stretching from $\tau_1^{i}$ to $\tau_2^{i}$ crosses a $j$--chord stretching between $\tau_1^{j}$ and $\tau_2^{j}$, and zero otherwise.\footnote{In the case where the matrix $\alpha_{ij}$ is noninvertible, one should take the generalized inverse. The propagator for fields with zero $\alpha_{ij}$ eigenvalue is then simply zero, which corresponds to this field decoupling from the theory. We will see this decoupling also when discussing specific examples for $K=2$, in \eqref{action degenerating}. As the decoupling field is free, it is independent of $\beta \bJ$ and only adds a numerical coefficient to the path integral.} We have introduced the shorthand notation $|i \cap j|$, which is 1 if the $i$-- and the $j$--chord cross, and 0 otherwise. 

We now prove that the partition function of the $n$--chord DSSYK agrees with \eqref{eq: perturbative expansion of n-Liouville}. 
As a first step, since we can relabel the times $\tau_{1,2}^{i}$ arbitrarily inside the integral, we can perform a multinomial expansion of the integrand to get
\begin{equation}
    \left\langle \prod_{i=1}^{n}\left(\sum_{a=1}^{K}\kappa_{a}^{2}e^{\sum_{b}\alpha_{ab}g_{b}(\tau_{1}^{i},\tau_{2}^{i})}\right)\right\rangle _{0}=\sum_{m_{1}+\dots m_{K}=n}\left(\begin{array}{c}
    n\\
    m_{1},\dots,m_{K}
    \end{array}\right)\left(\prod_{a=1}^{K}\kappa_{a}^{2m_{a}}\right)\left\langle e^{\mathcal{O}_{\{m_a\}}}\right\rangle _{0} \, ,
\end{equation}
where $\{m_a\}=\{m_1,...,m_K\}$, the equality is to be understood as equality inside the integral, and the operator $\mathcal{O}_{\{m_a\}}$ is given as
\begin{equation}
    \begin{split}
        \mathcal{O}_{\{m_a\}}= &\sum_{p=1}^{m_{1}}\sum_{b=1}^{K}\alpha_{1b}g_b\left(\tau_{1}^{p},\tau_{2}^{p}\right)+\sum_{p=m_1+1}^{m_1 + m_{2}}\sum_{b=1}^{K}\alpha_{2b}g_b\left(\tau_{1}^{p},\tau_{2}^{p}\right)+\dots\\  = & \sum_{a=1}^{K} \sum_{p=\Tilde{m}_{a-1}+1}^{\Tilde{m}_{a}} \sum_{b=1}^{K}\alpha_{ab}g_b\left(\tau_{1}^{p},\tau_{2}^{p}\right)\, ,
    \end{split}
\end{equation}
with $m_0=0$ and $\Tilde{m}_a = \sum_{i=0}^{a} m_a$. Since the theory is free, $\langle e^{\mathcal{O}_{\{m_a\}}} \rangle_0 = e^{\frac{1}{2} \langle \mathcal{O}_{\{m_a\}} \mathcal{O}_{\{m_a\}} \rangle_0 }$. The 2-point function evaluates to 
\begin{equation}
    \langle \mathcal{O}_{\{m_a\}} \mathcal{O}_{\{m_a\}} \rangle  
    = - 2 \sum_{\substack{a,a'=1 \\ a'<a}}^K \, \sum_{p=\Tilde{m}_{a-1}+1}^{\Tilde{m}_a} \, \sum_{\substack{p'=\Tilde{m}_{a'-1}+1 \\ p'<p}}^{\Tilde{m}_{a'}} \lambda_{a a'} |p\cap p'|,
\end{equation}
where as before $|p\cap p'|$ is $1$ if the chords stretching between $\tau_1^p, \tau_2^p$ and $\tau_1^{p'},\tau_2^{p'}$ cross, and zero otherwise. The partition function can then be rewritten as
\begin{equation}\label{partition function step x}
    \begin{split}
        Z=& \sum_{n=0}^{\infty}\frac{\bJ^{2n}}{\lambda^{n}n!}
        \sum_{m_{1}+\dots m_{K}=n}\left(\begin{array}{c}
        n\\
        m_{1},\dots,m_{K}
        \end{array}\right)\left(\prod_{a=1}^{K}\kappa_{a}^{2m_{a}}\right)\\ & \times \iint_{\Delta_{1}}\dots\iint_{\Delta_{n}}  \exp \left(- \sum_{\substack{a,a'=1 \\ a'<a}}^K \, \sum_{p=\Tilde{m}_{a-1}+1}^{\Tilde{m}_a} \, \sum_{\substack{p'=\Tilde{m}_{a'-1}+1 \\ p'<p}}^{\Tilde{m}_{a'}} \lambda_{a a'} |p\cap p'|\right)\, .
    \end{split}
\end{equation}
This expression contains a sum over chord diagrams with the correct exponential weights and the correct prefactors of the Hamiltonian insertions. It remains to check that the combinatorics work out, i.e. whether the chord diagrams appear with the correct numerical prefactor. We thus need to work out the integrations.

The integrand only depends on the pattern of crossings between the chords, and not on their precise locations. But each of the patterns is equally likely, as for a given choice of endpoints (set of $\tau$'s) we can choose how to contract and `color' the chords such that they give any pattern. Hence the prefactor of each chord configuration is given by the total area of integration divided by the number of possible diagrams. Each integration region $\Delta_p$ has an area of $\beta^2/2$. For the $n$-th term in the expansion, the total area covered is thus $\frac{\beta^{2n}}{2^n}$. The total number of chord configurations for a partition $\{m_a\}$ is equal to the number of indistinguishable Wick contractions of $2n$ insertions, $(2n-1)!!=\frac{(2n)!}{n! 2^n}$, multiplied by the multinomial coefficient which counts the different ways we can `color' the chord, $\binom{n}{m_1,\cdots,m_K}$. Then, evaluating \eqref{partition function step x} gives exactly \eqref{eq:moments as crossings}, finishing our proof.

In Section \ref{generic deformations}, we focus on some special cases when we have two types of chords $(K=2)$, and use the general parameterization 
\begin{equation}
    \alpha_{ij} = \begin{pmatrix}
        1 & \Delta \\
        \Delta & \alpha_{22} \\
    \end{pmatrix}.
\end{equation}
In this notation
\begin{itemize}
\item  $\alpha_{22}=\Delta^2$ is the familiar case of deforming by an SYK operator of a different length $\Tilde{p}=p\Delta$. After introducing the field 
\begin{equation}
    l = g_1 + \Delta g_2,
\end{equation}
the action becomes 
\begin{equation}\label{action degenerating}
    I[l] = - \frac{1}{4\lambda} \int_0^{\beta} d\tau_1 \int_0^{\beta} d\tau_2 \Bigg[ \frac{1}{2} l \partial_1 \partial_2 l  + 2 \kappa_1^2 \bJ^2 e^{l} + 2 \kappa_2^2 \bJ^2 e^{\Delta l} \Bigg].
\end{equation}
The action degenerates and only depends on one field, $l$.\footnote{A slightly more rigorous way to obtain this action is to start with $\alpha_{22}=\Delta^2-\epsilon$ and then take $\epsilon \rightarrow 0$. In this way, after canonically normalizing, one sees that the second field $g_2$ properly decouples.} It has also been derived from $G\Sigma$ methods \cite{Jiang:2019pam,Anninos:2020cwo,Anninos:2022qgy}. In Section \ref{subsec: spectrum at kappa^2}, we analyze the low--energy quadratic spectrum of these models at $\mathcal{O}(\kappa_2^2)$, using the IR description identified in Section \ref{sec: Schwarzian computation}.

\item $\alpha_{22}$=0 is the case of an integrable deformation of the SYK model discussed in \cite{Berkooz:2024ofm,Berkooz:2024evs}. In general, $\alpha_{22}<\Delta^2$ corresponds to a deformation by a polarized operator. We analyze the on--shell action of these generic deformations in Section \ref{subsec: the role of gz}.
\item $\Delta=1$ and $\alpha_{22}\neq 1$ is a case where we deform by an operator of the same length as the original one -- a naively marginal operator. Were $\alpha_{22}=1$, this would only change the normalization of the system, but this is not true for $\alpha_{22}\not=1$. Also in Section \ref{subsec: kappa^4 quadratic spectrum}, we analyze the quadratic spectrum of these special deformations. 
\end{itemize}

\subsection{Generating functionals, the bilocal, and the fake disk}
\label{sec: gen functionals}

In a very similar fashion, one can derive a Liouville theory expression for the generating functional of $n$-point functions of some operator $\mathcal{O}$. Consider a generalized DSSYK theory, with Hamiltonian $H = \sum_{i=1}^K \kappa_i H_i$ and intersection weights given by $\alpha_{ij}$. Let $\alpha_{i,K+1}=\alpha_{K+1,i}$ be the intersection weights of some operator $\mathcal{O}$ with $H_i$ and $\alpha_{K+1,K+1}$ be the self-intersection weight of $\mathcal{O}$. 
In the microscopic description, the generating functional for $n$-point functions of $\mathcal{O}$ is given by\footnote{A generating function in the  case $\lambda=0$ was discussed in \cite{Almheiri:2024xtw}.}
\begin{equation}
    Z[\chi]=\left\langle \Tr\left( \mathcal{T} \Bigl\{ e^{-\int_0^{\beta} d\tau \mathcal{H}_{\chi}(\tau)} \Bigl\} \right) \right\rangle_J, 
\end{equation}
with $\mathcal{T}$ denoting the time ordering operator and $\mathcal{H}_{\chi}(\tau)$ given as $\mathcal{H}_{\chi}(\tau) = H + \chi(\tau) \mathcal{O}$. Comparing this to a uniformly smeared out deformation (which was the case so far), we see that each power of $\mathcal{O}$ is accompanied by a factor of $\int_0^{\beta} d\tau \chi(\tau)$ instead of just $\beta$. On the other hand, on the Liouville theory side, each insertion of a $\mathcal{O}$-chord corresponds to bringing down an exponential $  e^{\sum_{j=1}^{K+1} \alpha_{K+1,j} g_j(\tau_1,\tau_2)}$ with $\tau_1,\tau_2$-integrations. 

In order to pass to the generating functional, we then simply have to replace the integrals by integrations weighed with $\chi(\tau)$:
\begin{equation}
    \begin{split}
        Z[\chi] = \int \left(\prod_{i=1}^{K+1} \mathcal{D}[ g_{i}]\right) \exp\Bigg[ &  \frac{1}{4\lambda}\int_0^\beta d\tau_{1}\int_0^{\beta}d\tau_{2}\Bigg[\sum_{i,j=1}^{K+1}\frac{1}{2}\alpha_{ij}g_{i}\partial_{1}\partial_{2}g_{j} + \sum_{i=1}^K 2 \kappa_{i}^2 \bJ^2 e^{\sum_{j=1}^{K+1}\alpha_{ij}g_{j}} \\ & + 2 \chi(\tau_1)\chi(\tau_2)\bJ^2 e^{\sum_{j=1}^{K+1} \alpha_{K+1,j} g_j} \Bigg] \Bigg] \,.
    \end{split}
\end{equation}
We can use this expression to derive $n$--point functions of the operator $\mathcal{O}$ in the Liouville language. For example, the 2-point function of the operator $\mathcal{O}$ is given as\footnote{The factor of $\lambda$ has to be supplied since in this choice of convention, the variance of the random couplings scales like $1/\lambda$ \eqref{eq: multi-SYK conventions}.}
\begin{equation}
    \langle \mathcal{O}(\tau_1) \mathcal{O}(\tau_2) \rangle = \lambda \frac{\delta}{\delta \chi(\tau_1)} \frac{\delta}{\delta \chi(\tau_2)}Z[\chi]\big|_{\chi=0} = \langle \bJ^2 e^{\sum_{j=1}^{K+1} \alpha_{K+1,j} g_j(\tau_1,\tau_2)} \rangle.
\end{equation}
We see that in order to compute an $n$-point function of some operator, it in general does not suffice to just compute some local correlator in the background theory; it is instead more economical to introduce an additional field. 

\paragraph{The bilocal and the fake disk} 
It is instructive to study the generating functional of a fully random operator with $\alpha_{22}= \Delta^2$. This is the case of an additional SYK-like operator. As before, only one linear combination of the two fields survive. The corresponding Liouville action is given by
\begin{equation}
    I = - \frac{1}{4\lambda} \int_0^{\beta} d\tau_1 \int_0^{\beta} d\tau_2 \left[ \frac{1}{2}l \partial_1 \partial_2 l + 2\bJ^2 e^{l} + 2\bJ^2 e^{\Delta l} \chi(\tau_1)\chi(\tau_2) \right].
\end{equation}
For $\chi=0$, the theory reduces to ordinary DSSYK, and it has the usual saddle and twisted reparametrizations that were observed in the previous section.

\begin{figure}
    \centering
    \includegraphics[scale=0.4]{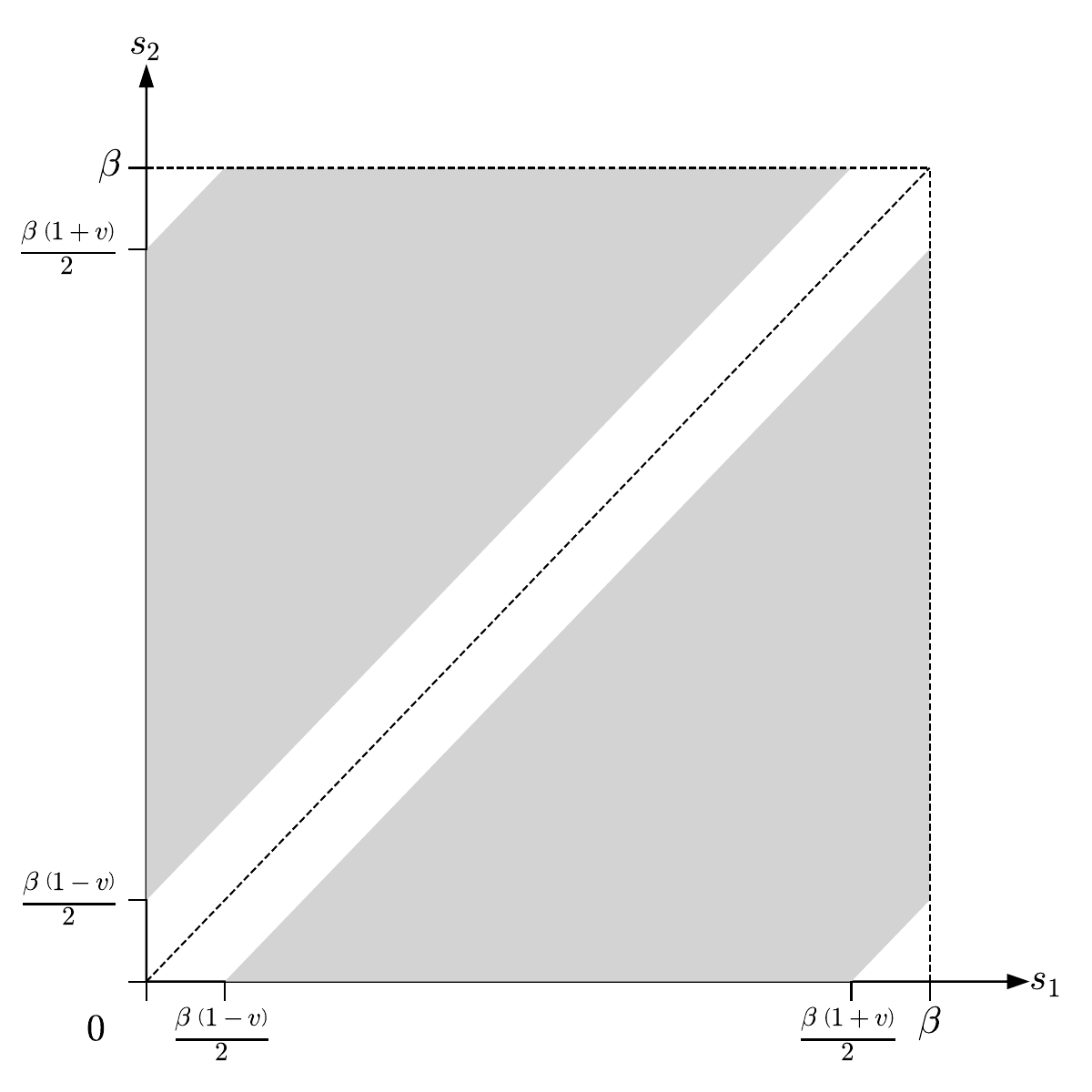}
    \caption{Integration region in the $\left( s_1, s_2 \right)$ coordinates, by reordering of Eq \eqref{tau1,tau2 in terms of s1,s2}. }
    \label{fig:fake disk reordering}
\end{figure}

On the soft mode manifold of twisted reparametrizations, $l$ is given as
\begin{equation}
    l_{\phi}(s_1,s_2)= l^{(0)}(\phi(s_1),\phi(s_2)) + \log \phi'(s_1) + \log \phi'(s_2).
\end{equation}
The source term in the Liouville action can then be written as\footnote{There is another effect that we ignore here: the deformation also induces a change of the soft mode manifold, encoded by a change of the saddle. We will discuss it in Section \ref{subsec: spectrum at kappa^2}.}
\begin{equation}
    \delta I[l_{\phi}] =  - \frac{\bJ^2}{2\lambda} \int_0^{\beta} d\tau_1 \chi(\tau_1) \int_0^{\beta} d\tau_2 \chi(\tau_2) \left(  \frac{\cos^2 (\frac{\pi v}{2} )\phi'(s_1) \phi'(s_2)}{\sin^2(\frac{\pi}{\beta}(\phi(s_1)-\phi(s_2)))} \right)^{\Delta} \,.
\end{equation}
Rewriting the integration region in terms of $s_1,s_2$ reveals
\begin{equation}\label{eq: bilocal}
    \delta I[l_{\phi}] =  - \frac{\bJ^2}{2\lambda v} \cos^{2\Delta}(\frac{\pi v}{2}) \int_{\mathcal{I}} ds_1 ds_2 \chi(\tau_1(s_1,s_2)) \chi(\tau_2(s_1,s_2)) \left(\frac{\phi'(s_1)\phi'(s_2)}{\sin^2 \frac{\pi}{\beta}(\phi(s_1)-\phi(s_2))}\right)^{\Delta},
\end{equation}
with
\begin{equation}\label{tau1,tau2 in terms of s1,s2}
    \begin{split}
        \tau_1 = \frac{1}{2}\left(1+\frac{1}{v}\right)s_1 + \frac{1}{2}\left(1-\frac{1}{v}\right)s_2 - \frac{\beta (1-v)}{4v} \\
        \tau_2 = \frac{1}{2}\left(1+\frac{1}{v}\right)s_2 + \frac{1}{2}\left(1-\frac{1}{v}\right)s_1 + \frac{\beta (1-v)}{4v}.
    \end{split}
\end{equation}
The integration region $\mathcal{I}$ can be reordered to the shape of Figure \ref{fig:fake disk reordering}.
Notice that the integrand looks just like the usual bilocal deformation \cite{Maldacena:2016upp}. The only peculiarity is the integration region. At low temperatures, as long as $\tau_1$ and $\tau_2$ are separated by a distance
\begin{equation}
    |\tau_1-\tau_2| \gg \frac{\beta (1-v)}{2}\approx \frac{1}{\bJ},
\end{equation}
we have $s_1 \approx \tau_1$, $s_2 \approx \tau_2$, and the result looks exactly like the JT bilocal deformation. When $s_1$ gets close to $s_2$, however, the result starts to deviate: from Figure \ref{fig:fake disk reordering}, it is apparent that $s_1$ and $s_2$ are always separated from each other in the integration region:
\begin{equation}
    |s_1-s_2| \geq \frac{\beta (1-v)}{2} \approx \frac{1}{\bJ}.
\end{equation}
Something like this is to be expected, as the JT expression for the bilocal deformation is UV divergent and DSSYK is UV finite; we thus expect that it provides its own UV regulator, as happens here. Moreover, the original distribution $\chi(\tau_1)\chi(\tau_2)$ is squashed to fit into the integration region of Figure \ref{fig:fake disk reordering}. This kind of integration region is reminiscent of the fake disk \cite{Lin:2023trc}: understanding the 2-point function as an overlap of an in-state with an out-state, it tells us that an operator insertion for the in-state is always separated by at least a distance of $1/\bJ$ from any operator insertion for the out-state. Recall that this separation was also the reason why \eqref{eq: twisted coordinate reparametrization of g} only softly violates the boundary conditions.

In the same way, the integration region can be understood as enforcing the length positivity constraint of \cite{Blommaert:2024ydx}: the holographically renormalized length between two points $s_1,s_2$ on the boundary circle is given by 
\begin{equation}
    L = 2 \log \left(\frac{\beta \bJ}{\pi}\right) + 2 \log \left[\sin \left(\frac{\pi}{\beta}|s_1-s_2|\right)\right].
\end{equation}
At low temperatures, this expression is positive as long as $|s_1-s_2|\geq \frac{\beta (1-v)}{2} \approx \frac{1}{\bJ}$, 
which is the same condition as the one arising from the integration region. 

There is also a heuristic microscopic explanation of the latter: at low temperatures, the average number of Hamiltonian insertions is given by\footnote{This can, for example, be seen by using the techniques of \cite{Berkooz:2024ofm}: the total number of insertions (which is two times the total number of chords) is given by 
\begin{equation}
    \frac{1}{\lambda} \int_0^{\beta} d\tau_1 \int_0^{\beta} d\tau_2 n(\tau_1,\tau_2) = \frac{1}{\lambda}\int_0^{\beta} d\tau_1 \int_0^{\beta} d\tau_2 \bJ^2 \frac{\cos^2 \left(\frac{\pi v}{2}\right)}{\cos^2 \left( \frac{\pi v}{2}(1-2\frac{|\tau_1-\tau_2|}{\beta}) \right)} =  2\frac{\beta \bJ}{\lambda} \sin \left(\frac{\pi v}{2}\right).
\end{equation}
}
\begin{equation}
    \# \text{Hamiltonian insertions} \sim \frac{\beta \bJ}{\lambda}.
\end{equation}
The average density of Hamiltonian insertions on the thermal circle is thus
\begin{equation}
    \sim \frac{\bJ}{\lambda}.
\end{equation}
On distances less than $1/\bJ$, there are then on average less than $1/\lambda$ insertions. The $q$-factors can then essentially be approximated to $1$; the matter operator can be swapped with the Hamiltonian insertions essentially `free of charge', and there is no nontrivial contribution to the 2-point function. 

\section{Low energy physics of generic deformations}
\label{generic deformations}

\begin{figure}
    \centering
    \begin{subfigure}{.5\textwidth}
        \centering
        \includegraphics{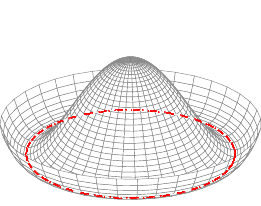}
    \end{subfigure}%
    \begin{subfigure}{0.5\textwidth}
        \centering 
        \includegraphics{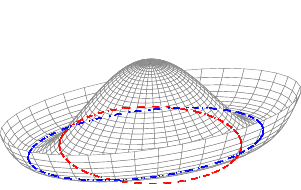}
    \end{subfigure}
    \caption{Left: schematic picture of the undeformed soft mode manifold. Right: the deformed soft mode manifold: both the location of the manifold (here encoded by the size of the circle) as well as the potential on it (here encoded by the tilt of the circle) can have an impact.}
    \label{fig:low energy def}
\end{figure}

Studying the low energy theory of deformations amounts to studying how physics on the Schwarzian soft mode manifold is modified. This can generically happen in two ways: both the potential on and the location of the soft mode manifold can change as a result of the deformation (see Figure \ref{fig:low energy def} for a schematic picture).

In previous works \cite{Maldacena:2016upp,Milekhin:2021cou,Milekhin:2021sqd} it wasn't known explicitly how the Schwarzian sector is embedded into DSSYK, hence the bilocal deformation \eqref{eq: bilocal} was only treated as changing the potential on the soft mode manifold rather than changing the location of the soft mode manifold itself. Here, we also address the latter effect, using our results of Section \ref{sec: multifield Liouville} to formulate deformations in the Liouville language, and using the techniques of Section \ref{sec: Schwarzian computation} to identify the IR theory inside the Liouville description. In general, both effects are important.

The main points of this Section are the following:
\begin{itemize}
   \item In Section \ref{subsec: spectrum at kappa^2}, we discuss how a generic deformation of strength $\kappa$ modifies the quadratic spectrum. We focus on effects at $\mathcal{O}(\kappa^2)$, the lowest order\footnote{Odd powers of $\kappa$ do not enter, as the random coupling associated to the deformation has zero mean.} where the deformation enters.
   In Section \ref{subsubsec: Delta greater 1.5} we discuss deformations with conformal dimension $\Delta > 3/2$. Such a perturbation decreases in strength in the IR. However, it can renormalize the coefficient in the leading effective action, which indeed happens, and we compute the shift in the coefficient of the Schwarzian. This is an example where the change of the saddle point matters. The shift can become large when $\Delta\rightarrow 3/2$. Previously, numerical evidence for such a shift was found in \cite{Anninos:2022qgy}.
   
   \item In Section \ref{subsubsec: Delta smaller 1.5} we discuss $\Delta < 3/2$. The usual treatment with a bi-local operator \cite{Maldacena:2016upp} indicates a large correction in the IR which is far from a Schwarzian. But by itself it has the wrong sign for the fluctuations, i.e. it destabilizes the original saddle point. We show that this instability is remedied by a shift in the saddle point to give an action which is far from a Schwarzian and, to the order checked, is the bilocal action with a correct sign. Numerical evidence for this sign was found in a model of coupled SYKs \cite{Milekhin:2021cou}.
    We also show that the strength of the deformation of the quadratic spectrum is not controlled by the same parameter as the deformation of the saddle point, allowing for a scenario where the 2-point function is essentially unchanged, but the spectrum strongly deviates from the Schwarzian. 
     
    \item In Section \ref{subsec: the role of gz}, we go beyond the bilocal deformation and discuss how the saddle point and the quadratic spectrum are modified by effects at $\mathcal{O}(\kappa^4)$. These are important when $\alpha_{22}<\Delta^2$ ($\Delta$ being the deforming operators conformal dimension, and  $\alpha_{22}$ being associated with its 4-point function -- see Section \ref{sec: multifield Liouville}) . 
     We first discuss the impact of $\alpha_{22}<\Delta^2$ on the IR--saddle point of the generating functional and, correspondingly, on the (classical) 4-point functions of the deformation operator (2-point functions are not modified, as $\alpha_{22}$ is only relevant for self--intersections of the chords corresponding to the deformation, which are at least proportional to $\kappa^4$). $\alpha_{22}$ thus influences the backreaction of the deformation onto itself. Correspondingly, $\alpha_{22}$ is encoded in a term at $\mathcal{O}(\kappa^4)$ which corresponds to a 4-point function. We explain how this term comes about and how to handle it. 
    We then specialize to a deformation with conformal dimension $1$ and generic $\alpha_{22}$. Such a deformation only modifies the theory at $\mathcal{O}(\kappa^4)$. We show that the quadratic spectrum at this order consists of two parts: one of them is a change of the potential on the soft mode manifold, encoded by reparametrizing the 4-point function term found in previous parts of this Section, and the other one once again corresponds to a change of location of the soft mode manifold. 
\end{itemize}

\subsection{The generic quadratic spectrum at $\mathcal{O}(\kappa^2)$}
\label{subsec: spectrum at kappa^2}

Recall that in Section \ref{sec: gen functionals}, we have seen that at $\mathcal{O}(\kappa^2)$, a deformation of DSSYK can be tracked into the IR as the usual bilocal operator. This bilocal operator modifies the spectrum in two ways: it changes the potential on the soft mode manifold, and it changes the location of the soft mode manifold in field space (i.e. the saddle point). Previous works \cite{Milekhin:2021cou,Milekhin:2021sqd,Maldacena:2016upp} focused mostly on the former effect. We show that both for $\Delta < 3/2$ and $\Delta > 3/2$, the changing location of the soft mode manifold is significant. For $\Delta > 3/2$, the deformation results at leading order in a modification of the Schwarzian coefficient. Only the change of the saddle point contributes to that; the potential on the soft mode manifold is only modified at subleading order in $1/ (\beta \bJ)$. For $\Delta<3/2$, we confirm that the Schwarzian is not the dominant contribution to the low--energy spectrum \cite{Milekhin:2021cou,Milekhin:2021sqd,Maldacena:2016upp}, however, the change of the saddle is also important here, as it ensures the positivity of the low--energy modes.

First, we show explicitly that the self--intersection weight $\alpha_{22}$ is not relevant at $\mathcal{O}(\kappa^2)$ (as expected since it encodes a backreaction of the matter operator onto itself) and discuss properties of the quadratic spectrum for generic $\Delta$. We then specialize to $\Delta>3/2$ and $\Delta<3/2$ at a natural point.

The action for a generic deformation of scaling dimension $\Delta$ and self-intersection weight $\alpha_{22}$ is given by (see \eqref{eq:multiLiouville})
\begin{equation}
    I = - \frac{1}{4\lambda} \int_0^{\beta} d\tau_1 \int_0^{\beta} d\tau_2 \Bigg[ \frac{1}{2} \begin{bmatrix}
        g_1 & g_2
    \end{bmatrix}
    \begin{bmatrix}
        1 & \Delta \\
        \Delta & \alpha_{22} 
    \end{bmatrix}\partial_1 \partial_2 \begin{bmatrix}
        g_1 \\
        g_2 
    \end{bmatrix} + 2 \kappa_1^2 \bJ^2 e^{g_1 + \Delta g_2} + 2 \kappa_2^2 \bJ^2 e^{\Delta g_1 + \alpha_{22} g_2}  \Bigg].
\end{equation}
For simplicity, we rewrite $l=g_1+\Delta g_2$, and set $\kappa_1=1$, $\kappa_2 = \kappa$ to get
\begin{equation}
    I = - \frac{1}{4\lambda} \int_0^{\beta} d\tau_1 \int_0^{\beta} d\tau_2 \Bigg[ 
    \frac{1}{2} l \partial_1 \partial_2 l - \frac{\Delta^2- \alpha_{22}}{2} g_2 \partial_1 \partial_2 g_2 + 2 \bJ^2 e^l + 2 \kappa^2 \bJ^2 e^{\Delta l -(\Delta^2 - \alpha_{22}) g_2}
    \Bigg].
\end{equation}
We write the saddle point in a $\kappa^2$--expansion, $l = l^{(0)} + \kappa^2 l^{(1)}+\dots$ and $g_2 = g_2^{(0)} + \kappa^2 g_2^{(1)}+\dots $.
The resulting equations of motion for $l^{(0)}, l^{(1)}, g_2^{(0)}, g_2^{(1)}$ are
\begin{equation}\label{eq: kappa^2 equation of motion}
    \begin{aligned}
        &\partial_1 \partial_2 l^{(0)} + 2\bJ^2 e^{l^{(0)}} = 0 \,, & & \partial_1 \partial_2 g_2^{(0)} = 0 \,,\\
        &\partial_1 \partial_2 l^{(1)} + 2\bJ^2 e^{l^{(0)}} l^{(1)}  + 2 \Delta \bJ^2 e^{\Delta l^{(0)}}  = 0 \,,
        & &\partial_1 \partial_2 g_2^{(1)} + 2 \bJ^2 e^{\Delta l^{(0)}} =0 \, .
    \end{aligned}
\end{equation}
By the boundary and symmetry condition, $g_2^{(0)}$ is zero. Then the quadratic action around the saddle point at $\mathcal{O}(\kappa^2)$ reads
\begin{equation}
    \begin{split}
        I = - \frac{1}{4\lambda} \int_0^{\beta} d\tau_1 \int_0^{\beta} d\tau_2 \Bigg[ 
        & \frac{1}{2} \delta l \partial_1 \partial_2 \delta l - \frac{\Delta^2- \alpha_{22}}{2} \delta g_2 \partial_1 \partial_2 \delta g_2 + \bJ^2 e^{l^{(0)}} \delta l^2  \\ + & \kappa^2 \bJ^2 e^{l^{(0)}} l^{(1)} \delta l^2 +  \kappa^2 \bJ^2 e^{\Delta l^{(0)}}(\Delta \delta l - (\Delta^2 -\alpha_{22})\delta g_2)^2
    \Bigg]\, .
    \end{split}
\end{equation}
We observe two properties:\\
(1) There is mixing between the $\delta l, \delta g_2$-modes, but only at $\mathcal{O}(\kappa^2)$. The spectrum at $\mathcal{O}(\kappa^2)$ is unaltered by these mixings, as familiar from standard QM perturbation theory.\\
(2) There is a scale separation between the $\delta l, \delta g_2$ modes: at $\mathcal{O}(\kappa^0)$, the quadratic spectrum of $g_2$ is found by solving the equation
\begin{equation}
    \partial_1 \partial_2 g_2 = \xi_2 g_2,
\end{equation}
with $\xi_2$ the eigenvalue of question. Going to $\tau_+,\tau_-$ coordinates, we find that the spectrum is given by $\xi_2^{n,m} = \pi^2(n^2 -m^2)/\beta^2 $, with $n$ and $m$ integers. As in \eqref{eq: high temp k}, the  boundary and symmetry conditions on $g_2$ then enforce the constraint that $n+m$ has to be odd. Therefore, any $\delta g_2$ excitation has eigenvalue $\sim (\beta \bJ)^0$, which at low temperatures is arbitrarily heavy compared to the Schwarzian modes with eigenvalue $1/(\beta \bJ)$. At $\mathcal{O}(\kappa^2)$, 
\begin{equation}\label{changeo f spectrum}
    \delta \lambda_n \equiv - \frac{\kappa^2 \beta^2 \bJ^2}{ 4 \pi v \lambda} \int_{-\frac{\pi v}{2}}^{\frac{\pi v}{2}} dx \delta l_{n}(x)^* (\Delta^2 e^{\Delta l^{(0)}}+l^{(1)}e^{l^{(0)}})(x) \delta l_n(x) \equiv \delta \lambda_n^1 + \delta \lambda_n^2
\end{equation}
therefore captures the change of eigenvalues of the Schwarzian modes. $\delta l_n$ is given by the modes of \eqref{eq: specific light modes}, where (as in \eqref{changeo f spectrum}), we have already integrated out the $y$ -- direction. By $\delta \lambda_n^1$ we denote the change of the spectrum independent of $l^{(1)}$, while $\delta \lambda_n^2$ denotes the change of the spectrum proportional to $l^{(1)}$.

We now analyze the individual terms in \eqref{changeo f spectrum}. The first term can be understood as coming from the bilocal deformation: using the fact that the $\delta l_n$'s are infinitesimal twisted reparametrizations \footnote{Up to a small piece which was discussed in Section \ref{sec: Schwarzian computation} and is irrelevant here.}, it is easy to see that we can understand $\delta \lambda_n^1$ as the bilocal deformation expanded to quadratic order in an infinitesimal twisted reparametrization. This term thus encodes the change of the potential on the soft mode manifold. Its magnitude can be estimated straightforwardly: 
$e^{\Delta l^{(0)}}$ is given by 
    $\left(\frac{\cos^2 \left(\frac{\pi v}{2}\right)}{\cos^2 x} \right)^{\Delta} = \frac{\pi^{2\Delta}}{(\beta \bJ)^{2\Delta} \cos^{2\Delta} x} \left( 1+ \mathcal
    O\left(\frac{1}{\beta \bJ}\right)\right)$,
and from \eqref{changeo f spectrum}, the correction to the spectrum associated with $\delta \lambda_n^1$ should scale as
\begin{equation}\label{eq:SimplCorrection}
    \delta \lambda_n ^{1} \sim  \kappa^2 (\beta \bJ)^{2-2\Delta},
\end{equation}
where we have used the fact that in the bulk of the integration, $1/\cos^{2\Delta} x$  is an order one positive number.\footnote{We can also estimate a potential boundary contribution, where $1/\cos^{2\Delta} (x)$ becomes large: a distance $\delta v$ away from the boundary, the modes have value $\sim \delta v^2$. $e^{l^{(0)}}$ is of order 1 there, such that the boundary contribution to the change of the spectrum can be estimated to be
\begin{equation}
    \delta \lambda_n^{1,bdy} \sim \kappa^2 (\beta \bJ)^2 \delta v (\delta v^2)^2 \sim \frac{\kappa^2}{\lambda} (\beta \bJ)^{2-2\Delta} (\beta \bJ)^{2\Delta-5},
\end{equation}
which is small compared to the bulk contribution as long as $\Delta < 5/2$.}

The second term in \eqref{changeo f spectrum} involving $l^{(1)}$ is a new effect. It is harder to estimate, as it involves a change of the saddle. For $\Delta>3/2$, this change turns out to be larger than the naive scaling result obtained by estimating $l^{(1)}$ from its equation of motion, the latter of which agrees with \eqref{eq:SimplCorrection}. Rather the correction scales like $\delta \lambda_n^2 \sim \frac{\kappa^2}{\beta \bJ}$, so it is dominant in this regime of $\Delta$. 

The reason for this is that $l^{(1)}$ consists of a homogeneous and a particular solution to the equations of motion \eqref{eq: kappa^2 equation of motion}. The homogeneous solution can be scaled arbitrarily and is fixed by imposing the boundary conditions $l^{(1)}(\pm \frac{\pi v}{2})=0$. If the asymptotic behaviour of the homogeneous and particular solution towards the boundary is different, then the coefficients of the homogeneous solution can be enhanced (or suppressed), changing the scaling behaviour of the saddle.

In order to investigate whether this happens, we have to solve the equation for $l^{(1)}$ explicitly, which reads
\begin{equation}
    \partial_1 \partial_2 l^{(1)}  + 2 \bJ^2 e^{l^{(0)}} l^{(1)} + 2  \Delta \bJ^2 e^{\Delta l^{(0)}} = 0.
\end{equation}
Going to the usual $x$--coordinate and plugging in $l^{(0)}$  gives the equation
\begin{equation}
    \left(\partial_x^2 - \frac{2}{\cos^2 x}\right) \Bigl[\cos^{2-2\Delta} \left(\frac{\pi v}{2}\right) l^{(1)}(x)\Bigl] = \frac{2\Delta }{\cos^{2\Delta}(x)}.
\end{equation}
We see that $\cos^{2\Delta-2}\left(\frac{\pi v}{2}\right) \sim (\beta \bJ)^{2-2\Delta}$ is the naive scaling of $l^{(1)}$. Solving the equation and imposing the boundary conditions gives
\begin{equation} \label{eq: l1_label}
    \begin{split}
        \cos^{2-2\Delta} \left(\frac{\pi v}{2}\right) l^{(1)}(x)=& \frac{x \tan x + 1}{\frac{\pi v}{2} \tan \frac{\pi v}{2} + 1} \frac{1}{2} \text{Re}\left[\frac{1}{\Delta}  \, _2 F_1 \left(1,\Delta-\frac{1}{2},\Delta+1,\sec^2 \left(\frac{\pi v}{2}\right) \right) \sec^{2\Delta}\left(\frac{\pi v}{2}\right)\right]\\ - & \frac{1}{2} \text{Re}\left[\frac{1}{\Delta}  \, _2 F_1 \left(1,\Delta-\frac{1}{2},\Delta+1,\sec^2 x\right) \sec^{2\Delta}(x)\right].
    \end{split}
\end{equation}
The second term (the particular solution) is indeed order $1$ as long as $x$ is not close to the boundary. The first term however, exactly due to the boundary conditions, is not: at low temperatures, it is given by
\begin{equation}
    (x \tan x + 1) (\#_1 \delta v^{3-2\Delta} + \#_2 \delta v^0),
\end{equation}
where $\#_1,\#_2$ are coefficients whose precise form is not relevant for now.

There are then two qualitatively different behaviours depending on $\Delta$: for $\Delta < 3/2$, the leading scaling of $l^{(1)}$ is
\begin{equation}\label{Delta < 3/2}
    l^{(1)} \sim (\beta \bJ)^{2-2\Delta},
\end{equation}
which agrees with the naive scaling, while for $\Delta >3/2$, we instead have
\begin{equation}\label{Delta > 3/2}
    l^{(1)} \sim \frac{1}{\beta \bJ},
\end{equation}
which is larger than the naive scaling suggests.
Correspondingly, we will now distinguish between deformations of $\Delta > 3/2$ and ones that are with $\Delta < 3/2$.

\subsubsection{$\Delta > \frac{3}{2}$}
\label{subsubsec: Delta greater 1.5}
For $\Delta > \frac{3}{2}$, we analyze how the deformation impacts the low--energy theory. We will find that due to the change of the saddle, the low--energy Schwarzian spectrum will be modified by a $\beta \bJ$--independent prefactor. We additionally support this statement with an analysis of the on--shell action.

\paragraph{Change of the fluctuations} 
From the change of the saddle \eqref{Delta > 3/2},  $\delta \lambda_n^2$ can be estimated from \eqref{changeo f spectrum} to be
\begin{equation}
    \delta \lambda_n^2 \sim \kappa^2(\beta \bJ)^{2-2\Delta} (\beta \bJ)^{2\Delta -3} \sim \frac{\kappa^2}{\beta \bJ}. 
\end{equation}
It thus changes the spectrum at the same order (at $\mathcal{O}(1/\beta \bJ)$) as the Schwarzian contribution, and it is larger than the contribution coming from $\delta \lambda_n^1$.\footnote{ Using the form of $l^{(1)}$, the contribution of the integral close to the boundary can be estimated to be $\delta \lambda_n^{2,bdy} \sim \delta v^3$,
and is thus always subleading compared to the bulk term.} 
Of course, irrelevant operators can renormalize the coefficients of remaining low dimension operators, but in this case the contribution can be large. Also, if we are expecting to renormalize the low energy parameters, we expect to get a Schwarzian with a different coefficient.     
In fact, we can compute the change of the spectrum explicitly: as discussed around \eqref{eq: l1_label}, for $\Delta > 3/2$, the leading contribution to $l^{(1)}$ is given by 
\begin{equation}
    \cos^{2-2\Delta} \left(\frac{\pi v}{2}\right) l^{(1)}(x)=\frac{x \tan x + 1}{\frac{\pi v}{2} \tan \frac{\pi v}{2} + 1} \frac{1}{2} \text{Re}\left[\frac{1}{\Delta} \left( \, _2 F_1 (1,\Delta-\frac{1}{2},\Delta+1,\sec^2 \frac{\pi v}{2}) \right)\right].
\end{equation}
Since the $x$--dependence is relatively simple, one can use \eqref{changeo f spectrum} and compute $\delta \lambda_n^2$ explicitly. This gives 
\begin{equation}
    \delta \lambda_n^2 = \frac{\kappa^2 \pi^2}{16\lambda \beta \bJ} \frac{1}{\Delta - \frac{3}{2}} n^2(n^2-1) + \{\text{subleading}\}. 
\end{equation}
The $n$--dependence is once again a Schwarzian.
Comparing to the unperturbed theory, we see that at leading order at low temperatures, the Schwarzian inverse coupling is modified to
\begin{equation}
    \frac{1}{2\lambda \beta \bJ} \mapsto \frac{1}{2\lambda \beta \bJ } \left( 1 + \frac{\kappa^2}{4(\Delta - \frac{3}{2})}\right).
\end{equation}
In \cite{Anninos:2022qgy}, some numerical evidence was presented that the inverse coupling of the Schwarzian is modified by such deformations, but the explicit coefficient was not known. 
It would be interesting to see if this picture persists beyond the quadratic order.

\paragraph{Correction to the on--shell action}

\begin{equation}\label{bilocal on saddle point}
    -\frac{\kappa^2 \bJ^2}{2\lambda} \iint e^{\Delta l^{(0)}} = -\frac{\kappa^2(\beta \bJ)^2 \cos^{2\Delta}(\frac{\pi v}{2})}{2\lambda \pi v} \int_{-\pi v/2}^{\pi v/2} dx \frac{1}{\cos^{2\Delta }x}.
\end{equation}
We now explain how the $\kappa^2/(\beta \bJ)$--contribution arises and argue why it hasn't appeared in \cite{Maldacena:2016upp}, where the authors also computed the on--shell action for generic $\Delta$.

Integrating explicitly, we find the expression
\begin{equation}\label{kappa^2 onshell action}
    \begin{split}
        I_{on-shell} = & I_0 - \frac{\kappa^2}{2\lambda} \pi v \cos^{2\Delta -2}\left(\frac{\pi v}{2}\right) \Gamma\left(\frac{1}{2}-\Delta\right) \Bigg( \frac{\sqrt{\pi}}{\Gamma\left(1-\Delta\right)} \\ - & \cos\left(\frac{\pi v}{2}\right)^{1-2\Delta}  \frac{\,_2F_1\left(\frac{1}{2},\frac{1}{2}-\Delta,\frac{3}{2}-\Delta,\cos (\frac{\pi v}{2})^2\right)}{\Gamma\left(\frac{3}{2}-\Delta\right)}\Bigg) + \mathcal{O}(\kappa^4),
    \end{split}
    \end{equation}
with $I_0$ denoting the DSSYK on--shell action.
At low temperatures, for $\Delta > 3/2$, this becomes
\begin{equation}
    I_{on-shell} = - \frac{2\beta \bJ}{\lambda}\left(1+ \frac{\kappa^2}{4(\Delta - \frac{1}{2})}\right) + \frac{\pi^2}{2\lambda} - \frac{\pi^2}{\lambda \beta \bJ}\left( 1 + \frac{\kappa^2}{4(\Delta - \frac{3}{2})} \right) + \{\text{subleading}\},
\end{equation}
where we have used \eqref{eq: action in the saddle} for the zeroth order on--shell action. The first term can be understood as a shift of the ground state energy, while the $\beta \bJ$--independent term changes the entropy by an overall constant. The (classical) thermodynamics of the Schwarzian theory originates from the term $\sim 1/(\beta \bJ)$; we see that its prefactor is changed by the same factor as the prefactor of the quadratic Schwarzian coefficient. 

How does that term arise?
Performing the integration of \eqref{bilocal on saddle point} gives two types of contributions: one arising from the bulk of the integration range, and a second arising from the divergent behaviour of the integrand near the boundaries. In UV divergent theories (like JT), one takes care of the latter contribution by introducing a UV cutoff, subtracting the divergent piece, and sending the 
cutoff to infinity. Here, since DSSYK is UV finite, the integration region is automatically regulated with a finite cutoff $\sim 1/\bJ$, and thus there are finite contributions. One of these contributions is the change of the $1/(\beta \bJ)$--coefficient (another is the change of the ground state energy).\footnote{This also explains why this feature hasn't been observed in the JT calculation of \cite{Maldacena:2016upp}.}

Both the change in the on--shell action and the change in the quadratic spectrum can then be traced to a similar origin: in one case, it is the UV boundary condition which leads to an enhancement of the naive scaling behaviour, while in the other case it is the fact that the theory is UV finite.  

\subsubsection{$\Delta<\frac{3}{2}$}
\label{subsubsec: Delta smaller 1.5}

Now we analyze the situation in which $\Delta < 3/2$.
In \cite{Maldacena:2016upp}, it was pointed out that in JT gravity, deformations corresponding to scaling dimension $1<\Delta < \frac{3}{2}$ dominate over the Schwarzian in the IR, at least at the level of the on--shell partition function. In \cite{Milekhin:2021cou,Milekhin:2021sqd}, this feature has been studied in coupled SYK models. We reanalyze the phenomenon from the perspective of the quadratic spectrum\footnote{Repeating the on--shell action analysis gives the same results as \cite{Maldacena:2016upp}, except for additional contributions due to the UV finiteness (as discussed in the previous Section), which do not change the qualitative result for $\Delta < 3/2$.}
and point out that also here, the change of location of the soft mode manifold plays an important role, ensuring positivity of the light spectrum.

For $\Delta < \frac{3}{2}$, we have from \eqref{Delta < 3/2} the scaling behaviours  
\begin{equation}\label{change of eigenvalues schematically}
    \delta \lambda_n^1 \sim - \kappa^2 (\beta \bJ)^{2-2\Delta}, \, \qquad \delta \lambda_n^2 \sim \kappa^2 (\beta \bJ)^{2-2\Delta}. 
\end{equation}
Explicit numerical integration shows that there is no accidental cancellation between the two.\footnote{Except for $\Delta=1$, i.e. a deformation by a same length operator, which doesn't change the system up to a normalization, and results in changing $\bJ$ to $\bJ\sqrt{1+\kappa^2}$.}

In order to see when the deviation from the Schwarzian spectrum becomes strong, we have to compare the result \eqref{change of eigenvalues schematically} to the unperturbed eigenvalues, which are of order $1/(\beta \bJ)$. The deviation from the Schwarzian is thus controlled by the parameter $\kappa^2 (\beta \bJ)^{3-2\Delta}$. At constant $\kappa^2$ and in the range $1<\Delta <3/2$, this parameter indeed becomes large as we go to low temperatures, dominating over the quadratic Schwarzian spectrum. Note the negative coefficient for $\delta \lambda_n^1$ in \eqref{changeo f spectrum}, which means that the saddle point $l^{(0)}$ is no longer reliable, and we need to keep track of its shift, encoded in $\delta \lambda_n^2$.

In general, the integrals of \eqref{changeo f spectrum} are not accessible analytically. Numerically, it can be confirmed that $\delta \lambda_n^1$ scales with $n$ as 
\begin{equation}
    \delta \lambda_n^1 \sim - \frac{\kappa^2 (\beta \bJ)^{3-2\Delta}}{\lambda\beta\bJ} n^{1+2\Delta}.
\end{equation}
This is just the bilocal for an infinitesimal reparametrization \cite{Maldacena:2016upp}.
The numerical result for $\delta \lambda_n^2$ is
\begin{equation}
    \delta \lambda_n^2 \sim \frac{\kappa^2 (\beta \bJ)^{3-2\Delta}}{\lambda \beta \bJ} n^{1+2\Delta}.
\end{equation}
This contribution comes with the opposite sign compared to $\delta \lambda_n^1$. Adding them both up shows that $\delta \lambda_n^2$ is of larger absolute value than $\delta \lambda_n^1$; the total result evaluates to 
\begin{equation}
    \delta \lambda_n = \delta \lambda_n^1 + \delta \lambda_n^2 = C \frac{\kappa^2(\beta \bJ)^{3-2\Delta}} {{\lambda \beta \bJ}} n^{1+2\Delta},
\end{equation}
with $C$ a positive order one number. The change of the saddle is thus important to get a positive correction to the Schwarzian eigenvalues, making the light spectrum bounded from below. At low energies, the light modes describing fluctuations around the saddle do not mix with the heavy ones\footnote{The heavy modes have mass of order $(\beta \bJ)^0$ or larger, and $\kappa^2 (\beta \bJ)^{2-2\Delta}$ is small compared to 1. Thus no perturbation is strong enough to make one of the heavy modes light or vice versa.} and the low--energy theory is a nonlocal theory of reparametrizations.

It is also instructive to note that $\kappa^2 (\beta \bJ)^{2\Delta -3}$ is \text{not} the same parameter that controls the change of the saddle point: from the naive scaling analysis (which is correct for $\Delta < 3/2$), $l^{(1)}\sim (\beta \bJ)^{2-2\Delta}$ for $\Delta < 3/2$, such that the perturbation series for the saddle point organizes itself as
\begin{equation}
    l = l^{(0)} + \kappa^2 \left(\frac{\pi}{\beta \bJ}\right)^{2\Delta -2} l^{(1)} + \mathcal{O}\left( \kappa^4 \right) .
\end{equation}
This perturbation series exhibits the more standard behaviour expected by naive scaling dimension analysis: the parameter grows as $\Delta < 1$, and shrinks as $\Delta >1$.
Then, when going to low temperatures, the relative change to the saddle is arbitrarily small compared to the relative change of the quadratic spectrum.\footnote{At this stage, one might be confused why the saddle is unchanged in this limit, but $l^{(1)}$ is still relevant for the change of the spectrum. This is because in the calculation for the spectrum, the contribution proportional to $l^{(1)}$ is compared to $\frac{1}{\beta \bJ}$, not to $1$.}   In that way, the spectrum strongly deviates from the Schwarzian spectrum of quadratic fluctuations, but the saddle point is not changed at leading order in $\beta \bJ$, at least not when $\tau_-$ is not close to $0,\beta$. In a holographic picture, we associate the two point function with a geodesic length (which is then also unchanged at low temperatures), so we would expect that the bulk space is still $AdS_2$. Since the spectrum of fluctuations on the field theory side is no longer governed by a Schwarzian (and instead by a bilocal deformation), the corresponding gravitational theory must have the same feature. The low--energy holographic picture is then a theory of fluctuations of an $AdS_2$ cutout with a nonlocal action for the repararametrization of the boundary.

\subsection{Polarized deformations}
\label{subsec: the role of gz}
In the previous section, we have seen that the self--intersection weight $\alpha_{22}$ and the field $g_2$ do not enter at $\mathcal{O}(\kappa^2)$. In this section, we study the impact of $\alpha_{22}$ and $g_2$, on the saddle point and the quadratic spectrum, at $\mathcal{O}(\kappa^4)$. Explicitly, we show that in the IR, $g_2$ is correcting the crossed 4-point function to account for the self-intersection number of the random operator. We show this at the level of the saddle and, for some specialization of parameters, the quadratic spectrum.

\paragraph{The saddle point}

We first evaluate the on--shell action at $\mathcal{O}(\kappa^4)$. We consider the generating functional for DSSYK perturbed by a single polarized operator. We will add the latter with sources $\chi(\tau)$ to clarify the relation with the 4-point function. From Section \ref{sec: multifield Liouville}, the relevant action is
\begin{equation}
    \begin{split}
        I = - & \frac{1}{4\lambda} \int_0^{\beta} d\tau_1 \int_0^{\beta}  d\tau_2 \Bigl[ \frac{1}{2} l \partial_1 \partial_2 l + 2 \bJ^2 e^{l} \\ - & \frac{\Delta^2-\alpha_{22}}{2} g_2 \partial_1 \partial_2 g_2  + 2 (\kappa \bJ)^2 e^{\Delta l  - (\Delta^2-\alpha_{22}) g_{2}}\chi(\tau_1)\chi(\tau_2)\Bigl]\, .
    \end{split}
\end{equation}
Analogously to \eqref{eq: kappa^2 equation of motion}, we write
$l = l^{(0)} + \kappa^2 l^{(1)}+ \mathcal{O}(\kappa^4)$, $g_2 = \kappa^2 g_2^{(1)}  + \mathcal{O}(\kappa^4)$\footnote{Notice that $l^{(2)}, g_2^{(2)}$ are not necessary for the on--shell action at $\mathcal{O}(\kappa^4)$ due to the equations of motion.} to get the equations
\begin{equation}\label{equations with chi}
    \begin{split}
        & \partial_1 \partial_2 l^{(0)} + 2\bJ^2 e^{l^{(0)}} = 0, \\
        & \partial_1 \partial_2 l^{(1)} + 2\bJ^2 e^{l^{(0)}} l^{(1)}  + 2 \Delta ( \bJ)^2 e^{\Delta l^{(0)}} \chi(\tau_1) \chi(\tau_2) = 0, \\
        & \partial_1 \partial_2 g_2^{(1)} + 2 \bJ^2 e^{\Delta l^{(0)}} \chi(\tau_1)\chi(\tau_2)=0.
    \end{split}
\end{equation}
$l^{(1)}$ schematically contains two parts: one part amounts to a deformation of the saddle point away from the soft mode manifold (which can be seen by setting $\chi$ to $1$; then $l^{(1)}$ would depend only on $\tau_-$). The second part is a reparametrization of the saddle $l^{(0)}$, i.e. a motion along the soft mode manifold. Integrating out $g_2$ using its saddle solution, we get
\begin{equation}\label{on-shell l g_2 with source}
    \begin{split}
        I = - \frac{1}{4\lambda}\int_0^{\beta} d\tau_1 \int_0^{\beta} d\tau_2 \Bigg[ \frac{1}{2} l^{(0)} \partial_1 \partial_2 l^{(0)} + & 2 \bJ^2 e^{l^{(0)}} + 2 (\kappa \bJ)^2 e^{\Delta l^{(0)}} \chi(\tau_1) \chi(\tau_2) \\  +  
        \frac{\Delta^2-\alpha_{22}}{2} \kappa^4 \iiiint_{(\tau_1,\tau_2)\cap (\tau_3,\tau_4)} \bJ^2 & e^{\Delta l^{(0)}}(\tau_1,\tau_2) \bJ^2 e^{\Delta l^{(0)}}(\tau_3,\tau_4) \chi(\tau_1) \chi(\tau_2) \chi(\tau_3) \chi(\tau_4) \\
        + \kappa^4 \bJ^2 e^{\Delta l^{(0)}} & \Delta l^{(1)} \chi(\tau_1) \chi(\tau_2) + \mathcal{O}(\kappa^6)
        \Bigg],
    \end{split}
\end{equation}
where we have introduced the notation $(\tau_1,\tau_2)\cap (\tau_3,\tau_4)$ to denote that we only integrate over the region where a chord from $\tau_1$ to $\tau_2$ crosses a chord from $\tau_3$ to $\tau_4$. This on--shell action has, compared to a deformation with $ \alpha_{22}=\Delta^2$, an additional piece given by the second line of \eqref{on-shell l g_2 with source}. Such a piece influences the (classical) 4-point function, which we compute via\footnote{The factor of $\lambda^2$ again comes from the fact that each variation of $\chi$ is accompanied by $\sqrt{\lambda}$ to compensate the $1/\sqrt{\lambda}$ normalization of the random couplings \eqref{eq: multi-SYK conventions}.}
\begin{equation}
    \langle O_{\Delta, \alpha_{22}}(\tau_1) O_{\Delta, \alpha_{22}}(\tau_2) O_{\Delta, \alpha_{22}}(\tau_3) O_{\Delta, \alpha_{22}}(\tau_4) \rangle = \lambda^2 \frac{\delta}{\delta \chi(\tau_1)} \frac{\delta}{\delta \chi(\tau_2)} \frac{\delta}{\delta \chi(\tau_3)} \frac{\delta}{\delta \chi(\tau_4)}
    e^{-I}|_{\chi = 0}.
\end{equation}
In this 4-point function, when we expand $e^{-I}$ to ${\mathcal O}(\kappa^4)$ in \eqref{on-shell l g_2 with source}, there will be a disconnected piece, an $\alpha_{22}$--independent connected piece (coming from the $\kappa^4$-term in the last line of \eqref{on-shell l g_2 with source}; recall that $l^{(1)}$ depends on $\chi$) encoding the leading Schwarzian backreaction, and a $\alpha_{22}$--dependent piece, which is a new feature compared to the usual SYK--like deformation. It is given by
\begin{equation}
    \kappa^4 \lambda (\Delta^2 -\alpha_{22}) \bJ^2 e^{\Delta l^{(0)}}(\tau_1,\tau_2) \bJ^2 e^{\Delta l^{(0)}}(\tau_3,\tau_4),
\end{equation}
with $(\tau_1,\tau_2)\cap (\tau_3,\tau_4)$. Notice that up to the prefactors, this term is the classical crossed 4-point function (i.e. the crossed 4-point function at zeroth order in $\lambda$) since the $e^{\Delta l^{(0)}}$ are just 2-point functions. 

What is the purpose of this contribution? When computing the 4-point function of some operator, the crossed part is multiplied by the factor $e^{-\alpha_{22}\lambda}$ due to self--intersection of the matter chords. The $\Delta^2 -\alpha_{22}$--independent part of the action computes the crossed 4-point function with a prefactor $e^{-\Delta^2 \lambda}$, as the $g_2$--independent part of the action corresponds to a deformation by a fully chaotic operator. At first order in $\lambda$, the term
\begin{equation}
    \kappa^4 \lambda (\Delta^2 -\alpha_{22}) \bJ^2 e^{\Delta l^{(0)}}(\tau_1,\tau_2) \bJ^2 e^{\Delta l^{(0)}}(\tau_3,\tau_4),
\end{equation}
is then responsible for replacing the prefactor of the crossed 4-point function
\begin{equation}
    1-\Delta^2 \lambda \mapsto 1-\alpha_{22}\lambda. 
\end{equation}
The role of $g_2$ is thus to correct the self--intersection weights of chords corresponding to non SYK--like operators.
The resulting term quartic in $\chi$ in the on--shell action then looks like a genuine 4-point function term, contrary to the ordinary Schwarzian, where the interaction happens via gravitational shockwaves \cite{Shenker:2013pqa}. Investigating possible holographic interpretations of this 4-point interaction is an interesting option for future work. 

It would also be interesting to go beyond the saddle point and formulate the role of $g_2$ as an all order statement, at least in the low energy coupling $\lambda \beta \bJ$, i.e. a statement that holds everywhere on the soft mode manifold. We now show that at least for small fluctuations around the saddle point, we retain the structure of the 4-point term. 

\paragraph{The quadratic spectrum}
\label{subsec: kappa^4 quadratic spectrum}
For the analysis of the quadratic spectrum, we will focus on the case of a same--length deformation, $\Delta =1$, with generic self-intersection number $\alpha_{22}$. The reason for such a choice is that for general setups, the number of terms in the quadratic spectrum at $\mathcal{O}(\kappa^4)$ proliferates, making the analysis less insightful. On the other hand, for a deformation of same length with $\alpha_{22}=1$ (which is a deformation by an SYK operator of the same length), however, nothing happens. Deforming $\alpha_{22}$ away from 1 only has an impact at $\mathcal{O}(\kappa^4)$, simplifying the analysis of the resulting quadratic spectrum significantly. For this case, we show explicitly that the quadratic spectrum can be understood as a sum of an infinitesimal reparametrization of the 4-point interaction of \eqref{on-shell l g_2 with source}, and a modification associated to a change of the saddle point. At least at quadratic order, we thus have a very simple picture: the 4-point term takes the place of the bilocal deformation, influencing the low--energy spectrum.

Following once again Section \ref{sec: multifield Liouville}, we consider the action 
\begin{equation}
    I = - \frac{1}{4\lambda} \int_0^{\beta} d\tau_1 \int_0^{\beta} d\tau_2 \Bigg[ \frac{1}{2} g_1 \partial_1 \partial_2 g_1 + g_1 \partial_1 \partial_2 g_2 +\frac{\alpha_{22}}{2} g_2 \partial_1 \partial_2 g_2 + 2 \kappa_1^2 \bJ^2 e^{g_1 + g_2} + 2 \kappa_2^2 \bJ^2 e^{g_1+\alpha_{22}g_2}  \Bigg].
\end{equation}
As before, we define $l=g_1 + g_2$ to get
\begin{equation}
    I = - \frac{1}{4\lambda} \int_0^{\beta} d\tau_1 \int_0^{\beta} d\tau_2 \Bigg[ \frac{1}{2} l \partial_1 \partial_2 l -\frac{1-\alpha_{22}}{2} g_2 \partial_1 \partial_2 g_2 + 2 \kappa_2^2 \bJ^2 e^{l} + 2 \kappa_1^2 \bJ^2 e^{l-(1-\alpha_{22})g_2}  \Bigg].
\end{equation}
We also impose the constraint $\kappa_1^2 + \kappa_2^2 =1$ such that for $\alpha_{22}=1$, the model reduces to just ordinary DSSYK. We also write $\kappa_2=\kappa$ to lighten some formulae. 

We once again expand the saddle point in a perturbation series around $\kappa^2 =0$. For the special case $\Delta =1$, the equations \eqref{eq: kappa^2 equation of motion} give at $\mathcal{O}(\kappa^2)$ the solutions $g_2^{(1)}=l^{(0)}, l^{(1)}=0$. For the $\mathcal{O}(\kappa^4)$-correction to the saddle point,  $l^{(2)}$, we have the equation

\begin{equation}
    \partial_1 \partial_2 l^{(2)} + 2 \bJ^2 e^{l^{(0)}} l^{(2)} - (1-\alpha_{22})2 \bJ^2 e^{l^{(0)}} l^{(0)} = 0.
\end{equation}
$g_2^{(2)}$ doesn't enter at $\mathcal{O}(\kappa^4)$. The quadratic action at $\kappa^4$ is then given by
\begin{equation}\label{eq: generic quadratic ation at kappa4}
    \begin{split}
        I_{quad} = & -  \frac{1}{4\lambda} \int_0^{\beta} d\tau_1 \int_0^{\beta} d\tau_2 \Bigg[ \frac{1}{2} \delta l \partial_1 \partial_2 \delta l - \frac{1-\alpha_{22}}{2} \delta g_2 \partial_1 \partial_2 \delta g_2 + \bJ^2 e^{l^{(0)}} \delta l^2 \\
        & - 2 \kappa^2 \bJ^2 e^{l^{(0)}} (1-\alpha_{22})  \delta l \delta g_2 + \kappa^2 \bJ^2 e^{l^{(0)}} (1-\alpha_{22})^2 \delta g_2^2 \\ 
        & + \kappa^4 \bJ^2 e^{l^{(0)}} l^{(2)} \delta l^2 - \kappa^4 \bJ^2 e^{l^{(0)}} l^{(0)} (1-\alpha_{22}) (\delta l - (1-\alpha_{22})\delta g_2)^2 
        \Bigg].
    \end{split}
\end{equation}
Out of this expression, we are interested in the change of the light spectrum at $\mathcal{O}(\kappa^4)$. Since mixing between light (i.e. Schwarzian) and heavy $\delta l$ modes occurs only at $\mathcal{O}(\kappa^4)$ via the last line of \eqref{eq: generic quadratic ation at kappa4}, it influences the spectrum only at higher order in $\kappa$. Since $\delta g_2$ also contains no light modes (see Section \ref{subsec: spectrum at kappa^2}), we can integrate out $\delta g_2$ safely and only keep the effective theory of light modes. 
The equation of motion for $\delta g_2$
\begin{equation}
    \partial_1 \partial_2 \delta g_2 + 2 \kappa^2 \bJ^2 e^{l^{(0)}} \delta l = 0,
\end{equation}
which is the linearized version of \eqref{equations with chi}.
The equation for $\delta g_2$ then has an interpretation as the reaction of $g_2$ when walking on the soft mode manifold. Integrating it out gives
\begin{equation}\label{quadratic action at kappa4}
    \begin{split}
        I_{quad} = & - \frac{1}{4\lambda} \int_0^{\beta} d\tau_1 \int_0^{\beta} d\tau_2  \Bigg[ \frac{1}{2} \delta l \partial_1 \partial_2 \delta l  + \bJ^2 e^{l^{(0)}} \delta l^2 + \kappa^4 \bJ^2 e^{l^{(0)}}(l^{(2)}-(1-\alpha_{22})l^{(0)}) \delta l^2 \Bigg] \\
        & - \frac{1-\alpha_{22}}{8\lambda} \kappa^4 \iiiint_{(1,2)\cap (3,4)} \bJ^2 e^{l^{(0) }(\tau_1,\tau_2)} \delta l (\tau_1,\tau_2) \bJ^2 e^{l^{(0) }(\tau_3,\tau_4)} \delta l (\tau_3,\tau_4).
    \end{split}
\end{equation}
Clearly, the second line has an interpretation as an quadratic fluctuation of the 4-point term in \eqref{on-shell l g_2 with source}, where we expanded each of the exponentials to first order in an infinitesimal reparametrization. Expanding one of the exponentials to quadratic order, one gets exactly the term proportional to $(1-\alpha_{22})l^{(0)}$ (after using the equations of motion for $l^{(0)}$).

The quadratic action at $\mathcal
O (\kappa^4)$ thus splits into two parts:
\begin{equation}\label{kappa4 result}
    \begin{split}
        I_{\kappa^4}= & - \frac{\kappa^4 }{4\lambda} \int_0^{\beta} d\tau_1 \int_0^{\beta} d\tau_2 \bJ^2 e^{l^{(0)}} l^{(2)} \delta l^2 \\ & + \text{quadratic fluctuations of } \left\{ - \frac{1-\alpha_{22}}{8\lambda} \kappa^4 \iiiint_{(\tau_1,\tau_2)\cap (\tau_3,\tau_4)} \bJ^2 e^{ l(\tau_1,\tau_2)} \bJ^2 e^{l(\tau_3,\tau_4)} \right\},
    \end{split}
\end{equation}
where by `quadratic fluctuations' we mean expanding infinitesimal (twisted) reparametrizations of $l^{(0)}$  to quadratic order.
\eqref{kappa4 result} rhymes nicely with the observed phenomena for the bilocal at $\mathcal{O}(\kappa^2)$: there is a term which corresponds to a reparametrization of the 4-point term in \eqref{on-shell l g_2 with source}, and there is a term that is caused by a change of the saddle point.
For generic $\Delta$, the bilocal does not vanish, and we expect the emerging picture to be a combination of the influence of the bilocal and the 4-point term. \\

\section*{Acknowledgments}

We would like to thank A. Blommaert, N. Brukner,  M. Bucca, Y. Jia and M. Mezei for useful discussions. OM is supported by the ERC-COG grant NP-QFT No. 864583
``Non-perturbative dynamics of quantum fields: from new deconfined
phases of matter to quantum black holes", by the MUR-FARE2020 grant
No. R20E8NR3HX ``The Emergence of Quantum Gravity from Strong Coupling
Dynamics", and by the INFN ``Iniziativa Specifica GAST''. The work of MB, RF and JS is supported in part by the Israel Science Foundation grant no. 2159/22, by the Minerva foundation, and by a German-Israeli Project Cooperation (DIP) grant "Holography and the Swampland". MB is an incumbent of the Charles and David Wolfson Professorial Chair of Theoretical Physics. The work of JS is supported by a research grant from the Chaim Mida Prize in Theoretical Physics at the Weizmann Institute of Science.

\appendix

\section{Reparametrizations at finite $\beta$}
\label{sec: finite temperatures}

In Section \ref{sec: effective boundary term}, we showed that we can regulate the twisted reparametrizations by a smoothing function $f$, such that $
\Tilde{g}_{\phi}$ in \eqref{eq: twisted coordinate reparametrization of g regularized} respects the boundary conditions. Here, we present an alternative, where we replace the derivatives in \eqref{eq: twisted coordinate reparametrization of g} by finite differences and symmetrize appropriately, giving\footnote{For $\delta v \ll 1$, the first two lines of \eqref{eq: finite temp reparametrization} reduces to the first term in \eqref{eq: twisted coordinate reparametrization of g}, while the last two lines, which are acting as a finite differences  with resolution proportional to $\delta v$, translate to the derivative terms in \eqref{eq: finite temp reparametrization}. In constructing \eqref{eq: finite temp reparametrization}, we symmetrized with respect to \eqref{eq:boundary conditions g after rearange}.}

\begin{equation}\label{eq: finite temp reparametrization}
    \begin{split}
        \Tilde{g}_{\phi} = & \frac{1}{2}\Bigg[\log \frac{\cos \frac{\pi v}{2}}{\sin(\frac{\pi}{\beta}(\phi(s_1 + \frac{\beta}{2}\delta v)-\phi(s_2 + \frac{\beta}{2}\delta v)))} + \log \frac{\cos \frac{\pi v}{2}}{\sin(\frac{\pi}{\beta}(\phi(s_1 - \frac{\beta}{2}\delta v)-\phi(s_2 - \frac{\beta}{2}\delta v)))} \\ + & 2\log \frac{\cos \frac{\pi v}{2}}{\sin(\frac{\pi}{\beta}(\phi(s_1)-\phi(s_2)))}\Bigg] \\
        + & \frac{1}{2}\Bigg[\log \frac{\sin (\frac{\pi}{\beta}(\phi(s_1) - \phi(s_1 - \frac{\beta}{2}\delta v))) \sin (\frac{\pi}{\beta}(\phi(s_2+ \frac{\beta}{2}\delta v) - \phi(s_2 )))}{\cos^2 \frac{\pi v}{2}}  \\ + & \log \frac{\sin (\frac{\pi}{\beta}(\phi(s_2) - \phi(s_2- \frac{\beta}{2}\delta v ))) \sin (\frac{\pi}{\beta}(\phi(s_1 + \frac{\beta}{2}\delta v) - \phi(s_1 )))}{\cos^2 \frac{\pi v}{2}}\Bigg].
    \end{split}
\end{equation}
This expression is well-defined for any $\beta$, obeys the symmetry and boundary conditions \eqref{eq:boundary conditions g after rearange}, and at low temperatures reproduces the infinitesimal Schwarzian spectrum \eqref{eq: quadratic schwarzian action} up to corrections of $\mathcal{O}(\delta v^3)$. We expect the finite version to also give the full Schwarzian. 

Contrary to regulating with a smoothing function $f$, as in \ref{sec: effective boundary term}, \eqref{eq: finite temp reparametrization} respects the $SL(2,R)$ symmetry at all orders in $\delta v$: applying an $SL(2)$--transformation
\begin{equation}
    \phi(\tau) = \frac{\beta}{\pi} \arctan \left( \frac{a \tan (\frac{\pi}{\beta}\tau)+b}{c \tan (\frac{\pi}{\beta}\tau)+d} \right), \quad ad-bc=1 ,
\end{equation}
to \eqref{eq: finite temp reparametrization} leads to $\Tilde{g}_{\phi}=g_0$.
Compared to the fake disk $SL(2,R)$ \cite{Lin:2023trc}, the $SL(2,R)$ in this manifestation has the advantage that it is embedded into a well-defined reparametrization action. It also, however, brings a different limitation with itself: \eqref{eq: finite temp reparametrization} has no obvious simple composition rule.

\section{Details of the Schwarzian calculation}
\label{sec: Schwarzian details}

In this Appendix, we calculate the potential and kinetic terms ($I_{\text{pot}}[g_{\phi}]$ and $I_{\text{kin}}[g_{\phi}]$) of Section \ref{sec: summerize_schwarzian_computaiton} in detail.

\subsection{Calculation of the potential term}
\label{subsec: potential term}

Here, we simplify the potential term (as defined in \eqref{eq: definition of potantial and kinetic terms})
\begin{equation}
    I_{pot}[g_{\phi}] = - \frac{\bJ^2}{2 \lambda v} \int_{\mathcal{I}}ds_{1}ds_{2}e^{g_{\phi}} = - \frac{\pi^2 v}{2 \lambda \beta^2}\int_{\mathcal{I}}ds_{1}ds_{2}\frac{1}{\sin^{2} \left (\frac{\pi}{\beta}(\phi(s_{1})-\phi(s_{2})\right)}\phi'\left(s_{1}\right)\phi'\left(s_{2}\right) .
\end{equation}
We change the integration variables to $\phi_{1/2} = \phi(s_{1/2})$, before redefining $\phi_{\pm} = \phi_{1} \pm \phi_{2}$ to get a simplified expression
\begin{equation} \label{eq: potential term step 1}
    I_{pot}[g_{\phi}] =  - \frac{\pi^2 v}{2 \lambda \beta^2}\int_{\phi\left(\mathcal{I}\right)}d\phi_{1}d\phi_{2}\frac{1}{\sin^{2} \left( \frac{\pi}{\beta}(\phi_{1}-\phi_{2})\right)} = - \frac{\pi^2 v}{4 \lambda \beta^2} \int_{\phi\left(\mathcal{I}\right)}\frac{1}{\sin^{2}\frac{\pi}{\beta}(\phi_{-})}d\phi_{-}d\phi_{+} ,
\end{equation}
over the reparametrized area $\phi (\mathcal{I})$, as illustrated in Figure \ref{fig: splus sminus range}.
The integral in \eqref{eq: potential term step 1} can be solved using Stokes' theorem. We denote $\omega$ (and its exterior derivative) by
\begin{equation}
    \omega\equiv - \frac{\beta}{\pi}\cot\left(\frac{\pi}{\beta}\phi_{-}\right)d\phi_{+} , \qquad d\omega=\frac{1}{\sin^{2}\left(\frac{\pi}{\beta}\phi_{-}\right)}d\phi_{-}\wedge d\phi_{+} ,
\end{equation}
and we see that
\begin{equation} \label{eq: potential term step 2}
    \int_{\phi\left(\mathcal{I}\right)}\frac{1}{\sin^{2}\left(\frac{\pi}{\beta}\phi_{-}\right)}d\phi_{+}d\phi_{-}=\int_{\phi\left(\mathcal{I}\right)}dw=\int_{\partial\phi\left(\mathcal{I}\right)}\omega=\int_{\partial\phi\left(\mathcal{I}\right)}-\frac{\beta}{\pi}\cot\left(\frac{\pi}{\beta}\phi_{-}\right)d\phi_{+}.
\end{equation}
The rest of the computation follows by parametrizing the boundary $\partial \phi (\mathcal{I})$ and plugging the results in \eqref{eq: potential term step 2}.

The boundary $\partial \phi (\mathcal{I})$ is given by the reparametrized location of the original boundary $\partial I$, which we split into 4 regions $\partial \phi (\mathcal{I}) = \Gamma_1 +\Gamma_2 +\Gamma_3+ \Gamma_4$. $\Gamma_1$ corresponds to the reparametrization of the lower edge of $\partial \mathcal{I}$, and $\Gamma_2, \Gamma_3, \Gamma_4$ corresponds to the parametrization of the right, upper and left edges of $\partial \mathcal{I} $ in counter--clockwise fashion. Due to the $2\beta$--periodicity of the $\tau_+$--coordinate, the contributions of $\Gamma_1$ and $\Gamma_3$ cancel each other.

To find the contribution of $\Gamma_2$, we parametrize the right edge of $\partial I$ by $s_{-}=\frac{\beta}{2}\left(1+v\right),s_{+}=\xi$ with $\xi\in\left[0,2\beta\right]$ so that $s_{1}=\frac{1}{2}\xi+\frac{\beta}{4}\left(1+v\right),s_{2}=\frac{1}{2}\xi-\frac{\beta}{4}\left(1+v\right)$. The corresponding boundary $\Gamma_2$ is given by
\begin{equation} \label{eq: gamma2 raparam}
        \begin{split}
        &\phi_{1}=\phi\left(\frac{1}{2}\xi+\frac{\beta}{4}\left(1+v\right)\right),\phi_{2}=\phi\left(\frac{1}{2}\xi-\frac{\beta}{4}\left(1+v\right)\right) ,\\
        &\phi_{\pm}  =\phi\left(\frac{1}{2}\xi+\frac{\beta}{4}\left(1+v\right)\right) \pm \phi\left(\frac{1}{2}\xi-\frac{\beta}{4}\left(1+v\right)\right) .
        \end{split}
\end{equation}
The parametrization of $\Gamma_4$ is found in a similar way, replacing $1+v$ by $1-v$ in \eqref{eq: gamma2 raparam}.
The end result is (moving to the variable $\tau = 2\xi $)
\begin{equation}
    \begin{split}
        I_{pot}[g_{\phi}]=-\frac{1}{2\lambda}\frac{\pi v}{\beta}\int_{\tau=0}^{\tau=\beta}& \cot\left(\frac{\pi}{\beta}\left(\phi\left(\tau+\frac{\beta}{4}\delta v\right)-\phi\left(\tau-\frac{\beta}{4}\delta v\right)\right)\right) \\
        &\times\left(\phi'\left(\tau+\frac{\beta}{4}\delta v\right)+\phi'\left(\tau-\frac{\beta}{4}\delta v\right)\right)d\tau .
    \end{split}
\end{equation}

This expression can be Taylor--expanded to leading order in $\delta v$, and we find
\begin{equation}
    I_{pot}\left[g_{\phi}\right]-I\left[g_{0}\right] =-\frac{1}{\lambda}\frac{v}{12\bJ}\int_{\tau=0}^{\tau=\beta}\left(\frac{\phi^{\left(3\right)}}{\phi'}-\left(\frac{2\pi}{\beta}\right)^{2}\phi'{}^{2}+\left(\frac{2\pi}{\beta}\right)^{2}\right)d\tau + \mathcal{O}(\delta v^3).
\end{equation}

\subsection{Calculation of the Kinetic Term}
\label{subsec: kinetic term}
Here, we simplify the kinetic term (as defined in \eqref{eq: definition of potantial and kinetic terms})
\begin{equation}\label{eq: kinetic term in s1,s2}
    I_{kin}[g_{\phi}]=\frac{1}{8\lambda v}\int_{\mathcal{I}}ds_{1}ds_{2}\Bigg[\frac{1}{2}(1+v^{2})\partial_{s_{1}}g\partial_{s_{2}}g+\frac{1}{4}(1-v^{2})\left(\partial_{s_{1}}g\partial_{s_{1}}g+\partial_{s_{2}}g\partial_{s_{2}}g\right)\Bigg].
\end{equation}
By denoting
\begin{equation}
    \begin{split}
        \partial_{s_{1}}g_{\phi} & =\underbrace{\phi'\left(s_{1}\right)\left[-2\frac{\pi}{\beta}\cot\left( \frac{\pi}{\beta}(\phi(s_{1})-\phi(s_{2}))\right)\right]}_{\left(1\right)}+\underbrace{\frac{\phi''\left(s_{1}\right)}{\phi'\left(s_{1}\right)}}_{\left(2\right)}\\
        \partial_{s_{2}}g_{\phi} & =\underbrace{\phi'\left(s_{2}\right) \left[+2\frac{\pi}{\beta}\cot\left(\frac{\pi}{\beta}(\phi(s_{1})-\phi(s_{2}))\right)\right] }_{\left(A\right)}+\underbrace{\frac{\phi''\left(s_{2}\right)}{\phi'\left(s_{2}\right)}}_{\left(B\right)} ,
    \end{split}
\end{equation}
we can express the kinetic term in \eqref{eq: kinetic term in s1,s2} as
\begin{equation}
    I_{kin}[g_{\phi}]=\frac{1}{8\lambda v}\int_{\mathcal{I}}ds_{1}ds_{2}\left(\begin{array}{c}
    \frac{1}{2}(1+v^{2})\left(\left(1\right)\left(A\right)+\left(2\right)\left(A\right)+\left(1\right)\left(B\right)+\left(2\right)\left(B\right)\right)\\
    +\frac{1}{4}(1-v^{2})\left(\left(1\right)^{2}+2\left(1\right)\left(2\right)+\left(2\right)^{2}+\left(A\right)^{2}+2\left(A\right)\left(B\right)+\left(B\right)^{2}\right)
    \end{array}\right).
\end{equation}
In the following, we examine the contribution of each term in leading order of $\delta v $.

\paragraph{$\left(1\right)\left(A\right)$ term}

Using the same change of variables as in the potential term, we find
\begin{equation}
    \begin{split}
    \int_{\mathcal{I}}ds_{1}ds_{2}\left(1\right)\left(A\right) &=-\int_{\mathcal{I}}ds_{1}ds_{2}\phi'\left(s_{1}\right)\phi'\left(s_{2}\right)\left(\left(2\frac{\pi}{\beta}\right)^{2}\cot^{2}\left(\frac{\pi}{\beta}(\phi(s_{1})-\phi(s_{2}))\right)\right) \\
    &= -\left(2\frac{\pi}{\beta}\right)^{2}\frac{1}{2}\int_{\phi\left(\mathcal{I}\right)}d\phi_{-}d\phi_{+}\left(\frac{1}{\sin^{2}\left(\frac{\pi}{\beta}(\phi_{1}-\phi_{2})\right)}-1\right).    
    \end{split}
\end{equation}
This can be evaluated using Stokes' theorem in the same way as the potential term. The sine term was already evaluated \ref{subsec: potential term}, while the $-1$ can be evaluated using the differential form
\begin{equation}
    w=\phi_{-}d\phi_{+} \rightarrow  dw=d\phi_{-}\wedge d\phi_{+} .
\end{equation}
We find that
\begin{equation}
    \begin{split}
        \int_{\phi\left(\mathcal{I}\right)}d\phi_{-}d\phi_{+}= 2\beta^{2}+2\intop_{\tau=0}^{\tau=\beta} & d\tau  \left(\phi\left(\tau-\frac{\beta}{4}\delta v\right)-\phi\left(\tau+\frac{\beta}{4}\delta v\right)\right) \\
        & \times \left(\phi'\left(\tau-\frac{\beta}{4}\delta v\right)+\phi'\left(\tau+\frac{\beta}{4}\delta v\right)\right).
    \end{split}
\end{equation}
The sum of the two contributions is
\begin{multline}
    \int_{\mathcal{I}}ds_{1}ds_{2}\left(1\right)\left(A\right) = -\frac{4\pi^{2}}{\beta^{2}}\Bigg( -\beta^{2} + \int_{\tau=0}^{\tau=\beta} \left(\phi'\left(\tau+\frac{\beta}{4}\delta v\right)+\phi'\left(\tau-\frac{\beta}{4}\delta v\right)\right) \times \\
     \times \left( \frac{\beta}{\pi}\cot\left| \frac{\pi}{\beta}\left(\phi\left(\tau+\frac{\beta}{4}\delta v\right)-\phi\left(\tau-\frac{\beta}{4}\delta v\right)\right)\right|  + \phi\left(\tau+\frac{\beta}{4}\delta v\right)-\phi\left(\tau-\frac{\beta}{4}\delta v\right) \right) d\tau \Bigg) ,
\end{multline}
and the contribution at leading order in $\delta v$ is
\begin{equation}
    \delta I \supset  -\frac{1}{12\lambda \bJ} \int_0^{\beta} d\tau (\frac{\phi'''}{\phi'} + 2 (\frac{2\pi}{\beta})^2 \phi'^2) + \mathcal{O}(\delta v^2).
\end{equation}

\begin{figure}[t]
\centering
\includegraphics[width=0.99\linewidth]{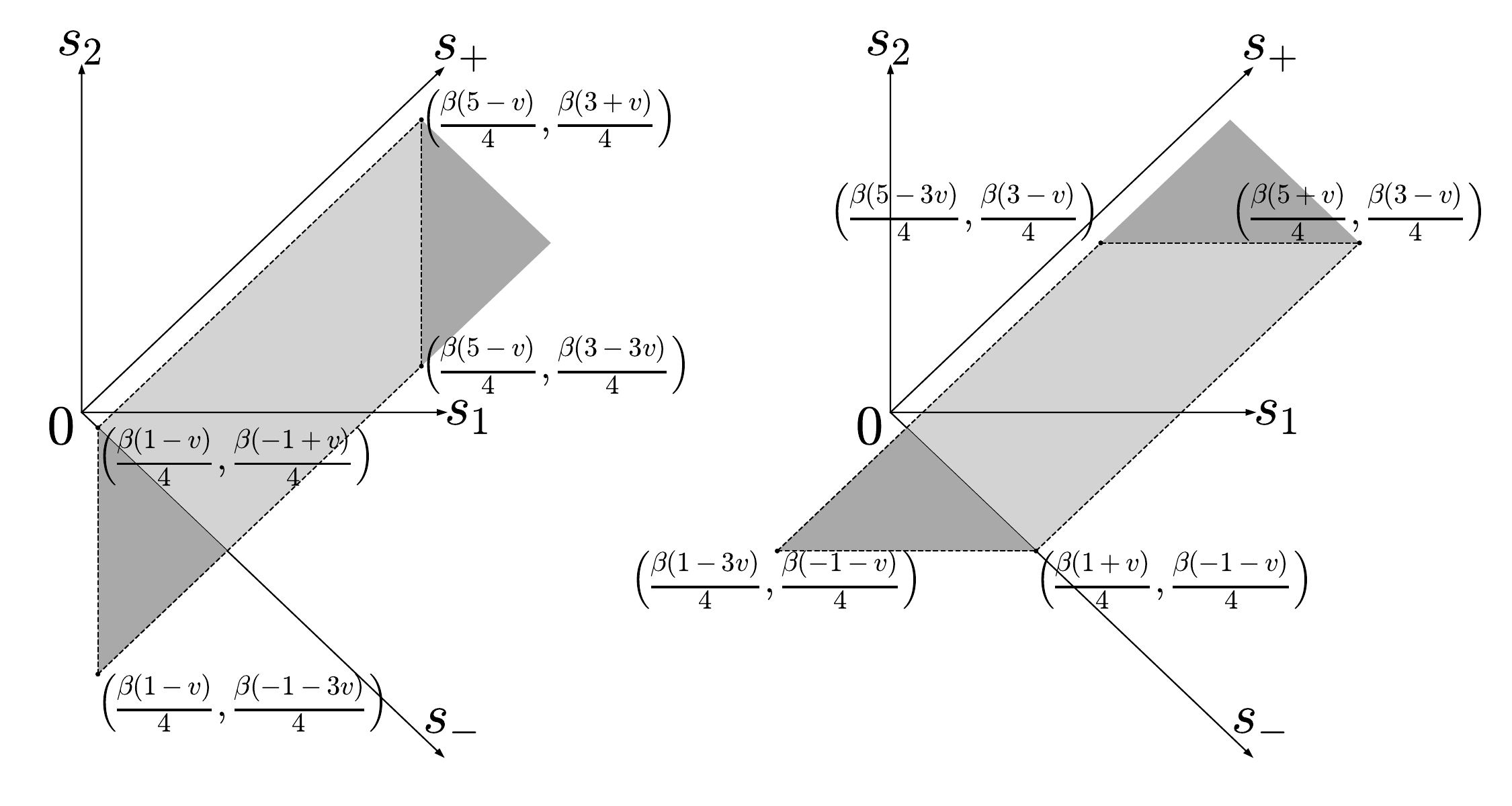}
    \caption{Two reorderings of the integration region into two different parallelograms. The integration region on the left is $s_{2}\in\left[s_{1}+\frac{-2-2v}{4}\beta,s_{1}+\frac{-2+2v}{4}\beta\right]$ for $s_{1}\in\left[\frac{\beta\left(1-v\right)}{4},\frac{\beta\left(5-v\right)}{4}\right]$. The integration region on the right is $s_{1}\in\left[s_{2}+\frac{2-2v}{4}\beta,s_{2}+\frac{2+2v}{4}\beta\right]$ for $s_{2}\in\left[\frac{\beta\left(-1-v\right)}{4},\frac{\beta\left(3-v\right)}{4}\right]$.}
    \label{fig:s_1 s_2 reordering}
\end{figure}

\paragraph{$\left(2\right)\left(A\right)$ and $\left(1\right)\left(B\right)$ terms}
We have
\begin{equation} \label{eq: kinetic term 1b}
\int_{\mathcal{I}}ds_{1}ds_{2}\left(1\right)\left(B\right)=\int_{\mathcal{I}}ds_{1}ds_{2}\phi'\left(s_{1}\right)\left(-2\frac{\pi}{\beta}\cot \frac{\pi}{\beta}(\phi(s_{1})-\phi(s_{2}))\right)\frac{\phi''\left(s_{2}\right)}{\phi'\left(s_{2}\right)}.
\end{equation}
Using $2\beta$--periodicity of $s_+$, the integration region can be rearranged into a parallelogram with $s_{2}\in\left[\frac{\beta\left(-1-v\right)}{4},\frac{\beta\left(3-v\right)}{4}\right]$ and $s_{1}\in\left[s_{2}+\frac{2-2v}{4}\beta,s_{2}+\frac{2+2v}{4}\beta\right]$, as shown in Figure \ref{fig:s_1 s_2 reordering}. This way, \eqref{eq: kinetic term 1b} can be integrated along $s_1$ to provide 
\begin{equation}
    \int_{\mathcal{I}} ds_{1}ds_{2}\left(1\right)\left(B\right) = \int_{\frac{\beta\left(-1-v\right)}{4}}^{\frac{\beta\left(3-v\right)}{4}}ds_{2}\left(-2\log\left(\sin\left|\frac{\pi}{\beta}\phi\left(s_{1}\right)-\phi\left(s_{2}\right)\right|\right) \right) \frac{\phi''\left(s_{2}\right)}{\phi'\left(s_{2}\right)}|_{s_{1}=s_{2}+\frac{2-2v}{4}\beta}^{s_{1}=s_{2}+\frac{2+2v}{4}\beta} .
\end{equation}
Changing the integration variable to $\tau$ and using $\beta$--periodicity, this term evaluates to

\begin{equation}
    \int_{\mathcal{I}}ds_{1}ds_{2}\left(1\right)\left(B\right) = 2\int_{0}^{\beta}d\tau\frac{\phi''\left(\tau\right)}{\phi'\left(\tau\right)}\log\left(\frac{\sin\frac{\pi}{\beta}\left|\phi(\tau+\frac{\beta}{2}\delta v)-\phi(\tau)\right|}{\sin\frac{\pi}{\beta}\left|\phi(\tau)-\phi(\tau-\frac{\beta}{2}\delta v)\right|}\right) .
\end{equation}
The same contribution is found from the $\left(2\right)\left(A\right)$ term. We find that in leading order of $\delta v $ those terms contribute to the action as
\begin{equation}
    \delta I \supset \frac{1}{2\lambda \bJ} \int_0^{\beta} d\tau (\frac{\phi''}{\phi'})^2 + \mathcal{O}(\delta v^2).
\end{equation}

\paragraph{$\left(2\right)\left(B\right)$ term}
We have:
\begin{equation}
\int_{\mathcal{I}}ds_{1}ds_{2}\left(2\right)\left(B\right)=\int_{\mathcal{I}}ds_{1}ds_{2}\partial_{s_{1}}\left(\log\phi'\left(s_{1}\right)\right)\frac{\phi''\left(s_{2}\right)}{\phi'\left(s_{2}\right)}.   
\end{equation}
The $s_1$ integration can be carried out in the same way as in the $\left(2\right)\left(A\right)$ term, providing
\begin{equation}
\begin{split}
    \int_{\mathcal{I}}ds_{1}ds_{2}\left(2\right)\left(B\right) &= \int_{\frac{\beta\left(-1-v\right)}{4}}^{\frac{\beta\left(3-v\right)}{4}}ds_{2}\log\frac{\phi'\left(s_{2}+\frac{2+2v}{4}\beta\right)}{\phi'\left(s_{2}+\frac{2-2v}{4}\beta\right)}\frac{\phi''\left(s_{2}\right)}{\phi'\left(s_{2}\right)} \\
    &=\int_{0}^{\tau}ds_{2}\log\frac{\phi'\left(\tau-\frac{\beta}{2}\delta v\right)}{\phi'\left(\tau+\frac{\beta}{2}\delta v\right)}\frac{\phi''\left(\tau\right)}{\phi'\left(\tau\right)},
\end{split}
\end{equation}
where in the last step we changed the integration variable to $\tau$ and used $\beta$--periodicity.
The contribution of this term to the action in leading order is thus
\begin{equation}
    \delta I \supset  - \frac{1}{4\lambda \bJ} \int_0^{\beta} d\tau (\frac{\phi''}{\phi'})^2 + \mathcal{O}(\delta v^2).
\end{equation}

\paragraph{$\left(1\right)^{2}+2\left(1\right)\left(2\right)$, $(A)^2+2(A)(B)$ terms}

The trick here is to observe that the sum of terms can be simplified to 
\begin{equation} \label{(1)^2+2(1)(2)}
    \begin{split}
        \int_{\mathcal{I}}\left(1\right)^{2}+2\left(1\right)\left(2\right) = &-4\left(\frac{\pi}{\beta}\right)^{2}\int_{\mathcal{I}}ds_{1}ds_{2}\phi'\left(s_{1}\right)^{2} \\
        & -4\frac{\pi}{\beta}\int_{\mathcal{I}}ds_{1}ds_{2}\partial_{s_{1}}\left(\phi'\left(s_{1}\right)\cot\left|\frac{\pi}{\beta}(\phi(s_{1})-\phi(s_{2}))\right|\right).
    \end{split}
\end{equation}

We can now repeat the integration region reordering as in Figure \ref{fig:s_1 s_2 reordering} and find: 
\begin{align}
&\int_{\mathcal{I}}ds_{1}ds_{2}\partial_{s_{1}}\left(\phi'\left(s_{1}\right)\cot\left|\frac{\pi}{\beta}(\phi(s_{1})-\phi(s_{2}))\right|\right) = \nonumber \\ &=   
    - \int_{0}^{\beta}d\tau\left(\phi'\left(\tau-\frac{\beta}{4}\delta v\right)+\phi'\left(\tau+\frac{\beta}{4}\delta v\right)\right)\cot\left(\frac{\pi}{\beta}\left(\phi\left(\tau+\frac{\beta}{4}\delta v\right)-\phi\left(\tau-\frac{\beta}{4}\delta v\right)\right)\right) . 
\end{align}
We also have
\begin{equation}
    \int_{\mathcal{I}}ds_{1}ds_{2}\phi'\left(s_{1}\right)^{2} = \beta v\int_{0}^{\beta}d\tau\phi'\left(\tau\right)^{2}.
\end{equation}
The $(A)^2+2(A)(B)$ term gives the same contribution, and so we find
\begin{align}\label{summary of (1)^2 + 2(1)(2)+(A)^2+2(A)(B)}
    & \int_{\mathcal{I}}ds_{1}ds_{2}[\left(1\right)^{2}+2\left(1\right)\left(2\right)+(A)^2+2(A)(B)]= -8\frac{\pi^{2}}{\beta}v\int_{0}^{\beta}d\tau\phi'\left(\tau\right)^{2} \nonumber + \\  +& 8\frac{\pi}{\beta}\int_{0}^{\beta}d\tau\left(\phi'\left(\tau+\frac{\beta}{4}\delta v\right)+\phi'\left(\tau+\frac{\beta}{4}\delta v\right)\right)\cot\left(\frac{\pi}{\beta}(\phi\left(\tau+\frac{\beta}{4}\delta v\right)-\phi\left(\tau-\frac{\beta}{4}\delta v\right)\right) ,
\end{align}
with the leading order contribution to the action
\begin{equation}
    \delta I \supset - \frac{1}{4\lambda \bJ} \int_0^{\beta} d\tau (\frac{2\pi}{\beta})^2 \phi'^2 + \mathcal{O}(\delta v^2).
\end{equation}

\paragraph{$\left(2\right)^{2}$ and $(B)^2$ terms}
We have 
\begin{equation}
\int_{\mathcal{I}}ds_{1}ds_{2}\left(2\right)^{2}=\int_{\mathcal{I}}ds_{1}ds_{2}\left(\frac{\phi''\left(s_{1}\right)}{\phi'\left(s_{1}\right)}\right)^{2} = \beta v\int_{0}^{\beta}d\tau\left(\frac{\phi''\left(\tau\right)}{\phi'\left(\tau\right)}\right)^{2} ,
\end{equation}
where the last equality is using the parallelogram reordering and $\beta$--periodicity. The $(B)^2$ term gives the same contribution, and the total contribution of $(2)^2+(B)^2$ to the action is 
\begin{equation}
    \delta I \supset \frac{1}{4\lambda \bJ} \int_0^{\beta} d\tau (\frac{\phi''}{\phi'})^2 + \mathcal{O}(\delta v^2).
\end{equation}

\paragraph{End result} Summarizing all the terms above, we get that the kinetic term is
\begin{equation}
    I_{kin}[g_{\phi}]-I_{kin}[g] = \frac{5}{12\lambda \bJ}\int_0^{\beta} d\tau [(\frac{\phi''}{\phi'})^2 -(\frac{2\pi}{\beta}\phi')^2+(\frac{2\pi}{\beta})^2]+ \mathcal{O}(\delta v^2).
\end{equation}

\section{Saddles from exact DSSYK spectrum}
\label{app:Other saddles from DSSYK}
The additional saddles described in the main text can be observed from the exact partition function and density of states in DSSYK. The derivation is similar to the one that appeared in \cite{Blommaert:2024whf}, which appeared as we were finalizing the current work. 

\begin{figure}
    \centering
    \includegraphics[width=0.6\linewidth]{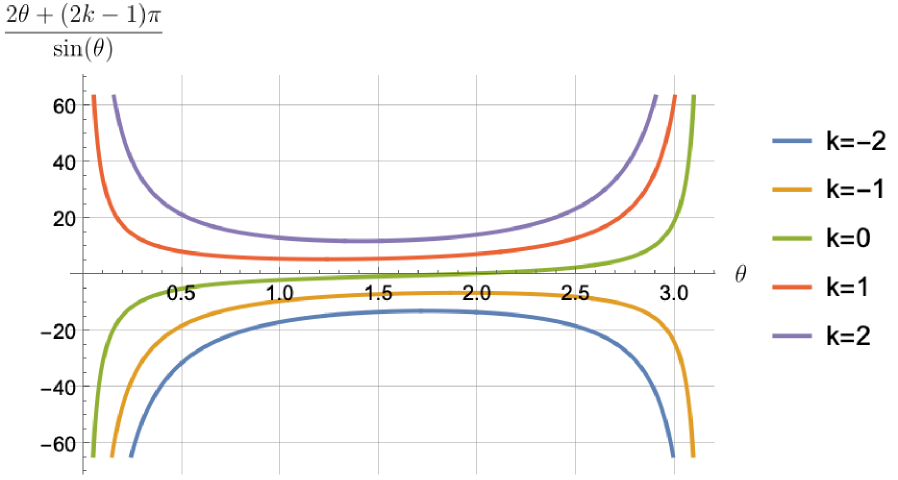}
    \caption{The function $\frac{2\theta+\left(2k-1\right)\pi}{\sin\left(\theta\right)}$ for different values of $k$.}
    \label{fig:saddles theta}
\end{figure}

The exact DSSYK partition function can be written as \cite{Berkooz:2018qkz, Berkooz:2024lgq,Verlinde:2024znh}
\begin{equation}
    Z\left(\beta\right) = \int_{0}^{\pi}d\theta\,\rho\left(\theta\right)e^{-\frac{2\beta\cos\left(\theta\right)}{\sqrt{\lambda\left(1-q\right)}}} \,, \quad \rho(\theta) = \frac{\sin\left(\theta\right)}{\pi q^{1/8}}\sqrt{\frac{2\pi}{\lambda}}\sum_{k=-\infty}^{\infty}\left(-1\right)^{k}e^{-\frac{2}{\lambda}\left(\theta+\pi\left(k-\frac{1}{2}\right)\right)^{2}}\,.
\end{equation}
In the limit $\lambda \to 0$ each summand can then be evaluated via a saddle point approximation. For the $k$-th term in the sum, the saddle point equations are
\begin{equation}
    \label{eq:saddle point equations theta}
    \frac{2\theta+\left(2k-1\right)\pi}{\sin\left(\theta\right)} = \beta\bJ\,,\qquad\theta\in\left[0,\pi\right] \,,
\end{equation}
These equations have various solutions. In Figure~\ref{fig:saddles theta} we plot the left hand side of the equation. Evidently, for positive\footnote{Note that for negative $\beta\bJ$ the situation is reversed. Since this is a system with a bounded spectrum and the density of states decreases starting from some energy, there is some sense in discussing its properties at negative temperatures.} and large enough $\beta\bJ$, there are two solutions for positive $k$, one solution for $k=0$ and no solutions for negative $k$.
We will find it convenient to introduce the variables $v_k$,
\begin{equation}
    \theta = \left(\frac{1}{2}-k\right)\pi+\left(-1\right)^{k}\frac{\pi v_{k}}{2} \,,
\end{equation}
which turns the equations into 
\begin{equation}
    \frac{\pi v_{k}}{\cos\left(\frac{\pi v_{k}}{2}\right)} = \beta\bJ \,, \qquad v_{k} \in \left[2(-1)^k k - 1, 2(-1)^k k + 1\right]\,.
\end{equation}
The contribution of each saddle to the partition function is (up to $O(\lambda^0))$
\begin{equation}
    Z_k = \frac{\cos\left(\frac{\pi v_{k}}{2}\right)}{\sqrt{1+\frac{\pi v_{k}}{2}\tan\left(\frac{\pi v_{k}}{2}\right)}}\exp\left[-\frac{1}{\lambda}\left(\frac{\pi^{2}v_{k}^{2}}{2}-2\pi v_{k}\tan\left(\frac{\pi v_{k}}{2}\right)\right)+\frac{1}{2}\pi v_{k}\tan\left(\frac{\pi v_{k}}{2}\right)\right].
\end{equation}

At low temperatures\footnote{Here we assume positive temperatures. For negative temperatures the same analysis can be carries out, with the viable solutions having non-negative $k$.}, $\beta\bJ\gg1$, there is a single solution to these equations when $k=0$, and two solutions when $k>0$, the first close to the lower edge of the spectrum and the second close to its upper edge,
\begin{align}
    \theta^{\text{lower}} &= \pi-\frac{\pi}{\beta\bJ}\left(1+2k\right)+\frac{2\pi}{(\beta\bJ)^{2}}\left(1+2k\right)-\frac{4(1+2k)\pi + \frac{\pi^{3}}{6} \left(1+2k\right)^{3}}{(\beta\bJ)^{3}}+\cdots\,, \\
    \theta^{\text{upper}} &= -\frac{\pi}{\beta\bJ}(1-2k)-\frac{2\pi}{(\beta\bJ)^{2}}(1-2k)-\frac{4\left(1-2k\right)\pi+\frac{\pi^{3}}{6}\left(1-2k\right)^{3}}{(\beta\bJ)^{3}}+\cdots \,, 
\end{align}
where the first applies for $k\ge0$ and the second for $k>0$. In terms of the $v_k$ variables gives, for the lower and upper edges respectively,
\begin{align}
    v_{k}^{\text{lower}} &= \left(-1\right)^{k}\left(1+2k\right)\left(1-\frac{2}{\beta\bJ}+\frac{4}{\left(\beta\bJ\right)^{2}}-\frac{8+\frac{\pi^{2}}{3}\left(1+2k\right)^{2}}{\left(\beta\bJ\right)^{3}}\right) \,, \\
    v_{k}^{\text{upper}} &= \left(-1\right)^{k+1}\left(1-2k\right)\left(1+\frac{2}{\beta\bJ}+\frac{4}{\left(\beta\bJ\right)^{2}}+\frac{8+\frac{\pi^{2}}{3}\left(1-2k\right)^{2}}{(\beta\bJ)^{3}}\right) \,.
\end{align}
The contributions of these saddles to the partition function is then 
\begin{align}
    Z_k^{\rm lower} &= \frac{\sqrt{2}\left(1+2k\right)\pi}{\left(\beta\bJ\right)^{3/2}}\exp\left[\phantom{-}\left(\frac{2}{\lambda}+\frac{1}{2}\right)\beta\bJ-\frac{\left(1+2k\right)^{2}\pi^{2}}{2\lambda}+\frac{\left(1+2k\right)^{2}\pi^{2}}{\lambda\beta\bJ}\right] \,, \qquad k\ge 0\,, \\ 
    Z_k^{\rm upper} &= \frac{\sqrt{2}\left(1-2k\right)\pi}{\left(-\beta\bJ\right)^{3/2}}\exp\left[-\left(\frac{2}{\lambda}+\frac{1}{2}\right)\beta\bJ-\frac{\left(1-2k\right)^{2}\pi^{2}}{2\lambda}-\frac{\left(1-2k\right)^{2}\pi^{2}}{\lambda\beta\bJ}\right] \,, \qquad k > 0 \,.
\end{align}
Evidently, the $k=0$ term dominates over the others in the sum. These contributions are exactly the ones associated to Schwarzians with conical defects, as described in Section~\ref{sec: other saddles}.

\bibliographystyle{JHEP}
\bibliography{SYK_and_AdS2}

\providecommand{\href}[2]{#2}\begingroup\raggedright\begin{thebibliography}{10}

\bibitem{sachdev1993}
S.~Sachdev and J.~Ye, \emph{Gapless spin-fluid ground state in a random quantum heisenberg magnet}, \href{https://doi.org/10.1103/PhysRevLett.70.3339}{\emph{Phys. Rev. Lett.} {\bfseries 70} (1993) 3339}.

\bibitem{sachdev2010}
S.~Sachdev, \emph{Holographic metals and the fractionalized fermi liquid}, \href{https://doi.org/10.1103/PhysRevLett.105.151602}{\emph{Phys. Rev. Lett.} {\bfseries 105} (2010) 151602}.

\bibitem{kitaev2015simple}
A.~Kitaev, \emph{A simple model of quantum holography (part 2)}, {\emph{Entanglement in Strongly-Correlated Quantum Matter} (2015) 38}.

\bibitem{Maldacena:2016hyu}
J.~Maldacena and D.~Stanford, \emph{{Remarks on the Sachdev-Ye-Kitaev model}}, \href{https://doi.org/10.1103/PhysRevD.94.106002}{\emph{Phys. Rev. D} {\bfseries 94} (2016) 106002} [\href{https://arxiv.org/abs/1604.07818}{{\ttfamily 1604.07818}}].

\bibitem{Maldacena:2015waa}
J.~Maldacena, S.H.~Shenker and D.~Stanford, \emph{{A bound on chaos}}, \href{https://doi.org/10.1007/JHEP08(2016)106}{\emph{JHEP} {\bfseries 08} (2016) 106} [\href{https://arxiv.org/abs/1503.01409}{{\ttfamily 1503.01409}}].

\bibitem{garcia2016}
A.M.~Garc\'ia-Garc\'ia and J.J.M.~Verbaarschot, \emph{Spectral and thermodynamic properties of the sachdev-ye-kitaev model}, \href{https://doi.org/10.1103/PhysRevD.94.126010}{\emph{Phys. Rev. D} {\bfseries 94} (2016) 126010}.

\bibitem{Cotler:2016fpe}
J.S.~Cotler, G.~Gur-Ari, M.~Hanada, J.~Polchinski, P.~Saad, S.H.~Shenker et~al., \emph{{Black Holes and Random Matrices}}, \href{https://doi.org/10.1007/JHEP05(2017)118}{\emph{JHEP} {\bfseries 05} (2017) 118} [\href{https://arxiv.org/abs/1611.04650}{{\ttfamily 1611.04650}}].

\bibitem{you2017sachdev}
Y.-Z.~You, A.W.~Ludwig and C.~Xu, \emph{Sachdev-ye-kitaev model and thermalization on the boundary of many-body localized fermionic symmetry-protected topological states}, {\emph{Physical Review B} {\bfseries 95} (2017) 115150}.

\bibitem{Maldacena:2016upp}
J.~Maldacena, D.~Stanford and Z.~Yang, \emph{{Conformal symmetry and its breaking in two dimensional Nearly Anti-de-Sitter space}}, \href{https://doi.org/10.1093/ptep/ptw124}{\emph{PTEP} {\bfseries 2016} (2016) 12C104} [\href{https://arxiv.org/abs/1606.01857}{{\ttfamily 1606.01857}}].

\bibitem{Saad:2018bqo}
P.~Saad, S.H.~Shenker and D.~Stanford, \emph{{A semiclassical ramp in SYK and in gravity}},  \href{https://arxiv.org/abs/1806.06840}{{\ttfamily 1806.06840}}.

\bibitem{Maldacena:2018lmt}
J.~Maldacena and X.-L.~Qi, \emph{{Eternal traversable wormhole}},  \href{https://arxiv.org/abs/1804.00491}{{\ttfamily 1804.00491}}.

\bibitem{Goel:2018ubv}
A.~Goel, H.T.~Lam, G.J.~Turiaci and H.~Verlinde, \emph{{Expanding the Black Hole Interior: Partially Entangled Thermal States in SYK}}, \href{https://doi.org/10.1007/JHEP02(2019)156}{\emph{JHEP} {\bfseries 02} (2019) 156} [\href{https://arxiv.org/abs/1807.03916}{{\ttfamily 1807.03916}}].

\bibitem{Jensen:2016pah}
K.~Jensen, \emph{{Chaos in AdS$_2$ Holography}}, \href{https://doi.org/10.1103/PhysRevLett.117.111601}{\emph{Phys. Rev. Lett.} {\bfseries 117} (2016) 111601} [\href{https://arxiv.org/abs/1605.06098}{{\ttfamily 1605.06098}}].

\bibitem{Polchinski:2016xgd}
J.~Polchinski and V.~Rosenhaus, \emph{{The Spectrum in the Sachdev-Ye-Kitaev Model}}, \href{https://doi.org/10.1007/JHEP04(2016)001}{\emph{JHEP} {\bfseries 04} (2016) 001} [\href{https://arxiv.org/abs/1601.06768}{{\ttfamily 1601.06768}}].

\bibitem{Sarosi:2017ykf}
G.~S\'arosi, \emph{{AdS$_{2}$ holography and the SYK model}}, \href{https://doi.org/10.22323/1.323.0001}{\emph{PoS} {\bfseries Modave2017} (2018) 001} [\href{https://arxiv.org/abs/1711.08482}{{\ttfamily 1711.08482}}].

\bibitem{Blommaert:2024whf}
A.~Blommaert, A.~Levine, T.G.~Mertens, J.~Papalini and K.~Parmentier, \emph{{An entropic puzzle in periodic dilaton gravity and DSSYK}},  \href{https://arxiv.org/abs/2411.16922}{{\ttfamily 2411.16922}}.

\bibitem{Blommaert:2024ydx}
A.~Blommaert, T.G.~Mertens and J.~Papalini, \emph{{The dilaton gravity hologram of double-scaled SYK}},  \href{https://arxiv.org/abs/2404.03535}{{\ttfamily 2404.03535}}.

\bibitem{Trunin:2020vwy}
D.A.~Trunin, \emph{{Pedagogical introduction to the Sachdev\textendash{}Ye\textendash{}Kitaev model and two-dimensional dilaton gravity}}, \href{https://doi.org/10.3367/UFNe.2020.06.038805}{\emph{Usp. Fiz. Nauk} {\bfseries 191} (2021) 225} [\href{https://arxiv.org/abs/2002.12187}{{\ttfamily 2002.12187}}].

\bibitem{erdHos2014phase}
L.~Erd\H{o}s and D.~Schr\"oder, \emph{{Phase Transition in the Density of States of Quantum Spin Glasses}}, \href{https://doi.org/10.1007/s11040-014-9164-3}{\emph{Math. Phys. Anal. Geom.} {\bfseries 17} (2014) 441} [\href{https://arxiv.org/abs/1407.1552}{{\ttfamily 1407.1552}}].

\bibitem{Berkooz:2018jqr}
M.~Berkooz, M.~Isachenkov, V.~Narovlansky and G.~Torrents, \emph{{Towards a full solution of the large N double-scaled SYK model}}, \href{https://doi.org/10.1007/JHEP03(2019)079}{\emph{JHEP} {\bfseries 03} (2019) 079} [\href{https://arxiv.org/abs/1811.02584}{{\ttfamily 1811.02584}}].

\bibitem{Berkooz:2018qkz}
M.~Berkooz, P.~Narayan and J.~Simon, \emph{{Chord diagrams, exact correlators in spin glasses and black hole bulk reconstruction}}, \href{https://doi.org/10.1007/JHEP08(2018)192}{\emph{JHEP} {\bfseries 08} (2018) 192} [\href{https://arxiv.org/abs/1806.04380}{{\ttfamily 1806.04380}}].

\bibitem{Berkooz:2024lgq}
M.~Berkooz and O.~Mamroud, \emph{{A Cordial Introduction to Double Scaled SYK}},  \href{https://arxiv.org/abs/2407.09396}{{\ttfamily 2407.09396}}.

\bibitem{Lin:2022rbf}
H.W.~Lin, \emph{{The bulk Hilbert space of double scaled SYK}}, \href{https://doi.org/10.1007/JHEP11(2022)060}{\emph{JHEP} {\bfseries 11} (2022) 060} [\href{https://arxiv.org/abs/2208.07032}{{\ttfamily 2208.07032}}].

\bibitem{Berkooz:2024evs}
M.~Berkooz, N.~Brukner, Y.~Jia and O.~Mamroud, \emph{{From Chaos to Integrability in Double Scaled Sachdev-Ye-Kitaev Model via a Chord Path Integral}}, \href{https://doi.org/10.1103/PhysRevLett.133.221602}{\emph{Phys. Rev. Lett.} {\bfseries 133} (2024) 221602} [\href{https://arxiv.org/abs/2403.01950}{{\ttfamily 2403.01950}}].

\bibitem{Berkooz:2024ofm}
M.~Berkooz, N.~Brukner, Y.~Jia and O.~Mamroud, \emph{{Path integral for chord diagrams and chaotic-integrable transitions in double scaled SYK}}, \href{https://doi.org/10.1103/PhysRevD.110.106015}{\emph{Phys. Rev. D} {\bfseries 110} (2024) 106015} [\href{https://arxiv.org/abs/2403.05980}{{\ttfamily 2403.05980}}].

\bibitem{Stanford-talk-kitp}
D.~Stanford, \emph{Talk at kitp},  2018.

\bibitem{Lin:2023trc}
H.W.~Lin and D.~Stanford, \emph{{A symmetry algebra in double-scaled SYK}},  \href{https://arxiv.org/abs/2307.15725}{{\ttfamily 2307.15725}}.

\bibitem{Goel:2023svz}
A.~Goel, V.~Narovlansky and H.~Verlinde, \emph{{Semiclassical geometry in double-scaled SYK}},  \href{https://arxiv.org/abs/2301.05732}{{\ttfamily 2301.05732}}.

\bibitem{Kitaev:2017awl}
A.~Kitaev and S.J.~Suh, \emph{{The soft mode in the Sachdev-Ye-Kitaev model and its gravity dual}}, \href{https://doi.org/10.1007/JHEP05(2018)183}{\emph{JHEP} {\bfseries 05} (2018) 183} [\href{https://arxiv.org/abs/1711.08467}{{\ttfamily 1711.08467}}].

\bibitem{Streicher:2019wek}
A.~Streicher, \emph{{SYK Correlators for All Energies}}, \href{https://doi.org/10.1007/JHEP02(2020)048}{\emph{JHEP} {\bfseries 02} (2020) 048} [\href{https://arxiv.org/abs/1911.10171}{{\ttfamily 1911.10171}}].

\bibitem{Choi:2019bmd}
C.~Choi, M.~Mezei and G.~S\'arosi, \emph{{Exact four point function for large $q$ SYK from Regge theory}}, \href{https://doi.org/10.1007/JHEP05(2021)166}{\emph{JHEP} {\bfseries 05} (2021) 166} [\href{https://arxiv.org/abs/1912.00004}{{\ttfamily 1912.00004}}].

\bibitem{Mukhametzhanov:2023tcg}
B.~Mukhametzhanov, \emph{{Large p SYK from chord diagrams}},  \href{https://arxiv.org/abs/2303.03474}{{\ttfamily 2303.03474}}.

\bibitem{Okuyama:2023bch}
K.~Okuyama and K.~Suzuki, \emph{{Correlators of double scaled SYK at one-loop}}, \href{https://doi.org/10.1007/JHEP05(2023)117}{\emph{JHEP} {\bfseries 05} (2023) 117} [\href{https://arxiv.org/abs/2303.07552}{{\ttfamily 2303.07552}}].

\bibitem{MezeiBucca}
M.~Bucca and M.~Mezei, \emph{{Nonlinear soft mode action for the large-$p$ SYK model}},  \href{https://arxiv.org/abs/2412.14799}{{\ttfamily 2412.14799}}.

\bibitem{Garcia-Garcia:2018fns}
A.M.~Garc\'\i{}a-Garc\'\i{}a, Y.~Jia and J.J.M.~Verbaarschot, \emph{{Exact moments of the Sachdev-Ye-Kitaev model up to order $1/N^2$}}, \href{https://doi.org/10.1007/JHEP04(2018)146}{\emph{JHEP} {\bfseries 04} (2018) 146} [\href{https://arxiv.org/abs/1801.02696}{{\ttfamily 1801.02696}}].

\bibitem{Garcia-Garcia:2017pzl}
A.M.~Garc\'\i{}a-Garc\'\i{}a and J.J.M.~Verbaarschot, \emph{{Analytical Spectral Density of the Sachdev-Ye-Kitaev Model at finite N}}, \href{https://doi.org/10.1103/PhysRevD.96.066012}{\emph{Phys. Rev. D} {\bfseries 96} (2017) 066012} [\href{https://arxiv.org/abs/1701.06593}{{\ttfamily 1701.06593}}].

\bibitem{Bagrets:2016cdf}
D.~Bagrets, A.~Altland and A.~Kamenev, \emph{{Sachdev\textendash{}Ye\textendash{}Kitaev model as Liouville quantum mechanics}}, \href{https://doi.org/10.1016/j.nuclphysb.2016.08.002}{\emph{Nucl. Phys. B} {\bfseries 911} (2016) 191} [\href{https://arxiv.org/abs/1607.00694}{{\ttfamily 1607.00694}}].

\bibitem{Stanford:2017thb}
D.~Stanford and E.~Witten, \emph{{Fermionic Localization of the Schwarzian Theory}}, \href{https://doi.org/10.1007/JHEP10(2017)008}{\emph{JHEP} {\bfseries 10} (2017) 008} [\href{https://arxiv.org/abs/1703.04612}{{\ttfamily 1703.04612}}].

\bibitem{Moitra:2021uiv}
U.~Moitra, S.K.~Sake and S.P.~Trivedi, \emph{{Jackiw-Teitelboim gravity in the second order formalism}}, \href{https://doi.org/10.1007/JHEP10(2021)204}{\emph{JHEP} {\bfseries 10} (2021) 204} [\href{https://arxiv.org/abs/2101.00596}{{\ttfamily 2101.00596}}].

\bibitem{Mertens:2022irh}
T.G.~Mertens and G.J.~Turiaci, \emph{{Solvable models of quantum black holes: a review on Jackiw\textendash{}Teitelboim gravity}}, \href{https://doi.org/10.1007/s41114-023-00046-1}{\emph{Living Rev. Rel.} {\bfseries 26} (2023) 4} [\href{https://arxiv.org/abs/2210.10846}{{\ttfamily 2210.10846}}].

\bibitem{Mertens:2019tcm}
T.G.~Mertens and G.J.~Turiaci, \emph{{Defects in Jackiw-Teitelboim Quantum Gravity}}, \href{https://doi.org/10.1007/JHEP08(2019)127}{\emph{JHEP} {\bfseries 08} (2019) 127} [\href{https://arxiv.org/abs/1904.05228}{{\ttfamily 1904.05228}}].

\bibitem{Gu:2016oyy}
Y.~Gu, X.-L.~Qi and D.~Stanford, \emph{{Local criticality, diffusion and chaos in generalized Sachdev-Ye-Kitaev models}}, \href{https://doi.org/10.1007/JHEP05(2017)125}{\emph{JHEP} {\bfseries 05} (2017) 125} [\href{https://arxiv.org/abs/1609.07832}{{\ttfamily 1609.07832}}].

\bibitem{Altland:2019lne}
A.~Altland, D.~Bagrets and A.~Kamenev, \emph{{Quantum Criticality of Granular Sachdev-Ye-Kitaev Matter}}, \href{https://doi.org/10.1103/PhysRevLett.123.106601}{\emph{Phys. Rev. Lett.} {\bfseries 123} (2019) 106601} [\href{https://arxiv.org/abs/1903.09491}{{\ttfamily 1903.09491}}].

\bibitem{Jiang:2019pam}
J.~Jiang and Z.~Yang, \emph{{Thermodynamics and Many Body Chaos for generalized large q SYK models}}, \href{https://doi.org/10.1007/JHEP08(2019)019}{\emph{JHEP} {\bfseries 08} (2019) 019} [\href{https://arxiv.org/abs/1905.00811}{{\ttfamily 1905.00811}}].

\bibitem{Anninos:2020cwo}
D.~Anninos and D.A.~Galante, \emph{{Constructing AdS$_{2}$ flow geometries}}, \href{https://doi.org/10.1007/JHEP02(2021)045}{\emph{JHEP} {\bfseries 02} (2021) 045} [\href{https://arxiv.org/abs/2011.01944}{{\ttfamily 2011.01944}}].

\bibitem{Anninos:2022qgy}
D.~Anninos, D.A.~Galante and S.U.~Sheorey, \emph{{Renormalisation Group Flows of the SYK Model}},  \href{https://arxiv.org/abs/2212.04944}{{\ttfamily 2212.04944}}.

\bibitem{Gao:2024lem}
P.~Gao, H.~Lin and C.~Peng, \emph{{D-commuting SYK model: building quantum chaos from integrable blocks}},  \href{https://arxiv.org/abs/2411.12806}{{\ttfamily 2411.12806}}.

\bibitem{Almheiri:2024xtw}
A.~Almheiri, A.~Goel and X.-Y.~Hu, \emph{{Quantum gravity of the Heisenberg algebra}}, \href{https://doi.org/10.1007/JHEP08(2024)098}{\emph{JHEP} {\bfseries 08} (2024) 098} [\href{https://arxiv.org/abs/2403.18333}{{\ttfamily 2403.18333}}].

\bibitem{Milekhin:2021cou}
A.~Milekhin, \emph{{Non-local reparametrization action in coupled Sachdev-Ye-Kitaev models}}, \href{https://doi.org/10.1007/JHEP12(2021)114}{\emph{JHEP} {\bfseries 12} (2021) 114} [\href{https://arxiv.org/abs/2102.06647}{{\ttfamily 2102.06647}}].

\bibitem{Milekhin:2021sqd}
A.~Milekhin, \emph{{Coupled Sachdev-Ye-Kitaev models without Schwartzian dominance}},  \href{https://arxiv.org/abs/2102.06651}{{\ttfamily 2102.06651}}.

\bibitem{Shenker:2013pqa}
S.H.~Shenker and D.~Stanford, \emph{{Black holes and the butterfly effect}}, \href{https://doi.org/10.1007/JHEP03(2014)067}{\emph{JHEP} {\bfseries 03} (2014) 067} [\href{https://arxiv.org/abs/1306.0622}{{\ttfamily 1306.0622}}].

\bibitem{Verlinde:2024znh}
H.~Verlinde, \emph{{Double-scaled SYK, Chords and de Sitter Gravity}},  \href{https://arxiv.org/abs/2402.00635}{{\ttfamily 2402.00635}}.

\end{thebibliography}\endgroup

\end{document}